\documentclass[12pt,a4paper]{article}

\usepackage{fullpage}
\usepackage{amsfonts}
\usepackage{amssymb}
\usepackage{amsmath}
\usepackage{dsfont}
\usepackage{mathtools}
\usepackage{bbm} 
\usepackage{multirow}
\usepackage{latexsym}
\usepackage{mcite}
\usepackage{cite}
\usepackage{xspace} 
\usepackage{xcolor}

\usepackage{hyperref}

\numberwithin{equation}{section}
\makeatletter
\g@addto@macro\bfseries{\boldmath}
\makeatother

\def\beq{\begin{equation}}
\def\eeq{\end{equation}}
\def\beqa{\begin{eqnarray}}
\def\eeqa{\end{eqnarray}}

\def\a{{\alpha}}
\def\b{{\beta}}

\def\g{{\gamma}}

\def\d{{\delta}}
\def\eps{{\varepsilon}}

\def\m{{\mu}}
\def\n{{\nu}}

\def\dm{{\dot{\mu}}}

\def\bfone{\relax{\rm 1\kern-.35em 1}}

\newcommand{\nt}{n_{\rm t}}

\newcommand{\be}{\begin{equation}}
\newcommand{\ee}{\end{equation}}
\newcommand{\ben}{\begin{displaymath}}
\newcommand{\een}{\end{displaymath}}
\newcommand{\bea}{\begin{eqnarray}}
\newcommand{\eea}{\end{eqnarray}}

\newcommand{\bean}{\begin{eqnarray*}}
\newcommand{\eean}{\end{eqnarray*}}

\newenvironment{formula}%
{\begin{equation}\begin{aligned}\relax}%
{\end{aligned}\end{equation}\ignorespacesafterend}

\DeclareMathOperator{\Tr}{Tr}

\newcommand{\nv}{{n_{\rm v}}}

\newcommand{\cC}{{\mathcal{C}}}
\newcommand{\cD}{{\mathcal{D}}}

\newcommand{\cF}{{\mathcal{F}}}

\newcommand{\cH}{{\mathcal{H}}}

\newcommand{\cK}{{\mathcal{K}}}
\newcommand{\cL}{{\mathcal{L}}}
\newcommand{\cM}{{\mathcal{M}}}
\newcommand{\cN}{{\mathcal{N}}}

\newcommand{\cP}{{\mathcal{P}}}
\newcommand{\cQ}{{\mathcal{Q}}}
\newcommand{\cR}{{\mathcal{R}}}
\newcommand{\cS}{{\mathcal{S}}}
\newcommand{\cT}{{\mathcal{T}}}

\newcommand{\bbR}{{\mathbb{R}}}

\newcommand{\bbZ}{{\mathbb{Z}}}

\newcommand{\Edone}{\ensuremath{\mathrm{E}_{d+1(d+1)}}\xspace}

\newcommand{\Eseven}{\ensuremath{\mathrm{E}_{7(7)}}\xspace}

\def\SL(#1){\ensuremath{\mathrm{SL}(#1)}}
\def\PSL(#1){\ensuremath{\mathrm{PSL}(#1)}}
\def\GL(#1){\ensuremath{\mathrm{GL}(#1)}}
\def\Sp(#1){\ensuremath{\mathrm{Sp}(#1)}}
\def\USp(#1){\ensuremath{\mathrm{USp}(#1)}}
\def\SU(#1){\ensuremath{\mathrm{SU}(#1)}}
\def\PSU(#1){\ensuremath{\mathrm{PSU}(#1)}}
\def\SUs(#1){\ensuremath{\mathrm{SU}^*\kern-1pt(#1)}}
\def\U(#1){\ensuremath{\mathrm{U}(#1)}}
\def\SO(#1){\ensuremath{\mathrm{SO}(#1)}}
\def\PSO(#1){\ensuremath{\mathrm{PSO}(#1)}}
\def\SOs(#1){\ensuremath{\mathrm{SO}^*\kern-1pt(#1)}}
\def\SOp(#1){\ensuremath{\mathrm{SO}^+\kern-1pt(#1)}}
\def\O(#1){\ensuremath{\mathrm{O}(#1)}}
\def\Op(#1){\ensuremath{\mathrm{O}^+\kern-1pt(#1)}}
\def\CSO(#1){\ensuremath{\mathrm{CSO}(#1)}}
\def\CSOs(#1){\ensuremath{\mathrm{CSO}^*\kern-1pt(#1)}}
\def\ISO(#1){\ensuremath{\mathrm{ISO}(#1)}}

\def\fsl(#1){\ensuremath{\mathfrak{sl}(#1)}}
\def\su(#1){\ensuremath{\mathfrak{su}(#1)}}
\def\sus(#1){\ensuremath{\mathfrak{su}^*\!(#1)}}
\def\sp(#1){\ensuremath{\mathfrak{sp}(#1)}}
\def\so(#1){\ensuremath{\mathfrak{so}(#1)}}
\def\sos(#1){\ensuremath{\mathfrak{so}^*\!(#1)}}
\def\cso(#1){\ensuremath{\mathfrak{cso}(#1)}}
\def\csos(#1){\ensuremath{\mathfrak{cso}^*\!(#1)}}
\def\iso(#1){\ensuremath{\mathfrak{iso}(#1)}}

\begin{document}


\begin{titlepage}

\vskip -10mm

\begin{flushright} 
\small
UUITP-31/16\\
YITP-16-137
\end{flushright}
\vspace*{1cm}

\bigskip

\begin{center}

\vskip -10mm

{\Large\bf Double Field Theory at SL(2) angles} \\[6mm]

\vskip 0.5cm

\textbf{Franz~Ciceri}$\,^{1}$,\ \textbf{Giuseppe~Dibitetto}$\,^{2}$,\ \textbf{J.~J.~Fernandez-Melgarejo}$\,^{3,4}$,\\[1ex] \textbf{Adolfo~Guarino}$\,^{5}$ and \textbf{Gianluca~Inverso}$\,^{6}$

\vskip 8mm

$\,^{1}${\em Nikhef Theory Group, Science Park 105, 1098 XG Amsterdam, The Netherlands}\\[1.5mm]
$\,^{2}${\em Institutionen f\"or fysik och astronomi, University of Uppsala, \\
Box 803, SE-751 08 Uppsala, Sweden}\\[1.5mm]
$\,^{3}${\em Yukawa Institute for Theoretical Physics, Kyoto University, Kyoto 606-8502 Japan}\\[1.5mm]
$\,^{4}${\em Jefferson Physical Laboratory, Harvard University, Cambridge, MA 02138, USA}\\[1.5mm]
$\,^{5}${\em Physique Th\'eorique et Math\'ematique, Universit\'e Libre de Bruxelles and \\International Solvay Institutes, ULB-Campus Plaine CP231, B-1050 Brussels, Belgium}\\[1.5mm]
$\,^{6}${\em Center for Mathematical Analysis, Geometry and Dynamical Systems, \\
Department of Mathematics, Instituto Superior Técnico,\\
Universidade de Lisboa, Av. Rovisco Pais, 1049-001 Lisboa, Portugal}

\vskip 25pt 

\vskip 0.8cm

\end{center}

\vskip .7cm
\begin{center}
{\bf Abstract}\\[3ex]
\begin{minipage}{13cm}
\small

An extended field theory is presented that captures the full $\SL(2)\times\O(6,6+n)$ duality group of four-dimensional half-maximal supergravities. The theory has section constraints whose two inequivalent solutions correspond to minimal $D=10$ supergravity and chiral half-maximal $D=6$ supergravity, respectively coupled to vector and tensor multiplets. The relation with $\O(6,6+n)$ (\mbox{heterotic}) double field theory is thoroughly discussed. Non-Abelian interactions as well as background fluxes are captured by a deformation of the generalised diffeomorphisms. Finally, making use of the $\SL(2)$ duality structure, it is shown how to generate gaugings with non-trivial de Roo--Wagemans angles via generalised Scherk--Schwarz ans\"atze. Such gaugings allow for moduli stabilisation including the $\textrm{SL}(2)$ dilaton.

\end{minipage}
\end{center}

\vfill
\end{titlepage}

\tableofcontents



\section{Introduction and outlook}

Recently, exceptional generalised geometries \cite{Coimbra:2011ky,Coimbra:2012af} and exceptional field theories (EFT) \cite{Hohm:2013pua,Hohm:2013vpa,Hohm:2013uia,Hohm:2014fxa} have been the stage of intense activity. These frameworks capture the degrees of freedom and gauge symmetries of maximal supergravities in a way that makes their exceptional $\,\Edone\,$ structures manifest, mirroring how $\,\textrm{O}(d,d+n)\,$ structures are reproduced in generalised geometry and double field theory (DFT) \cite{Siegel:1993xq,Siegel:1993th,Hull:2009mi,Hohm:2010jy}. Not only do these frameworks give a better understanding of how duality structures determine the geometrical and physical properties of maximal supergravities, but they also provide the necessary tools to study solutions, dimensional reductions and consistent truncations on non-trivial backgrounds \cite{Hohm:2014qga,Baguet:2015sma}.

While most of the recent research has been focused on exploiting the manifest duality structures of DFT and EFT, it must be possible to introduce generalised geometries and extended field theories associated to groups different from those of the $\,{\textrm{O}(d,d+n)}\,$ and $\,\Edone\,$ series. For instance, several generalised geometries were introduced in \cite{Strickland-Constable:2013xta}, in particular examples based on a $\,\mathrm{Spin}(d,d)\,$ structure. In \cite{Lee:2014mla} it was proven that any $d$-dimensional sphere is (generalised) parallelisable in an appropriate $\,\mathrm{GL}^+(d+1)\,$ generalised geometry. One can look for other relevant structures in the series of duality groups of supergravity theories. A particularly interesting case is the series of duality symmetries of half-maximal supergravities which, for specific dimensions (see Table~\ref{Table:duality_groups}), contains groups larger than the $\,\O(d,d+n)\,$ captured by DFT.\footnote{Interesting results on reproducing (Heterotic) DFT from $\,D=7\,$ EFT have recently appeared in \cite{Malek:2016vsh}.}
One example arises from the reduction of ten-dimensional $\,\cN=1\,$ supergravity coupled to $\,\nv=n\,$ gauge vectors \cite{Chamseddine:1980cp,Bergshoeff:1981um} down to $\,D=4\,$. This yields an $\,\textrm{SL}(2) \times \textrm{O}(6,6+n)\,$ duality group which is larger than the $\,\textrm{O}(6,6+n)\,$ symmetry of DFT. A further reduction to $\,D=3\,$ gives $\,\textrm{O}(8,8+n)\,$ thus containing the $\,\textrm{O}(7,7+n)\,$ captured by DFT. Also notable is the $\,\textrm{O}(5,n_{\rm t})\,$ duality symmetry of $\,\cN=(2,0)\,$ supergravity in six dimensions coupled to $\,n_{\rm t}\,$ tensors \cite{Townsend:1983xt,Awada:1985er,Romans:1986er}. Upon subsequent reduction to $\,D<3\,$, these duality symmetries would become infinite-dimensional reaching up to $\,{\rm D}^{+++}_n\,$ and $\,{\rm B}^{+++}_n\,$ very extended Kac--Moody algebras \cite{Schnakenburg:2004vd,Bergshoeff:2007vb} analogous to the $\,\rm E_{11}\,$ of the maximal supergravities \cite{West:2001as}. It is therefore natural to construct extended field theories based on the duality groups of half-maximal supergravities for $\,D=4\,$ and $\,D=3\,$, in the same fashion as exceptional field theory for the maximal cases \cite{Hohm:2013uia,Hohm:2014fxa}.

In this paper we investigate the $\,D=4\,$ case and construct the extended field theory whose associated duality group is $\,\textrm{SL}(2) \times \textrm{O}(6,6+n)\,$. Apart from the theoretical motivation of understanding the similarities and differences between the latter and the DFT with $\,\textrm{O}(6,6+n)\,$ symmetry, having an enhancement of the duality group with an $\,\textrm{SL}(2)\,$ factor is also phenomenologically relevant. This becomes manifest, for example, when studying the issue of moduli stabilisation in the lower-dimensional gauged supergravities arising from generalised Scherk--Schwarz (SS) reductions of the extended field theories. In particular, generalised SS reductions of DFT down to $\,D=4\,$ can only produce \textit{electric} gaugings of $\,\cN=4\,$ (half-maximal) supergravity, even when allowing for locally non-geometric twists that violate the section constraint \cite{Geissbuhler:2011mx,Grana:2012rr}. Such electric gaugings are subject to the no-go result by de Roo--Wagemans (dRW) \cite{deRoo:1985jh} stating the impossibility of stabilising the $\,\textrm{SL}(2)\,$ dilaton of the $\,\cN=4\,$ theory. A crucial ingredient for stabilising such a scalar in half-maximal $\,D=4\,$ supergravity is the presence of non-trivial $\,\textrm{SL}(2)\,$ angles, known as dRW phases, in the gauge group. In the framework of the embedding tensor which allows to systematically investigate $\,\cN=4\,$ gaugings \cite{Schon:2006kz}, the presence of non-trivial dRW phases requires non-vanishing embedding tensor components which are $\,\textrm{SL}(2)\,$ rotated with respect to each other. Various maximally symmetric solutions compatible with four-dimensional $\,\cN=4\,$ gaugings of this type were discussed in \cite{deRoo:2002jf,deRoo:2003rm}.

\begin{table}[t!] 
\renewcommand{\arraystretch}{1.25}
\begin{center}
\scalebox{0.90}[1]{
\begin{tabular}{|c|c|c|c|c|c|c|}
\hline 
$D$ & Maximal sugra / EFT & Half-maximal sugra  & DFT\\[0.8mm]
\hline 
$9$ & $\mathbb{R}^{+} \times \textrm{SL}(2)$ & $\mathbb{R}^{+} \times \textrm{O}(1,1+n)$ & $\mathbb{R}^{+} \times \textrm{O}(1,1+n)$ \\
$8$ & $\textrm{SL}(2)\times \textrm{SL}(3)$ & $\mathbb{R}^{+} \times \textrm{O}(2,2+n)$ & $\mathbb{R}^{+} \times \textrm{O}(2,2+n)$ \\
$7$ & $\textrm{SL}(5)$ & $\mathbb{R}^{+} \times \textrm{O}(3,3+n)$ & $\mathbb{R}^{+} \times \textrm{O}(3,3+n)$ \\
$6$ & $\textrm{SO}(5,5)$ & $\mathbb{R}^{+} \times \textrm{O}(4,4+n)$ & $\mathbb{R}^{+} \times \textrm{O}(4,4+n)$ \\
$5$ & $\textrm{E}_{6(6)}$ & $\mathbb{R}^{+} \times \textrm{O}(5,5+n)$ & $\mathbb{R}^{+} \times \textrm{O}(5,5+n)$ \\
$4$ & $\textrm{E}_{7(7)}$ & $\textrm{SL}(2) \times \textrm{O}(6,6+n)$ & $\mathbb{R}^{+} \times \textrm{O}(6,6+n)$ \\
$3$ & $\textrm{E}_{8(8)}$ & $\textrm{O}(8,8+n)$ & $\mathbb{R}^{+} \times \textrm{O}(7,7+n)$ \\
\hline 
\end{tabular}
}
\caption{Relevant duality groups in maximal and half-maximal supergravity as well as in extended field theory. Only the non-chiral $\cN=(1,1)$ supergravity in $D=6$ is displayed. 
The $\bbR^+$ factor in the duality structure of DFT is actually a combination of an internal $\bbR^+$ contained in the second column and a trombone rescaling.\label{Table:duality_groups}} 
\end{center}
\end{table}

It thus becomes crucial to have access to the $\,\textrm{SL}(2)\,$ factor of the duality group in the half-maximal extended field theory in order to generate $\,\cN=4\,$ gaugings that may stabilise the moduli upon reduction to a $\,D=4\,$  gauged supergravity. One systematic manner of obtaining $\,{\cN=4}\,$ gaugings at $\,\textrm{SL}(2)\,$ angles is by $\mathbb{Z}_{2}$-truncating gaugings of $\,\cN=8\,$ supergravity \cite{Dibitetto:2011eu} for which moduli stabilisation is known to occur, \textit{e.g.} the $\,\textrm{CSO}(p,q,r)\,$ gaugings ($p+q+r=8$) of maximal supergravity \cite{Hull:1984vg,Hull:1984qz,DallAgata:2011aa,Dibitetto:2012ia}. Some of these gaugings arise from consistent reductions of string/M-theory with fluxes\footnote{See \cite{Hohm:2014qga} (and references therein) for a unified account of electric gaugings, as well as \cite{Guarino:2015jca,Guarino:2015vca,Guarino:2015tja} for dyonic ones \cite{DallAgata:2012bb,DallAgata:2014ita,Guarino:2015qaa}.}, and without extra spacetime-filling sources. However, from a phenomenological point of view, these gaugings are not yet fully satisfactory because they cannot arise from \mbox{\textit{compactifications}} (without boundaries) and, at the same time, produce Minkowski or de Sitter (dS) solutions due to the no-go theorem of \cite{Maldacena:2000mw} (see also \cite{Hertzberg:2007wc}). In order to circumvent this no-go theorem, one may add sources (branes, orientifold planes, \mbox{KK-monopoles}, ...) and/or introduce non-geometric fluxes \cite{Shelton:2005cf,Aldazabal:2006up} whose higher-dimensional origin is not yet well understood. The resulting four-dimensional supergravity is no longer compatible with maximal supersymmetry but still can preserve some fraction thereof if the sources and fluxes are judiciously distributed over the internal space. When they are set to preserve $\,\cN=4\,$ supersymmetry, no example of a perturbatively stable dS vacuum in $\,D=4\,$ has been found\footnote{The only examples of stable dS vacua in half-maximal gauged supergravity have recently appeared in $\,D=7\,$ \cite{Dibitetto:2015bia}. In the context of $\,\cN=1\,$ supergravity in $\,D=4\,$ including sources and non-geometric fluxes, the first examples were found in \cite{deCarlos:2009fq,deCarlos:2009qm} and further investigated in \cite{Danielsson:2012by}.}. More strikingly, while $\,\cN=4\,$ gaugings can arise from either reductions of Type~I/Heterotic supergravity \cite{Kaloper:1999yr,deRoo:2005be} or from orientifold reductions of Type~II theories \cite{Roest:2009dq,Dall'Agata:2009gv,Dibitetto:2010rg,Dibitetto:2011gm}, an analysis based on the embedding tensor formulation of gauged supergravities shows that the vast majority of such gaugings lacks a higher-dimensional string/M-theory interpretation. For this reason, much of the recent activity in the field has been directed towards assessing to what extent gaugings induced by non-geometric fluxes may have an extended field theory origin\footnote{An interesting analysis was carried out in \cite{Lee:2015xga} within the context of exceptional generalised geometry and \Eseven-EFT in order to reproduce the family of maximal \SO(8) gaugings in $\,D=4\,$ of \cite{DallAgata:2012bb}, also giving an alternative origin for the family of half-maximal \SO(4) gaugings in $\,D=7\,$ of \cite{Dibitetto:2012rk}.}.

The above discussion motivates us to construct the $\,\SL(2)\times\O(6,6+n)\,$ extended field theory with the aim of obtaining $\,\cN=4\,$ gaugings at non-trivial SL(2) angles upon generalised Scherk--Schwarz reductions to four dimensions. In this extended field theory, an $\,\mathbb{R}^{+} \times \, \textrm{O}(6,6\,+\,n)\,$ symmetry corresponds to the one captured by Heterotic DFT where the internal coordinates are extended to fill the vector $\,\textbf{12 + $n$}\,$ representation. 
To accommodate for the enhanced $\,\SL(2)\,$ factor in the duality group, a further doubling of these coordinates is necessary to fill the $(\textbf{2},\textbf{12 + $n$})$ representation.  
We will refer to this theory as half-maximal extended field theory or $\,\SL(2)$-DFT. 
The algebra of generalised diffeomorphisms follows the general structure described in \cite{Berman:2012vc}. 
Moreover, in order to supplement the $\,\O(6,6+n)\,$ structure with the $\,\SL(2)\,$ one, a hierarchy of tensor fields must be introduced in analogy with that of gauged supergravities and EFT's \cite{deWit:2008ta,deWit:2008gc,Hohm:2013vpa}. 
The SL(2)-DFT is restricted by two section constraints which admit a maximal solution that keeps two internal coordinates and allows to capture a six-dimensional theory with $\,\O(5,\nt)\,$ duality symmetry, matching $\,{\cN=(2,0)}\,$ supergravity in six dimensions coupled to $\,{\nt=5+n}\,$ tensor multiplets. 
An inequivalent maximal solution of the section constraints, unique up to duality transformations, keeps six internal coordinates and thus corresponds to the ten-dimensional half-maximal supergravity coupled to $\,\nv=n\,$ vector multiplets\footnote{We are counting vector multiplets from ten dimensions but the general structure of our results applies also to general $\SL(2)\times\O(6,\tilde n)$ groups. Of course no link to ten dimensions is available when $\tilde n<6$, but the chiral $D=6$ theory is captured for any $\tilde n>0$.}. Importantly, one can also recover the standard formulation of DFT in \cite{Hohm:2013nja} (with four external dimensions) by dualising away certain fields. In this process, no physical degrees of freedom are truncated but $\,\SL(2)\,$ covariance is inevitably lost.
Gauge groups for the $\,\nv=n\,$ ten-dimensional vectors can be accommodated in the same way as in Heterotic DFT \cite{Hohm:2011ex} (see also gauged DFT \cite{Grana:2012rr}).  
In fact, more general deformations are compatible with the ten-dimensional solution of the section constraints.
This is the half-maximal counterpart of the $X$ deformation introduced in \cite{Ciceri:2016dmd} for \Eseven-EFT. 
However, unlike in Heterotic/gauged DFT and $X$-deformed EFT, an additional constraint first mentioned in \cite{Ciceri:2016dmd} plays a prominent role in guaranteeing consistency and restricting the allowed deformations.

Equipped with the SL(2)-DFT, we investigate generalised twisted torus reductions that reproduce $\,\cN=4\,$ gaugings at non-trivial SL(2) angles. More concretely, we find that taking any two instances of DFT reductions to $\,D=7\,$ without warping, they can be assembled into a $D=4$ reduction that violates the section constraints but introduces dRW phases in the final gauge group. As a prominent example of this feature we reproduce families of $\,\textrm{SO}(3)^{(4-p)} \times \textrm{U}(1)^{3p}\,$ gaugings of $\,\cN=4\,$ supergravity with $\,{p=0,...,4}\,$. The case $\,p=0\,$ reproduces the most general family of $\,\SO(4)\times\SO(4)\,$ gaugings of half-maximal supergravity recently classified in \cite{Inverso:2015viq}, in terms of a twisted quadruple torus reduction ($n=0$). These gaugings include as a special case the ones obtained from a $\mathbb{Z}_{2}$-truncation of the one-parameter families of $\,\SO(8)\,$ and $\,\SO(4,4)\,$ gauged maximal supergravities of \cite{DallAgata:2012bb,DallAgata:2014ita}, but also include other $\,\cN=4\,$ gaugings which are not permitted by $\,\cN=8\,$ supersymmetry.

The paper is organised as follows. In section~\ref{sec:SL(2)-DFT} we construct the $\,{\SL(2)\times\O(6,6)}\,$ extended field theory ($\,n=0\,$) as a truncation of $\textrm{E}_{7(7)}$-EFT. We present the generalised Lie derivative, tensor hierarchy and bosonic (pseudo-) action and discuss the solution of the corresponding section constraints. Various checks in the limit of trivial $\textrm{SL}(2)$ phases are performed where the action and generalised Lie derivative reduce to those of standard DFT. We also discuss the embedding of Type II orientifolds within the degrees of freedom of $\textrm{SL}(2)$-DFT and identify the set of physical coordinates in cases which are relevant to the $4+6$ splitting of ten-dimensional Type IIB supergravity. In section~\ref{sec:gauge_vectors} we generalise the results to include $\,2\times n\,$ extra gauge vectors. First we study the Abelian case and then consider non-Abelian deformations of the generalised Lie derivative, both in the gauge and the gravity sectors, and connect them to the embedding tensor of $\,\cN=4\,$ gauged supergravity. In section~\ref{sec:Scherk-Schwarz} we investigate the $\textrm{SL}(2)$-DFT origin of classes of $\,\cN=4\,$ gaugings at SL(2) angles that admit full moduli stabilisation. Finally we collect some technical results in the appendix~\ref{app:Z2-truncation}.

\section{$\SL(2)\times\O(6,6)$ extended field theory}
\label{sec:SL(2)-DFT}

The extended field theory featuring an $\SL(2)\times\O(6,6)$ duality group ($n=0$) can be obtained by modding out the $\textrm{E}_{7(7)}$-EFT by a discrete $\mathbb{Z}_{2}$ subgroup of $\textrm{E}_{7(7)}$. In the supergravity context, the same prescription was applied in \cite{Dibitetto:2011eu} to truncate the four-dimensional maximal supergravity to a half-maximal one coupled to six vector multiplets. In the following we focus on the main results of such a truncation of the $\textrm{E}_{7(7)}$-EFT. The technical details and conventions are gathered in the appendix~\ref{app:Z2-truncation}.

\subsection{Generalised diffeomorphisms}

The $\,\SL(2)\times\O(6,6)\,$ extended field theory lives on an extended space-time that consists of an external space-time with coordinates $\,x^{\mu}\,$ and an internal space with coordinates $\,y^{\alpha M}\,$. The latter sit in the $\,(\textbf{2},\textbf{12})\,$ representation of $\,{\SL(2)\times\O(6,6)}\,$ with $\,\alpha=+,-\,$ and $\,{M=1,...,12}\,$ being SL(2) and O(6,6) fundamental indices, respectively. In addition to the usual internal coordinates in DFT dual to momentum and winding, the theory contains a second copy of such coordinates which are needed to fill the $\,(\textbf{2},\textbf{12})\,$ representation of the duality group. When acting on covariant $\,\mathbb{R}^{+}\times\SL(2)\times\O(6,6)\,$ tensors, the generalised diffeomorphisms act with a generalised Lie derivative $\,\mathbb{L}_{\Lambda}\,$. For a vector field $\,U^{\a M}\,$ of weight $\,\lambda(U)=\lambda_{U}\,$, the action of latter reads
\be
\label{L-QFT_0}
\begin{array}{ccl}
\mathbb{L}_{\Lambda} U^{\a M} &=& \Lambda^{\b N} \partial_{\b N} U^{\a M} - U^{\b N} \partial_{\b N} \Lambda^{\a M} + Y^{\a M \b N}{}_{\g P \d Q} \, \partial_{\b N} \Lambda^{\g P} \, U^{\d Q} \\[1.5mm]
&+& (\lambda_{U}-\omega) \partial_{\b N} \Lambda^{\b N} U^{\a M}  \ ,
\end{array}
\ee
where $\,\Lambda^{\a M}(x,y)\,$ is the generalised gauge parameter and $\,\omega=\frac{1}{2}\,$. As in $\textrm{E}_{7(7)}$-EFT, all generalised diffeomorphism parameters carry weight $\,\lambda=\omega\,$. The generalised Lie derivative (\ref{L-QFT_0}) is expressed in terms of an invariant structure tensor
\begin{equation}
\label{Y_bos-QFT}
\begin{array}{llll}
Y^{\a M \b N}{}_{\g P \d Q} &=&  \delta_{\d}^{\a} \, \delta_{\g}^{\b}  \, \eta^{MN} \, \eta_{PQ} + 2 \,  \varepsilon^{\a \b} \, \varepsilon_{\g \d} \, \delta^{MN}_{PQ}  & .
\end{array}
\end{equation}
The relative coefficient between the two terms in (\ref{Y_bos-QFT}) follows from the $\mathbb{Z}_2$-truncation of the structure tensor of $\textrm{E}_{7(7)}$-EFT (see appendix~\ref{app:Z2-truncation}). Substituting (\ref{Y_bos-QFT}) into (\ref{L-QFT_0}) one finds
\begin{equation}
\label{L-QFT}
\begin{array}{ccl}
\mathbb{L}_{\Lambda} U^{\a M} &=& \Lambda^{\b N} \partial_{\b N} U^{\a M} - U^{\b N} \partial_{\b N} \Lambda^{\a M} + \eta^{MN} \, \eta_{PQ} \, \partial_{\b N} \Lambda^{\b P} \, U^{\a Q} \\[2mm] &+& 2 \, \varepsilon^{\a \b} \, \varepsilon_{\g \d} \, \partial_{\b N} \Lambda^{\g[M} \, U^{|\d |N]}  + (\lambda_{U}-\omega) \partial_{\b N} \Lambda^{\b N} U^{\a M}  \ .
\end{array}
\end{equation}
The first line and the density term can be seen as the SL(2) generalisation of the generalised Lie derivative of DFT. The term with $\,\varepsilon^{\alpha \beta}\,$ is intrinsic to SL(2)-DFT and does not contribute when restricting the coordinate dependence of all fields and parameters to $y^{M}\equiv y^{+M}$, or equivalently setting $\partial_{-M}=0$ (`DFT limit' in the following).

The algebra of the generalised Lie derivative must close for consistency of the SL(2)-DFT. This  condition can be expressed as
\begin{equation}
\label{closure_eq}
\big[ \mathbb{L}_{\Lambda} , \mathbb{L}_{\Sigma} \big] W^{\a M} = \mathbb{L}_{[\Lambda,\Sigma]_{\textrm{S}}} W^{\a M} \ ,
\end{equation}
where the SL(2) generalisation of the C(ourant)-bracket of DFT (denoted here S-bracket) is defined as
\begin{equation}
\label{SL2_C-bracket}
\big[ \Lambda , \Sigma \big]_{\textrm{S}}^{\alpha M} \equiv \frac12\big(\mathbb{L}_{\Lambda} \Sigma^{\alpha M} - \mathbb{L}_{\Sigma} \Lambda^{\alpha M}) \ 
\end{equation}
for any two vectors $\,\Lambda\,$ and $\,\Sigma\,$ of weight $\lambda=1/2$. As in DFT/EFT, the closure condition (\ref{closure_eq}) requires to impose a so-called \textit{section constraint}. There are two such constraints in SL(2)-DFT which read
\begin{equation}
\label{sec_cons-QFT_2}
\eta^{MN} \, \partial_{\a M} \otimes  \partial_{\b  N} = 0
\hspace{10mm} \textrm{and} \hspace{10mm}
\varepsilon^{\a \b}  \, \partial_{\a [M |} \otimes  \partial_{\b | N]} = 0
 \ ,
\end{equation}
and which restrict the dependence of fields and parameters on the internal coordinates $\,y^{\a M}\,$. The first constraint in (\ref{sec_cons-QFT_2}) is identified with the SL(2) generalisation of the section constraint of DFT that forbids simultaneous dependence on a momentum coordinate and its dual winding. The second constraint is again a genuine feature of SL(2)-DFT and forbids the dependence on more than one coordinate of type $\,+\,$ and its SL(2) duals (of type $\,-\,$). This constraint is therefore trivially satisfied in the DFT limit.

The SL(2) generalisation of the C-bracket in (\ref{SL2_C-bracket}) fails to satisfy the Jacobi identity. This issue is commonly resolved by noticing that the Jacobiator can be expressed as a symmetric bracket defined as
\begin{align}
\{\Lambda ,\Sigma \}^{\alpha M}_{\textrm{S}} \equiv\ &  \tfrac12\big(\mathbb{L}_{\Lambda} \Sigma^{\alpha M}+\mathbb{L}_{\Sigma} \Lambda^{\alpha M}) \nonumber\\[2mm]
=\ & \varepsilon^{\alpha\beta} \, \eta^{MP} \, \eta^{NQ} \, \partial_{\beta N}\big(\varepsilon_{\gamma\delta} \, \Lambda^\gamma{}_{[P} \, \Sigma^{\delta}{}_{Q]} \big) + \tfrac12 \, \varepsilon^{\alpha\gamma} \, \varepsilon^{\beta\delta} \, \eta^{MN} \, \partial_{\beta N} \big(\eta_{PQ} \, \Lambda_{(\gamma}{}^P \, \Sigma_{\delta)}{}^Q\big) \nonumber\\[2mm]
&-\tfrac14 \, \varepsilon^{\alpha\beta} \, \eta^{MN} \, \big( \Sigma^{\gamma P}\partial_{\beta N} \, \Lambda_{\gamma P} + \Lambda^{\gamma P}\partial_{\beta N} \, \Sigma_{\gamma P} \big) \ .
\label{eq:sym_bra-QFT}
\end{align}
Each of the three terms in (\ref{eq:sym_bra-QFT}) is a trivial gauge parameter so that $\,\mathbb{L}_{\{\Lambda,\Sigma\}_{\textrm{S}}}\,$ vanishes identically. Indeed, using the section constraints (\ref{sec_cons-QFT_2}), it can be shown that the following parameters do not generate generalised diffeomorphisms 
\begin{equation}
\label{eq:trivparacov}
\Lambda^{\alpha M}=\varepsilon^{\alpha\beta}\eta^{MP}\eta^{NQ}\partial_{\beta N} \chi_{PQ}
\hspace{3mm} , \hspace{3mm}
\Lambda^{\alpha M}= \varepsilon^{\alpha\gamma}\varepsilon^{\beta\delta}\eta^{MN}\partial_{\beta N}\chi_{\gamma\delta}
\hspace{3mm} \textrm{ and } \hspace{3mm}
\Lambda^{\alpha M}=\varepsilon^{\alpha\beta}\eta^{MN}\chi_{\beta N} \ .
\end{equation}
Here $\,\chi_{PQ}=-\chi_{QP}\,$ and $\,\chi_{\g \d}=\chi_{\d \g}\,$ are respectively in the $\,(\mathbf{1,66})\,$ and $\,(\mathbf{3,1})\,$ representations of the duality group and carry weight $\,1\,$, whereas $\,\chi_{\beta N}\,$ is in the $\,(\mathbf{2,12})\,$, carries weight $\,1/2\,$ and is covariantly constrained as
\begin{equation}
\label{eq:cov_cons-QFT}
\big(\mathbb{P}_{(\mathbf{1,1})+(\mathbf{1,66})+(\mathbf{3,1})}\big)^{\alpha M\beta N}\chi_{\alpha M}\,\partial_{\beta N} \,\, = \,\, 0 \,\, = \,\, \big(\mathbb{P}_{(\mathbf{1,66})+(\mathbf{3,1})}\big)^{\alpha M\beta N}\chi_{\alpha M}\,\chi_{\beta N} \ ,
\end{equation}
where $\,\mathbb{P}\,$ denotes the projector onto the displayed representations. In particular, it can be shown that the bracket in the last term of \eqref{eq:sym_bra-QFT} satisfies the above constraints. The necessity for the class of trivial parameters in the $\,(\mathbf{2,12})\,$ becomes apparent when facing the task of constructing a gauge covariant field strength for the vectors $\,A_{\mu}{}^{\alpha M}\,$, as we will see next.

\subsection{Yang--Mills sector and tensor hierarchy}
\label{sec:Yang--Mills}

Generalised diffeomorphisms with parameters $\,\Lambda^{\a M}(x,y)\,$ depending on the external space-time coordinates $\,x^{\mu}\,$ require the customary covariantisation in extended field theories of the external derivative with gauge connections $\,A_{\mu}{}^{\a M}(x,y)\,$, namely
\begin{equation}
\mathcal{D}_{\mu} = \partial_{\mu} - \mathbb{L}_{A_{\mu}} \ .
\end{equation}
The vectors $\,A_\mu{}^{\alpha M}\,$ carry weight $\,\lambda(A_{\mu})=\frac{1}{2}\,$ and are chosen to transform as
\begin{equation}
\delta_{\Lambda}A_{\mu}{}^{\alpha M}=\,\mathcal{D}_\mu \,\Lambda^{\alpha M}=\,\big(\partial_{\mu}-\mathbb{L}_{A_{\mu}}\big)\Lambda^{\alpha M} \ .
\end{equation}
Due to the non-vanishing Jacobiator, the naive expression for the associated field strength $\,F_{\mu\nu}=\,2\,\partial_{[\mu}A_{\nu]}-[A_{\mu},A_{\nu}]_{\textrm{S}}\,$ fails to transform covariantly under generalised diffeomorphisms. To cure this, a set of tensor fields is introduced whose variations precisely cancel the non-covariant terms. The modified field strengths read
\begin{equation}
\label{F-modified}
\mathcal{F}_{\mu\nu}{}^{\alpha M}= F_{\mu\nu}{}^{\alpha M}+2\,\varepsilon^{\alpha\beta}\eta^{MP}\eta^{NQ}\partial_{\beta N} B_{\mu\nu\,PQ}+\varepsilon^{\alpha\gamma}\varepsilon^{\beta\delta}\eta^{MN}\partial_{\beta N}B_{\mu\nu\,\gamma\delta}-\frac12 \, \varepsilon^{\alpha\beta}\eta^{MN}B_{\mu\nu\,\beta N} \ ,
\end{equation}
where the tensor fields are in the same representations and carry the same weights as the trivial parameters \eqref{eq:trivparacov}, and where $\,B_{\mu\nu \beta N}\,$ is subject to the covariant constraints \eqref{eq:cov_cons-QFT}. A general variation of the modified field strength (\ref{F-modified}) yields
\begin{equation}
\begin{array}{lll}
\delta\mathcal{F}_{\mu\nu}{}^{\alpha M} &=& 2\,\mathcal{D}_{[\mu}\delta A_{\nu]}{}^{\alpha M}+2\,\varepsilon^{\alpha\beta}\eta^{MP}\eta^{NQ}\partial_{\beta N} \Delta B_{\mu\nu\,PQ} \\[2mm]
&+& \varepsilon^{\alpha\gamma}\varepsilon^{\beta\delta}\eta^{MN}\partial_{\beta N}\Delta B_{\mu\nu\,\gamma\delta} - \dfrac12 \, \varepsilon^{\alpha\beta}\eta^{MN}\Delta B_{\mu\nu\,\beta N} \ ,
\end{array}
\end{equation}
where we have defined the covariant variations 
\begin{equation}
\begin{array}{lll}
\Delta B_{\mu\nu\,PQ} &=& \delta B_{\mu\nu\,PQ}+\varepsilon_{\gamma\delta}\,A_{[\mu}{}^{\gamma}{}_{[P}\,\delta A_{\nu]}{}^{\delta}{}_{Q]} \ , \\[2mm]
\Delta B_{\mu\nu\,\gamma\delta} &=& \delta B_{\mu\nu\,\gamma\delta}+\eta_{PQ}\,A_{[\mu(\gamma}{}^P\,\delta A_{\nu]\delta)}{}^Q \ , \\[2mm]
\Delta B_{\mu\nu\,\beta N} &=& \delta B_{\mu\nu\,\beta N}+\delta A_{[\nu}{}^{\gamma P}\partial_{\beta N}A_{\mu]\gamma P}+A_{[\mu}{}^{\gamma P}\partial_{\beta N}\delta A_{\nu]\gamma P} \ .
\end{array}
\end{equation}
We finally choose the following vector (\textit{i.e.} generalised diffeomorphism) and tensor gauge transformations
\begin{equation}
\label{gauge_transfs}
\begin{array}{lll}
\delta A_{\mu}{}^{\alpha M} &=& \mathcal{D}_\mu\, \Lambda^{\alpha M}-2\,\varepsilon^{\alpha\beta}\eta^{MP}\eta^{NQ}\partial_{\beta N} \Xi_{\mu PQ}-\varepsilon^{\alpha\gamma}\varepsilon^{\beta\delta}\eta^{MN}\partial_{\beta N}\Xi_{\mu \gamma\delta}+\dfrac12 \, \varepsilon^{\alpha\beta}\eta^{MN}\Xi_{\mu\beta N} \ , \\[2mm]
\Delta B_{\mu\nu\,PQ} &=& 2\,\mathcal{D}_{[\mu}\,\Xi_{\nu]PQ} + \varepsilon_{\gamma\delta}\,\Lambda^\gamma{}_{[P}\,\mathcal{F}_{\mu\nu}{}^\delta{}_{Q]} \ ,\\[2mm]
\Delta B_{\mu\nu\,\gamma\delta} &=& 2\,\mathcal{D}_{[\mu}\,\Xi_{\nu]\gamma\delta} + \eta_{PQ}\,\Lambda_{(\gamma}{}^P\,\mathcal{F}_{\mu\nu\,\delta)}{}^Q \ , \\[2mm]
\Delta B_{\mu\nu\,\beta N} &=& 2\,\mathcal{D}_{[\mu}\Xi_{\nu]\beta N}  + \mathcal{F}_{\mu\nu}{}^{\gamma P}\partial_{\beta N}\Lambda_{\gamma P}+\Lambda^{\gamma P}\partial_{\beta N}\mathcal{F}_{\mu\nu\,\gamma P} \\[1.5mm]
&& + \,  8\,  \eta^{SP}  \,(\partial_{\beta N}\,\partial_{\gamma S}\,A_{[\mu}{}^{\gamma R})\,\Xi_{\nu] PR}+4\, \eps^{\delta \xi}   \,(\partial_{\beta N}\,\partial_{\xi P}\,A_{[\mu}{}^{\lambda P})\,\Xi_{\nu]\lambda \delta} \ ,
\end{array}
\end{equation}
where the tensor gauge parameters $\,\Xi_{\mu}{}_{PQ}=-\Xi_{\mu}{}_{QP}\,$, $\,\Xi_{\mu}{}_{\alpha\beta}=\Xi_{\mu}{}_{\beta\alpha}\,$ and $\,\Xi_{\mu\beta N}\,$ lie in the $(\textbf{1},\textbf{66})$, $(\textbf{3},\textbf{1})$ and $(\textbf{2},\textbf{12})$ as the corresponding tensor fields and also carry weights $1$, $1$ and $1/2$, respectively. After some algebra along the lines of the $\textrm{E}_{7(7)}$-EFT case, it can be proven that the modified field strengths (\ref{F-modified}) transform as $\,\mathbb{R}^{+}\times\SL(2)\times\O(6,6)\,$ vectors of weight $\lambda(\mathcal{F}_{\mu\nu}{}^{\alpha M})=1/2$ under generalised diffeomorphisms and are invariant under tensor gauge transformations, namely 
\begin{equation}
\delta_\Lambda\mathcal{F}_{\mu\nu}{}^{\alpha M}=\mathbb{L}_\Lambda \mathcal{F}_{\mu\nu}{}^{\alpha M}
\hspace{8mm} \textrm{ and }\hspace{8mm}
\delta_\Xi\mathcal{F}_{\mu\nu}{}^{\alpha M}=0\ .
\end{equation}

\subsection{Bosonic pseudo-action}
\label{sec:pseudo}

We now present the pseudo-action governing the dynamics of the theory. It can be derived by $\mathbb{Z}_{2}$-truncating the pseudo-action of $\textrm{E}_{7(7)}$-EFT \cite{Hohm:2013uia}, as described in the appendix~\ref{app:Z2-truncation}, and must be supplemented with the twisted self-duality relations
\begin{equation}
\label{eq:tsd-equ}
\mathcal{F}_{\mu\nu}{}^{\alpha M} = - \frac12\,e\,\varepsilon_{\mu\nu\rho\sigma}\eta^{MN}\varepsilon^{\alpha\beta}\,\mathcal{M}_{\b N \g P} \,\mathcal{F}^{\rho\sigma\,\gamma P} \ ,
\end{equation}
where $\,e\,$ is the determinant of the vierbein and $\,\mathcal{M}^{\alpha M\beta N}\equiv \,M^{\alpha\beta}M^{MN}\,$ is a symmetric matrix parameterising the scalar manifold. The dynamics of the theory is completely specified by imposing the above twisted self-duality equations after varying the pseudo-action
\begin{formula}
\label{eq:action-QFT}
S_{\text{\tiny{SL(2)-DFT}}} =  \displaystyle\int d^4 x\,d^{24} y \,e\,\big[\,\hat{R}\,+\,\tfrac{1}{4} \, g^{\mu\nu}\,\mathcal{D}_\mu M^{\a\b}\,\mathcal{D}_{\nu} M_{\a\b}  + \tfrac{1}{8} \, g^{\mu\nu} \, \mathcal{D}_\mu M^{MN}\,\mathcal{D}_{\nu} M_{MN}\ &\\
-\tfrac18 \, M_{\a \b} \, M_{MN} \,\mathcal{F}^{\mu\nu \,  \a M}\mathcal{F}_{\mu\nu}{}^{\b N}  + \, e^{-1}\,\mathcal{L}_{\text{top}}-V_{\text{\tiny{SL(2)-DFT}}}(M,g)\,\big]& \, .
\end{formula}
The gauge invariance of this pseudo-action is guaranteed by the fact that the section constraints \eqref{sec_cons-QFT_2} are in one-to-one correspondence with the truncation of the $\mathrm{E}_{7(7)}$-EFT section constraint. Nevertheless, gauge invariance can be checked explicitly using the fact that the vierbein and the scalar matrix $\mathcal{M}^{\alpha M\beta N}$ transform under generalised diffeomorphisms as a scalar density and as a symmetric tensor of weight $\lambda(e_\mu{}^a)=1/2$ and $\lambda(\mathcal{M}^{\alpha M\beta N})=0$, respectively. This implies\footnote{There is an ambiguity in how to distribute the density term between the transformation of $\,M^{\a \b}\,$ and the one of $\,M^{MN}\,$. Note however that this is irrelevant for the gauge invariance of the pseudo-action \eqref{eq:action-QFT}. In order to recover later on the correct transformation of $\,M^{MN}\,$ in DFT, we have chosen here to move the whole density term to the transformation of $\,M^{\a \b}\,$.} in particular
\begin{equation}
\label{gen_diff_e_M}
\begin{array}{rll}
\delta_\Lambda e_\mu{}^a&=& \Lambda^{\g P} \, \partial_{\g P}\, e_\mu{}^a+\tfrac12\,\partial_{\g P} \Lambda^{\g P} e_\mu{}^a\ , \\[2mm]
\delta_{\Lambda} M^{\a\b} &=&   \Lambda^{\g P} \, \partial_{\g P}  M^{\a\b} - 2 \, M^{\g (\a} \, \partial_{\g P} \Lambda^{\b) P} + M^{\a \b} \, \partial_{\g P} \Lambda^{\g P} \ ,  \\[2mm] 
\delta_{\Lambda} M^{MN} &=&  \Lambda^{\g P} \, \partial_{\g P}  M^{MN} - 2 \, M^{P (M} \, \partial_{\g P} \Lambda^{ | \g | N)} + 2 \,  \eta^{P (M} \, M^{N )R}  \, \partial_{\g P} \Lambda^{\g Q}   \, \eta_{QR}  \ .
\end{array}
\end{equation}
Equipped with these formulae  and the transformations \eqref{eq:tsd-equ}, it is then possible to verify that each term in the pseudo-action is invariant under generalised diffeomorphisms and tensor gauge transformations. The relative coefficients between the various term can be fixed by requiring invariance under external diffeomorphisms but this computation is more involved and we expect it to follow the same steps as in $\textrm{E}_{7(7)}$-EFT.
\\

\noindent \textbf{The kinetic terms:} In line with the structure of extended field theories, the Einstein-Hilbert term is constructed from a modified Riemann tensor 
\begin{equation}
\hat{R}_{\mu\nu}{}^{ab} = R_{\mu\nu}{}^{ab}[\omega]+\mathcal{F}_{\mu\nu}{}^{\a M} \,e^{a\rho}\,\partial_{\a M} e_\rho{}^b \ ,
\end{equation}
where $\,R_{\mu\nu}{}^{ab}[\omega]\,$ is the curvature of the spin connection in the external space-time and carries weight $\lambda(R_{\mu\nu}{}^{ab}[\omega])=0$. The corresponding modified Ricci scalar then transforms as scalar of weight $\lambda(\hat{R})=-1$ under generalised diffeomorphims.

The second, third and fourth terms respectively correspond to the kinetic terms for the $\,M_{\alpha \beta} \in \textrm{SL}(2)/\textrm{SO}(2)\,$ scalars, the $\,M_{MN} \in \textrm{SO}(6,6)/(\textrm{SO}(6) \times \textrm{SO}(6))\,$ scalars and the vector fields in the theory. Furthermore, we will parameterise $\,M_{\alpha \beta}\,$ and its inverse as
%
%
\begin{equation}
\label{SL(2)-param}
M_{\a \b} = \frac{1}{\textrm{Im}S} \,
\left(
\begin{array}{cc}
|S|^2  &  \textrm{Re}S \\[2mm]
\textrm{Re}S  &  1
\end{array}
\right)
\hspace{4mm} \textrm{ and } \hspace{4mm}
M^{\a \b} = \frac{1}{\textrm{Im}S} \,
\left(
\begin{array}{cc}
1  &  -\textrm{Re}S \\[2mm]
-\textrm{Re}S  &  |S|^2
\end{array}
\right) \ ,
\end{equation}
where $\,S(x,y) \equiv \chi_{0}+ i \, e^{-\phi}\,$ is the complex axion-dilaton of SL(2)-DFT. In particular, the rigid SL(2) symmetry acts linearly on $\,M_{\a\b}\,$ and as a fractional linear transformation on the complex field $\,S\,$. The specific parameterisation of $\,M_{MN}\,$ will not play any role in this work.
\\

\noindent \textbf{The topological term:} The topological term is obtained from the one of \Eseven-EFT and takes the form of a surface term in five dimensions
\begin{equation}
S_{\textrm{top}} = -\frac{1}{24} \int_{\Sigma^{5}} d^{5}x \, d^{24} y \, \varepsilon^{\mu\nu\rho\sigma\tau} \, \varepsilon_{\b\a} \, \eta_{MN}  \, \mathcal{F}_{\mu\nu}{}^{\a M}\,\mathcal{D}_{\rho}\mathcal{F}_{\sigma\tau}{}^{\b N}  \ .
\end{equation}

\noindent \textbf{The potential:} The potential resulting from the truncation of the \Eseven-EFT expression takes the following form
\begin{equation}
\label{eq:potQFT}
\begin{array}{lll}
V_{\text{\tiny{SL(2)-DFT}}}(M,g) &=& 
M^{\a \b} M^{MN} \big[ - \frac{1}{4} \,  (\partial_{\a M} M^{\g\d})(\partial_{\b N} M_{\g\d}) -  \frac{1}{8} \,  (\partial_{\a M} M^{PQ})(\partial_{\b N} M_{PQ})    \\[2mm]
&& \quad\quad\quad\quad\quad\,\, + \, \tfrac{1}{2} \,  (\partial_{\a M} M^{\g\d})(\partial_{\d N} M_{\b\g})   +  \tfrac{1}{2} (\partial_{\a M} M^{PQ})(\partial_{\b Q} M_{NP}) \big]  \\[2mm]
&+& \tfrac{1}{2}  \, M^{MN} M^{PQ}  (\partial_{\a M} M^{\a \d})(\partial_{\d Q} M_{NP})  +  \tfrac{1}{2} \, M^{\a \b} M^{\g \d}  (\partial_{\a M} M^{MQ})(\partial_{\d Q} M_{\b \g}) \\[2mm]

&-& \frac14 \, M^{\a \b} \, M^{MN} \,  \big[ g^{-1} (\partial_{\a M} g) \, g^{-1} (\partial_{\b N} g) \, + \, (\partial_{\a M} g^{\mu\nu}) \, (\partial_{\b N} g_{\mu\nu}) \big] \\[2mm]
&-& \frac12 \,g^{-1} \, (\partial_{\a M} g) \,\partial_{\b N} ( M^{\a \b} M^{MN} ) \ ,
\end{array}
\end{equation}
and depends on both $\,\SL(2)\,$ and $\,\SO(6,6)\,$ scalars.
\\

\noindent \textbf{Vector and tensor field equations:} The field equations for the vectors $\,A_{\mu}{}^{\a M}\,$ can be derived by varying the Lagrangian \eqref{eq:action-QFT}
\begin{equation}
\label{eq:vec-eom1}
\delta_A\mathcal{L} = \Big[\frac14\,\mathcal{D}_\mu\big(\,2\,e\,M_{\alpha\beta}M_{MN}\mathcal{F}^{\mu\nu\,\beta N}+\varepsilon^{\mu\nu\rho\sigma}\mathcal{F}_{\rho\sigma\,\alpha M}\big)+e\,\hat{\mathcal{J}}^{\nu}{}_{\alpha M}+e\,\mathcal{J}^{\nu}{}_{\alpha M}\Big]\,\delta A_\nu{}^{\alpha M} \ ,
\end{equation}
where the first and second terms come from the variation of the kinetic and topological term\footnote{This variation is once again easily derived by truncating the expression of $\mathrm{E}_{7(7)}$-EFT.}, respectively. The currents $\,\hat{\mathcal{J}}\,$ and $\,\mathcal{J}\,$ in (\ref{eq:vec-eom1}) are defined by 
\begin{align}
\label{eq:currents}
\delta\mathcal{L}_{\textrm{EH}}= e\,\hat{\mathcal{J}}^{\nu}{}_{\alpha M}\,\delta A_\nu{}^{\alpha M}
\hspace{8mm} \textrm{ and } \hspace{8mm}
\delta\mathcal{L}_{\tiny{\textrm{kin. scal}}}= e\,\mathcal{J}^{\nu}{}_{\alpha M}\,\delta A_\nu{}^{\alpha M} \ ,
\end{align}
and are associated to the Einstein-Hilbert term and the kinetic terms for the scalars, respectively. Using the twisted self-duality equation \eqref{eq:tsd-equ}, the field equations for the vectors \eqref{eq:vec-eom1} become 
\begin{equation}
\label{eq:fieldeq-cov}
\delta_A\mathcal{L}=\delta A_\nu{}^{\alpha M}\,\Big[\tfrac12\varepsilon^{\mu\nu\rho\sigma}\,\mathcal{D}_\mu\mathcal{F}_{\rho\sigma\,\alpha M}+e\,\hat{\mathcal{J}}^{\nu}{}_{\alpha M}+e\,\mathcal{J}^{\nu}{}_{\alpha M}\Big] \ .
\end{equation}

The variation of the Lagrangian \eqref{eq:action-QFT} with respect to the tensor fields yields the twisted self-duality equations \eqref{eq:tsd-equ} projected under internal derivatives. It is important to emphasise the role of the twisted self-duality equations (\ref{eq:tsd-equ}). They allow for the manifest duality covariance of this formulation and reflect the on-shell relations between dual degrees of freedom. As previously mentioned, they can be derived only \textit{partially} as field equations for the tensor fields and must be imposed on top of the vector field equations derived from the pseudo-action (\ref{eq:action-QFT}).

\subsection{Section constraints and string embedding}
\label{sec:orientifold}

We now investigate the solutions of the section constraints (\ref{sec_cons-QFT_2}). Let us consider them acting on any single field $\,\Phi(x^{\mu},y^{\a M})\,$ of the theory, namely
\begin{equation}
\label{sec_cons_QFT_+-}
\partial_{\alpha}{}^M\partial_{\beta\,M}\Phi = 0
\hspace{10mm} \textrm{and} \hspace{10mm}
\partial_{+[M}\partial_{-N]}\Phi = 0 \ .
\end{equation}
The first equation imposes that any internal coordinate that $\,\Phi\,$ depends on must be null with respect to the O(6,6) metric $\,\eta_{MN}\,$. We now look for a set of coordinates that satisfies the above constraints. Let us use $\,\SL(2,\bbR)\times\O(6,6)\,$ to fix the choice of one first coordinate: we can choose $\,y^{+1}\,$ without loss of generality. Then the second equation combined with this choice restricts the dependence on the other internal coordinates as
\begin{formula}
	\partial_{+[1} \partial_{-N]} \Phi  =  0 
	\quad \Rightarrow \quad \partial_{-N} \Phi = 0\ \quad \forall N\neq 1\ .
\end{formula}
One thus finds two possible solutions of the section constraints (\ref{sec_cons_QFT_+-}):
\begin{itemize}

\item[$i)$] We may take $\,y^{-1}\,$ as another coordinate independent from $\,y^{+1}\,$. In this case, no extra coordinate dependence is allowed and we have a two-dimensional solution of the section constraints. Imposing the above coordinate dependence on all fields and parameters, we obtain a six-dimensional theory. There is an $\,\O(5,5)\times\bbR^+\,$ residual duality symmetry, where $\,\bbR^+\,$ acts as a trombone in the entire six-dimensional spacetime. On the two coordinates $\,y^{\alpha1}\,$ there is an action of the $\,\GL(2,\bbR)\,$ structure group for the internal manifold obtained from \SL(2,\bbR) and an $\bbR$ subgroup of $\,\O(6,6)\,$. This leads us to identify this case with a $\,4+2\,$ dimensional split of six-dimensional chiral $\,\cN=(2,0)\,$ half-maximal supergravity coupled to five tensor multiplets \cite{Romans:1986er}.

\item[$ii$)] The other independent solution is obtained by only allowing for a dependence on $\,y^{+M}\,$ coordinates. Then the section constraints in (\ref{sec_cons_QFT_+-}) reduce to those of DFT, and  a dependence on up to six mutually null coordinates is allowed. Up to \O(6,6) transformations, we can restrict to $\,y^{+1,\ldots,d}\,$ with $\,d\le 6\,$. A $\GL(d)$ subgroup of \O(6,6) acts as structure group of the internal manifold, and global (continuous) symmetries are broken to $\,\bbR^{+} \times \O(6-d,6-d)\,$. The theory is identified with half-maximal $(4+d)$-dimensional supergravity coupled to $\,\nv=6-d\,$ vector multiplets. If $\,d=2\,$ the non-chiral $\,\cN=(1,1)\,$ six-dimensional supergravity \cite{Giani:1984dw,Romans:1985tw} coupled to four vector multiplets is recovered in a $4+2$ split. The (maximal) $\,d=6\,$ solution is identified with a $\,4+6\,$ dimensional split of ten-dimensional $\,\cN=1\,$ half-maximal supergravity \cite{Chamseddine:1980cp,Bergshoeff:1981um} without vector multiplets. 

\end{itemize}

\subsubsection*{Type IIB orientifolds and physical coordinates}

The $\,\mathbb{Z}_{2}\,$ discrete group we have used to truncate \Eseven-EFT and obtain SL(2)-DFT can be identified with applying an orientifold projection in Type IIB string theory. This amounts to modding out the Type IIB theory by the worldsheet orientation-reversal transformation $\Omega_{\textrm{p}}$, the fermion number projector for left-moving fermions $(-1)^{F_L}$ and an internal space involution $\,\sigma_{\textrm{O}p}\,$ which must be an isometry of the internal space and is induced by an O$p$-plane. Here we are interested in the behaviour of the six physical internal coordinates (upon solving the section constraints) under the orientifold involution $\,\sigma_{\textrm{O}p}\,$ in the presence of an \mbox{O$p$-plane}. The group theoretical decomposition of the $\,\textbf{56}\,$ generalised coordinates of \Eseven-EFT under ordinary SL(6) internal diffeomorphisms that is relevant to discuss Type IIB orientifolds reads
\begin{equation}
\label{Embedding_IIB}
\begin{array}{llcll}
\textrm{E}_{7(7)} & \supset &  \textrm{SL}(2)_{S} \times \textrm{SO}(6,6)  & \supset &  \textrm{SL}(2)_{S} \times \textrm{SL}(6) \times \mathbb{R}_{T}^{+}   \\[2mm]
\textbf{56} & \rightarrow  & \textbf{(2,12)} & \rightarrow & \textbf{(2,6)}_{(+\frac{1}{2})} + \textbf{(2,6')}_{(-\frac{1}{2})}  \\[2mm]
            &   & \textbf{(1,32)} & \rightarrow & \textbf{(1,6')}_{(+1)} + \textbf{(1,20)}_{(0)} + \textbf{(1,6)}_{(-1)} \\[4mm]
&&            & \supset & \textrm{SL}(6) \times \mathbb{R}_{S}^{+}  \times \mathbb{R}_{T}^{+}  \\[2mm]
&&          & \rightarrow &  
  \textbf{6}_{(+\frac{1}{2},+\frac{1}{2})} 
+ \textbf{6}_{(-\frac{1}{2},+\frac{1}{2})}
+ \textbf{6'}_{(+\frac{1}{2},-\frac{1}{2})} 
+ \textbf{6'}_{(-\frac{1}{2},-\frac{1}{2})}   \\[2mm]  
&&& \rightarrow &    
  \underbrace{\textbf{6'}_{(0,+1)}}_{\partial_m^{\rm O3}}  
+ \textbf{20}_{(0,0)} + \textbf{6}_{(0,-1)}  \ .
\end{array}
\end{equation}
For the sake of clarity, we have attached a label $\,S\,$ to the SL(2) factor of the duality group of SL(2)-DFT which acts as fractional linear transformations on the axion-dilaton~$\,S\,$.

When considering an O3-plane in Type IIB, the six internal coordinates are reflected by $\,\sigma_{\textrm{O}3}\,$ implying that they are parity-odd. Then the element $\,\textbf{6'}_{(0,+1)}\,$ must be identified with the six internal derivatives $\,\partial_m^{\rm O3}\,$, the $\,\SL(2)_S\,$ factor of the duality group corresponds to Type IIB S-duality\footnote{This implies that \O(6,6) is not identified with the Type IIB T-duality in this case.} and the scalar field $\,\mathrm{Im}S\,$ is the Type IIB dilaton \cite{Dibitetto:2014sfa}. The $\,\bbR^+_T\,$ charge is then identified with the combination of the rescaling of the coordinates of the internal space $\,\mathcal{M}_{6}\,$ and of the ten-dimensional metric that leaves the $\,D=4\,$ Einstein frame metric invariant.
We can thus write
\begin{equation}
\partial_m^{\rm O3} \neq 0\,:\quad
\bbR^+_{S} = \bbR^+_{\phi_{\rm IIB}}
\hspace{5mm} \textrm{ and } \hspace{5mm}
\bbR^+_{T} = \bbR^+_{\cM_6\,\mathrm{scaling}} \ . 
\end{equation}
Note that the physical coordinates descend from the spinor representation $\,(\textbf{1},\textbf{32})\,$ in order to flip sign under the orientifold action and therefore are projected out by the $\mathbb{Z}_{2}$-truncation. As a result, SL(2)-DFT does \textit{not} capture Type IIB backgrounds with O3-planes, neither does ordinary DFT\footnote{We are not considering DFT supplemented with an additional ``layer" of Ramond-Ramond (RR) potentials in the $\,\textbf{32'}\,$ of O(6,6) needed to formulate the Type IIB theory \cite{Hohm:2011dv}. Even in this case, our identification of physical derivatives $\,\partial_m^{\rm O3}\,$ holds.}. This clarifies some confusion in the literature.

When considering an O9-plane in Type IIB, the six internal coordinates are left invariant by $\,\sigma_{\textrm{O}9}\,$ implying that they are parity-even. Recalling that only the coordinates descending from the $\,(\textbf{2},\textbf{12})\,$ are $\,\mathbb{Z}_{2}$-even, one must select one of the $\,\textbf{6'}$'s coming from this representation to be the physical derivatives $\,\partial_m^{\rm O9}\,$. Up to $\,\SL(2)_S\,$ rotations, we can select the $\,\textbf{6'}_{(-\frac{1}{2},-\frac{1}{2})}\,$ without loss of generality. The $\,\bbZ_2$-truncation will now be interpreted as the truncation of the Type IIB theory to the pure supergravity sector of the Type I theory, equivalently Type IIB with O9-plane. However, since the physical derivatives are not singlets under the $\,\textrm{SL}(2)_{S}\,$ factor of the duality group, the latter can no longer be identified with the S-duality of Type IIB. An alternative interpretation of the same physical derivatives is in terms of the Heterotic ones $\,\partial_m^{\rm Het}\,$. The distinction between the Type~I and Heterotic pictures turns out to be a matter of conventions. First of all, the axion $\,\mathrm{Re}S\,$ is associated with either the internal $\,C_6\,$ of Type IIB or $\,B_6\,$ of Heterotic depending on the conventions. On the other hand, $\,\bbR^+_{S}\,$ is a combination of the Type~IIB dilaton scaling $\,\bbR^+_{\phi_\textrm{IIB}}\,$ and the scaling of the internal space $\,\bbR^+_{\cM_6\,\mathrm{scaling}}\,$. The correct matching of charges is given by
\begin{equation}
\label{IIB_charge_redefinitions}
\partial_m^{\textrm{Type I/Het}}\neq0\,:\quad
\begin{pmatrix}
\bbR^+_{\phi_{\textrm{IIB}}}\\
\bbR^+_{\cM_6\,\mathrm{scaling}}
\end{pmatrix}
 = 
\begin{pmatrix}
+\tfrac12 & -\tfrac12 \\
-\tfrac32 & -\tfrac12
\end{pmatrix}
\begin{pmatrix}
\bbR^+_{S}\\
\bbR^+_{T}
\end{pmatrix} \ .
\end{equation}
We see that the charge assignment that reflects the interpretation of the SL(2)-DFT in terms of its Type I/Heterotic origin has now changed to
\begin{equation}
\label{Embedding_IIB_new_charges}
\begin{array}{llcll}
\textrm{E}_{7(7)} & \supset &  \textrm{SL}(2)_{S} \times \textrm{SO}(6,6)  & \supset &  \textrm{SL}(6) \times \mathbb{R}_{\phi_{\rm IIB}}^{+}  \times \mathbb{R}_{\cM_6 \textrm{\;scaling}}^{+}   \\[2mm]
\textbf{56} & \rightarrow & \textbf{(2,12)} & \rightarrow  &         
  \textbf{6}_{(0,-1)} 
+ \textbf{6}_{(-\frac{1}{2},+\frac{1}{2})}
+ \textbf{6'}_{(+\frac{1}{2},-\frac{1}{2})} 
+ \underbrace{\textbf{6'}_{(0,+1)}}_{\partial_m^{\textrm{Type I/Het}}}   \\  
& \rightarrow & \textbf{(1,32)} & \rightarrow  &    
  \textbf{6'}_{(-\frac12,-\frac12)} 
+ \textbf{20}_{(0,0)} + \textbf{6}_{(+\frac12,+\frac12)}  \ .
\end{array}
\end{equation}
This charge assignment shows that the internal physical coordinates are invariant under shifts of the ten-dimensional dilaton. In fact they are invariant under the full $\,\SL(2)_{\rm IIB}\,$, though it is broken by the $\bbZ_2$-projection. 
Applying an $\,\SL(2)_{\rm IIB}\,$ transformation will exchange representations with opposite $\,\bbR^+_{\phi_{\rm IIB}}\,$ charges in \eqref{Embedding_IIB_new_charges}. 
This translates into the mixing of representations coming from the $\,(\textbf{2},\textbf{12})\,$ and the $\,(\textbf{1},\textbf{32})\,$. 
Indeed, the $\,\bbZ_2\,$ action does not commute with $\,\SL(2)_{\rm IIB}\,$.
We stress that the physical coordinates are by definition always $\,\SL(2)_{\rm IIB}\,$ singlets. 
Since the dictionary between \Eseven-EFT fields and Type IIB ones is also fixed only up to $\,\SL(2)_{\rm IIB}\,$ transformations, it is entirely a matter of conventions whether the truncation to the $\,(\mathbf2,\mathbf{12})\,$ indicated in \eqref{Embedding_IIB_new_charges} with $\,\textbf{6'}_{(0,+1)}\,$ as physical coordinates is to be identified with the action of an O9-plane, and hence with the supergravity sector of Type I, or with its $\textrm{S}_{\rm IIB}$-dual giving the supergravity sector of Heterotic. 
The \O(6,6) factor in the duality group of SL(2)-DFT is then interpreted as the T-duality of Type I or of Heterotic supergravity. 

Finally, under $\,\SL(2)_S\,$, the $\,{\partial_m^{\textrm{Type I/Het}} \equiv \frac\partial{\partial y^{+m}}}\,$ derivatives in the $\,\textbf{6'}_{(0,+1)}\,$ are rotated into the $\,\frac\partial{\partial y^{-m}}\,$ in the $\,\textbf{6'}_{(+\frac{1}{2},-\frac{1}{2})}\,$. 
Notice that there is no simple ten-dimensional interpretation for this dualisation: in terms of its action on fields, this duality mixes metric degrees of freedom with $\,C_6\,$ ones (or $B_6$), and $\,C_2\,$ (or $B_2$) degrees of freedom with the dual graviton. 
As already emphasised, such a dualisation has nothing to do with the $\textrm{S}_{\rm IIB}$-duality relating Type I and Heterotic.

Summarising, only the Type I/Heterotic theories retain physical coordinates which are all ``bosonic" inside $\,\textrm{E}_{7(7)}\,$ and thus survive the $\,\mathbb{Z}_{2}\,$-truncation halving \Eseven-EFT to \mbox{SL(2)-DFT}.  They belong to the unique orbit of six-dimensional solutions of the section constraints of SL(2)-DFT which, in turn, corresponds to the unique half-maximal supergravity in ten dimensions.  It is known that full moduli stabilisation cannot be achieved neither in Type~I nor in Heterotic compactifications without invoking non-geometric fluxes \cite{Shelton:2005cf,Aldazabal:2006up} that activate non-trivial $\,\textrm{SL}(2)_{S}\,$ de Roo--Wageman angles \cite{deRoo:1985jh}. We will show that these can be obtained from generalised Scherk--Schwarz \cite{Scherk:1979zr} reductions of SL(2)-DFT that necessarily violate the section constraints in (\ref{sec_cons-QFT_2}), \textit{e.g.}, by including dependence on coordinates related to each other by $\SL(2)_S$ dualisation. As we stressed above, despite the conventional name this is \emph{not} the string theory S-duality evident in Type IIB, and in particular does \emph{not} exchange Type I and Heterotic degrees of freedom.

\subsection{SL(2)-DFT in the electric frame}
\label{sec:DFT-limits}

The main advantage of the SL(2)-DFT pseudo-action we have provided is that invariance under generalised diffeomorphisms is manifest term by term except for the scalar potential.
However, it requires to treat vector fields and their duals in a democratic approach and to impose \eqref{eq:tsd-equ} on top of the field equations.
In this section we provide a true\footnote{Note that in order to actually perform integration in the internal space it is still generally necessary to first solve the section constraint and restrict the integration measure accordingly.} action in a symplectic frame where only the $\,A_{\mu}{}^{+M}\,$ vectors are treated as propagating and have a kinetic term.
This has the double purpose of allowing for a more direct comparison with the gauged supergravity literature \cite{Schon:2006kz} where usually such an action is used, and facilitate the discussion of the connection between our theory and the formulation of DFT provided in \cite{Hohm:2013nja}.
Indeed, in the latter an action with true kinetic terms for the physical vector fields is provided and the appropriate gauge-fixing and dualisation procedures that we will need to carry out are much simpler if we also start with true kinetic terms.
In such an action, the manifest \SL(2) covariance is broken in the vector kinetic terms and in the topological term.

\subsubsection*{$\O(6,6)$ covariant electric frame}

We choose an Sp(24) symplectic frame where the twelve vectors $\,A_{\mu}{}^{+M}\,$ are identified as physical electric vectors. This by no means implies that the $\,A_{\mu}{}^{-M}\,$ vectors disappear from the Lagrangian. They become non-dynamical but still enter the theory via the covariant derivatives $\,\mathcal{D}_{\mu}\,$, the non-Abelian structure of the S-bracket and a new topological term $\,\widetilde{\mathcal{L}}_{\textrm{top}}\,$. Similarly to what happens in gauged supergravity, the Yang--Mills and topological terms lose their manifest SL(2) duality covariance. However, the field equations derived from such an action, denoted as $\,\widetilde{S}_{\text{\tiny{SL(2)-DFT}}}\,$, remain SL(2)-covariant and reproduce those of the original SL(2)-DFT formulation presented in section \ref{sec:pseudo}. After moving to the electric frame, the action is given by
\begin{equation}
\label{eq:action-QFT_2}
\begin{array}{lll}
\widetilde{S}_{\text{\tiny{SL(2)-DFT}}} &=&  \displaystyle\int d^4 x\,d^{24} y \,e\,\big[\,\hat{R}\,-\,\dfrac{1}{2 (\textrm{Im}S)^2} \, g^{\mu\nu}\,\mathcal{D}_\mu S \, \mathcal{D}_{\nu}\bar{S} + \tfrac{1}{8} \, g^{\mu\nu} \, \mathcal{D}_\mu M^{MN}\,\mathcal{D}_{\nu} M_{MN} \\[3mm]
&&  \hspace{25mm} + \,\,  \widetilde{\mathcal{L}}_{V}+ \, e^{-1}\,\widetilde{\mathcal{L}}_{\text{top}}-V_{\text{\tiny{SL(2)-DFT}}}(M,g)\,\big] \ ,
\end{array}
\end{equation}
resembling the one of $\,\cN=4\,$ gauged supergravity \cite{Schon:2006kz}. In this formulation, only a subgroup $\,\textrm{SO}(1,1) \times \textrm{O(6,6)}\,$ is realised off-shell. The potential remains unaffected by the choice of symplectic frame and is still given by the expression in (\ref{eq:potQFT}). We also chose to rewrite the kinetic term for the SL(2) scalars in terms the complex field $S$. This kinetic term can be further decomposed to make the dilaton and the axion appear explicitly
\begin{equation}
\label{DSDSb}
-\,\dfrac{1}{2 (\textrm{Im}S)^2} \, g^{\mu\nu}\,\mathcal{D}_\mu S \, \mathcal{D}_{\nu}\bar{S}=\tfrac12\,g^{\mu\nu}\mathcal{D}_\mu(e^\phi)\mathcal{D}_\nu(e^{-\phi})-\tfrac12\,e^{2\phi}\,g^{\mu\nu}\mathcal{D}_\mu\chi_0\,\mathcal{D}_\nu\chi_0 \ .
\end{equation}
Note in passing that \eqref{gen_diff_e_M} implies  
\begin{equation}
\label{eq:trans-dil}
\delta_\Lambda e^{-\phi}=\,\Lambda^{\alpha M}\partial_{\alpha M}e^{-\phi}+e^{-\phi}\partial_{\alpha M}\Lambda^{\alpha M} \ .
\end{equation}

In the electric frame, the kinetic term and the generalised $\theta$-term for the dynamical vectors $\,A_{\mu}{}^{+M}\,$ take the form
\begin{equation}
\label{eq:realkinYM}
\widetilde{\mathcal{L}}_{V}=-\frac14 \, e \,\, \textrm{Im}S \,\, M_{MN}\,\mathcal{F}_{\mu\nu}{}^{+M}\mathcal{F}^{\mu\nu\,+N}-\frac18 \,\, \textrm{Re}S \,\,\varepsilon^{\mu\nu\rho\sigma}\eta_{MN}\,\mathcal{F}_{\mu\nu}{}^{+M}\mathcal{F}_{\rho\sigma}{}^{+N} \ .
\end{equation}
In order to discuss how the choice of electric frame affects the structure of the theory, we introduce a symplectic vector $\,\mathcal{G}_{\mu\nu}{}^{\alpha M}=(\mathcal{G}_{\mu\nu}{}^{+M},\mathcal{G}_{\mu\nu}{}^{-M})\,$ defined as
\begin{equation}
\label{G-def}
\begin{array}{lll}
\mathcal{G}_{\mu\nu}{}^{+M} &\equiv& \mathcal{F}_{\mu\nu}{}^{+M} \ ,  \\[2mm]
\mathcal{G}_{\mu\nu}{}^{-M} &\equiv&  -\eta^{MN}\varepsilon_{\mu\nu\rho\sigma}\,\frac{\partial \mathcal{L}_{V}}{\partial\mathcal{F}_{\rho\sigma}{}^{+N}}=\frac12\,e\,\varepsilon_{\mu\nu\rho\sigma}\, \textrm{Im}S \, \eta_{NP} \, M^{MN} \,\mathcal{F}^{\rho\sigma\,+P}+ \textrm{Re}S\,\mathcal{F}_{\mu\nu}{}^{+M} \ ,
\end{array}
\end{equation}
where we use a ``mostly plus" external spacetime metric and where $\,\varepsilon_{0123}=+1\,$. Therefore $\mathcal{G}_{\mu\nu}{}^{-M}$ denotes the dual of the electric field strength. Following the construction of gauge invariant Lagrangians in the presence of electric and magnetic charges \cite{deWit:2005ub}, the new transformations of the various fields under generalised diffeomorphisms are now given by
\begin{equation} 
\label{gen_diff_non-cov}
\begin{array}{lll}
\delta_{\Lambda} A_{\mu}{}^{\alpha M} &=& \mathcal{D}_\mu\, \Lambda^{\alpha M} \ , \\[2mm]
\Delta_\Lambda B_{\mu\nu\,PQ} &=& \varepsilon_{\gamma\delta}\,\Lambda^\gamma{}_{[P}\,\mathcal{G}_{\mu\nu}{}^\delta{}_{Q]} \ , \\[2mm]
\Delta_\Lambda B_{\mu\nu\,\gamma\delta}  &=& \eta_{PQ}\,\Lambda_{(\gamma}{}^P\,\mathcal{G}_{\mu\nu\,\delta)}{}^Q  \ , \\[2mm]
\Delta_\Lambda B_{\mu\nu\,\beta N}  &=& \mathcal{G}_{\mu\nu}{}^{\gamma P}\partial_{\beta N}\Lambda_{\gamma P}+\Lambda^{\gamma P}\partial_{\beta N}\mathcal{G}_{\mu\nu\,\gamma P} \ ,
\end{array}
\end{equation}
which in turn induce modifications in the transformation of the field strengths \eqref{F-modified}. By comparing (\ref{gen_diff_non-cov}) and (\ref{gauge_transfs}) one sees that only the transformations of the tensor fields under generalised diffeomorphisms are modified. In order to ensure gauge invariance of the Lagrangian under generalised diffeomorphisms, which is spoiled by the new $\,\widetilde{\mathcal{L}}_{V}\,$ term in \eqref{eq:realkinYM}, the following topological term is needed
\begin{align}
\label{Ltilde_top}
\widetilde{\mathcal{L}}_{\text{top}}=\,\varepsilon^{\mu\nu\rho\sigma}\,\Big[& \,\,\,\,\,\,\,\frac13\,[A_\mu,A_\nu]_{\textrm{S}}^{-M} \,\eta_{MN}\,\big(\,\partial_\rho A_\sigma{}^{+N}-\frac14\,[A_\rho,A_\sigma]_{\textrm{S}}^{+N} \,\big)\nonumber\\
&+\frac16\,[A_\mu,A_\nu]_{\textrm{S}}^{+M}\,\eta_{MN}\,\big(\,\partial_\rho A_\sigma{}^{-N}-\frac14\,[A_\rho,A_\sigma]_{\textrm{S}}^{-N} \,\big)\nonumber\\
&-\frac14\,\big(\,2\,\eta^{MP}\eta^{NQ}\partial_{-N}B_{\mu\nu\,PQ}+\eta^{MN}\partial_{+N}B_{\mu\nu\,--}\nonumber\\
&\;\;\;\;\;\;\;\;\;\;\;-\eta^{MN}\partial_{-N}B_{\mu\nu\,-+}-\frac{1}{2}\eta^{MN}B_{\mu\nu\,-N}\big)\,\eta_{MR}\,F_{\rho\sigma}{}^{-R}\nonumber\\
&-\frac12\,\eta^{MN}\eta^{PQ}\,\partial_{-M}B_{\mu\nu\,NP}\,\big(\,\eta^{RS}\,\partial_{+R}B_{\rho\sigma\,QS}-\frac12\,B_{\rho\sigma\,+Q}\,\big)\,\Big] \ .
\end{align}
Note the dependence of the above expression on the magnetic vectors $\,A_{\mu}{}^{-M}\,$. This will be relevant later on when recovering ordinary DFT.

The tensor gauge transformations are not affected by the choice of electric frame and can still be read off from (\ref{gauge_transfs}). To check the invariance of the Lagrangian under such transformations it is convenient to first compute the general variation of $\,\widetilde{\mathcal{L}}_{V}\,$ and $\,\widetilde{\mathcal{L}}_{\text{top}}\,$ with respect to the various fields
\begin{formula}
\label{eq:kin-top-var}
\delta_{A,B}\,\widetilde{\mathcal{L}}_V &= \phantom{-} \tfrac12\,\varepsilon^{\mu\nu\rho\sigma}\,\eta_{MN}\,\,\mathcal{G}_{\mu\nu}{}^{-M} \, \mathcal{D}_{\rho}\delta A_\sigma{}^{+N}+\tfrac14\,\varepsilon^{\mu\nu\rho\sigma}\eta_{MN}\,\, \mathcal{G}_{\mu\nu}{}^{-M}\,\widehat{\partial}^{+ N}\big[\Delta B_{\rho\sigma}\big] \ ,    \\[2mm]
\delta_{A,B}\,\widetilde{\mathcal{L}}_{\text{top}}  &=  -\tfrac12\,\varepsilon^{\mu\nu\rho\sigma}\,\eta_{MN}\,\mathcal{F}_{\mu\nu}{}^{+M}\,\mathcal{D}_{\rho}\delta A_{\sigma}{}^{-N}-\tfrac14\,\varepsilon^{\mu\nu\rho\sigma}\eta_{MN}\,\mathcal{F}_{\mu\nu}{}^{-M}\,\widehat{\partial}^{+ N}\big[\Delta B_{\rho\sigma}\big] \ , 
\end{formula}
where we have introduced the following notation for the projection onto the space of trivial gauge parameters
\begin{equation}
\begin{array}{lll}
\hspace{-1mm} \widehat{\partial}^{\alpha M} \, \big[\Delta B_{\mu\nu}\big] & \hspace{-2mm} \equiv 
& \hspace{-2mm} 2 \, \varepsilon^{\alpha\beta} \eta^{MP} \eta^{NQ}\partial_{\beta N}\Delta B_{\mu\nu\,PQ} + \eta^{MN} \varepsilon^{\alpha\gamma} \varepsilon^{\beta\delta}\partial_{\beta N}\Delta B_{\mu\nu\,\gamma\delta}-\frac{1}{2}\varepsilon^{\alpha\beta}\eta^{MN}\Delta B_{\mu\nu\,\beta N} \, . 
\end{array}
\end{equation}
This projection plays an important role and has appeared, for example, in the form of a St\"uckelberg coupling in the expression of the covariant field strengths $\,\mathcal{F}_{\mu\nu}{}^{\alpha M}\,$ in (\ref{F-modified}). In particular, it can be shown using \eqref{gauge_transfs}, that $\widehat{\partial}^{\alpha M}[\Delta B_{\mu\nu}]=\,2\,\mathcal{D}_{[\mu}\,\widehat{\partial}^{\alpha M}[\Xi_{\nu]}]$. From \eqref{eq:kin-top-var}, it is possible to verify that both $\,\widetilde{\mathcal{L}}_{V}\,$ and $\,\widetilde{\mathcal{L}}_{\textrm{top}}\,$ are invariant under tensor gauge transformations (up to total derivatives for the latter). This requires the use of the section constraints\footnote{In particular, it can be shown that terms of the form $\,\varepsilon_{\alpha\beta}\,\eta_{MN}\,\widehat{\partial}^{\alpha M}[\bullet]\,\widehat{\partial}^{\beta N}[\bullet]\,$ reduce to a total derivative by virtue of the section constraints \eqref{sec_cons-QFT_2}.} and of a Bianchi identity of the form
\begin{equation}
\label{eq:Bianchi}
3\,\mathcal{D}_{[\mu}\,\mathcal{F}_{\nu\rho]}{}^{\alpha M} = \widehat{\partial}^{\alpha M}\big[\mathcal{H}_{\mu\nu\rho}\big] \ ,
\end{equation}
where the field strengths $\,\mathcal{H}_{\mu\nu\rho\,PQ}\,$, $\,\mathcal{H}_{\mu\nu\rho\,\gamma\delta}\,$ and $\,\mathcal{H}_{\mu\nu\rho\,\beta N}\,$ associated to the tensor fields $\,B_{\mu\nu\,PQ}\,$, $\,B_{\mu\nu\,\gamma\delta}\,$ and $\,B_{\mu\nu\,\beta N}\,$ are defined up to terms that vanish upon projection with~$\, \widehat{\partial}^{\alpha M}\,$. Of particular relevance will be the expression for the three-form field strengths in the $\,(\textbf{3},\textbf{1})\,$ representation
\begin{equation}
\label{H3-SL(2)}
\mathcal{H}_{\mu\nu\rho\,\gamma\delta}=3 \, \Big(  \mathcal{D}_{[\mu}B_{\nu\rho]\,\gamma\delta}-\eta_{PQ}\,A_{[\mu(\gamma}{}^P\,\partial_{\nu} A_{\rho]\delta)}{}^Q + \frac13\,\eta_{PQ}\,A_{[\mu(\gamma}{}^{P}\,[A_{\nu},A_{\rho]}]_{{\textrm{S}}\,\,\,\delta)}{}^{Q} \Big) \ ,
\end{equation}
which displays a generalised Chern-Simons like modification based on the S-bracket. This is the SL(2) analog of the structure found in DFT \cite{Hohm:2013nja}.

The general variation of the Lagrangian (\ref{eq:action-QFT_2}) with respect to the various vector and tensor fields reads\footnote{Up to total derivatives and terms that vanish as a result of the field equations for tensors.} 
\begin{equation}
\label{eq:fieldeq-noncov}
\begin{array}{lll}
\delta_{A^+,A^-,B}\,\widetilde{\mathcal{L}}_{\textrm{\tiny{SL(2)-DFT}}} &=&\delta A_\nu{}^{+ M}\,\Big[- \tfrac12\,\eta_{MN}\,\varepsilon^{\mu\nu\rho\sigma}\,\mathcal{D}_\mu\mathcal{G}_{\rho\sigma}{}^{- N}+e\,\hat{\mathcal{J}}^{\nu}{}_{+ M}+e\,\mathcal{J}^{\nu}{}_{+ M}\Big]  \\[2mm]
&+& \delta A_\nu{}^{- M}\,\Big[ \,\,\,\,\,\,\, \tfrac12\,\eta_{MN}\,\varepsilon^{\mu\nu\rho\sigma}\,\mathcal{D}_\mu\,\mathcal{G}_{\rho\sigma}{}^{+N}+e\,\hat{\mathcal{J}}^{\nu}{}_{- M}+e\,\mathcal{J}^{\nu}{}_{- M}\Big]   \\[2mm]
&-& \frac14\,\varepsilon^{\mu\nu\rho\sigma}\,\widehat{\partial}^{+ M} \big[\Delta B_{\mu\nu}\big]\,\eta_{MN}\,\Big[\mathcal{F}-\mathcal{G}\Big]_{\rho\sigma}{}^{-N} \ ,
\end{array}
\end{equation}
where the currents $\,\hat{\mathcal{J}}\,$ and $\,\mathcal{J}\,$ were defined in \eqref{eq:currents}. The variation of the Lagrangian with respect to the tensor fields thus yields a projected duality relation between electric and magnetic vectors while the variation with respect to the magnetic vectors gives the duality relation between the tensor fields and the scalars. Observe that the combined field equations can be written covariantly as
\begin{align}
\label{EOM_vec+tens}
& \frac12\,\varepsilon^{\mu\nu\rho\sigma}\,\mathcal{D}_{\nu}\mathcal{G}_{\rho\sigma}{}^{\alpha M}=\,e\,\varepsilon^{\alpha\beta}\,\eta^{MN}\,\big[\hat{\mathcal{J}}^{\mu}{}_{\beta N}+\mathcal{J}^{\mu}{}_{\beta N}\big]\,,\nonumber\\
&\varepsilon^{\mu\nu\rho\sigma}\,\widehat{\partial}^{\alpha N}\big[\Delta B_{\mu\nu}\big]\,\varepsilon_{\alpha\gamma}\,\eta_{MP}\,\big[\mathcal{F}-\mathcal{G}\big]_{\rho\sigma}{}^{\gamma N}=0\,,
\end{align}
and correctly reproduce the field equations in \eqref{eq:fieldeq-cov} for the vectors obtained from the manifestly SL(2) covariant pseudo-action of SL(2)-DFT. 

Let us finally point out that when taking all the fields to be independent of the internal generalised coordinates $\,y^{+ M}\,$ and $\,y^{-M}\,$, the action \eqref{eq:action-QFT_2} reduces to the one of ungauged $\,\cN=4\,$ supergravity in four dimensions \cite{Schon:2006kz}. In particular, all the magnetic vectors and tensors drop out of the Lagrangian except for two remainders that come from the topological term and the kinetic term for the electric vectors and that combine into
\begin{equation}
\tfrac18\,\epsilon^{\mu\nu\rho\sigma} B_{\mu\nu\,-M}\,\big[F - G \big]_{\rho\sigma}{}^{-M} \ ,
\end{equation} 
where $\,G_{\mu\nu}{}^{-M}\,$ denote the duals of the Abelian electric field strengths (as defined in \eqref{G-def}). The field equation for the tensors then simply reflects the vector-vector duality in four dimensions.

\subsection{DFT limit and $\,\chi_0 \leftrightarrow B_{\mu\nu}\,$ dualisation}

Our goal now is to make contact with the formulation of DFT in \cite{Hohm:2013nja}.
As already mentioned, SL(2)-DFT must be equivalent to DFT when fields and parameters only depend on $\,y^{M} \equiv y^{+M}\,$ coordinates, namely
\begin{equation}
\label{DFT_limit-partial}
(\partial_{+M} , \partial_{-M})\equiv(\partial_{M} , 0) \ .
\end{equation}
The DFT action of \cite{Hohm:2013nja} contains a dynamical tensor field $\,B_{\mu\nu} \equiv [t^{++}]^{--} \, B_{\mu\nu}{}_{--}\,$ while the axion $\,\chi_0\,$ is absent. In contrast, both fields appear in the action (\ref{eq:action-QFT_2}) of SL(2)-DFT although only $\,\chi_0\,$ has a kinetic term (\ref{DSDSb}). The two fields are dual to each other with their duality relation being enforced by the field equations for the magnetic vectors in (\ref{EOM_vec+tens}). By an appropriate use of the duality relations and after gauge fixing, we will dualise away the dynamical axion $\,\chi_0\,$ from the action (\ref{eq:action-QFT_2}) in favor of a dynamical $\,B_{\mu\nu}\,$ tensor field, thus recovering the DFT formulation of \cite{Hohm:2013nja}. In the process, the topological term $\,\widetilde{\mathcal{L}}_{\textrm{top}}\,$ will be absorbed into the kinetic term for $\,B_{\mu\nu}\,$.

Let us start by applying the DFT limit (\ref{DFT_limit-partial}) to the equations of motion of the magnetic vectors in (\ref{EOM_vec+tens}). In this case it is easy to verify that
\begin{equation}
\begin{array}{lll}
e\,\hat{\mathcal{J}}^{\mu}{}_{-M} &=& 0 \ , \\[2mm]
e\,\mathcal{J}^{\mu}{}_{-M} &=& \partial_{\beta N}\big[\,e\,\mathcal{D}^{\mu}(M^{\beta \gamma}M^{NP})\,M_{-\gamma}M_{MP}\,\big]=\,\partial_{M}[\,e\,e^{2\phi}\,\mathcal{D}^\mu \chi_0\,\big] \ .
\end{array}
\end{equation} 
Using now the definition of the symplectic vector (\ref{G-def}) in combination with the Bianchi identity \eqref{eq:Bianchi}, the field equations for the magnetic vectors reduce to
\begin{equation}
\label{eq:duality-two-form}
 \partial_{M}\Big( \, \tfrac16\,\varepsilon^{\mu\nu\rho\sigma} \, \mathcal{H}_{\nu\rho\sigma}  \, + \, e\,e^{2\phi}\,\mathcal{D}^\mu \chi_0 \, \Big) = 0 \ ,
\end{equation}
with $\,\mathcal{H}_{\nu\rho\sigma} \equiv [t^{++}]^{--} \, \mathcal{H}_{\nu\rho\sigma\,--}\,$ and where the expression of the three-form field strength $\,\mathcal{H}_{\mu\nu\rho}\,$ can be obtained from (\ref{H3-SL(2)}) and reads
\begin{equation}
\label{H3--}
\mathcal{H}_{\mu\nu\rho}= 3 \, \Big( \mathcal{D}_{[\mu}B_{\nu\rho]}{} \, +\,A_{[\mu \, }{}_{-}{}^{N}\,\partial_{\nu} A_{\rho]}{}_{-N} \, - \, \frac13 \, A_{[\mu\,}{}_{-}{}^{N}\,[A_{\nu},A_{\rho]}]_{{\textrm{S}}}{}_{\,\,-N} \Big) \ .
\end{equation}
Note that, in the DFT limit (\ref{DFT_limit-partial}), $[\Lambda,\Sigma]^{+M}_{\rm S}$ reduces to the C-bracket and that therefore (\ref{H3--}) matches the corresponding expression in \cite{Hohm:2013nja}.

We continue with the gauge fixing of the axion $\,\chi_0=\textrm{Re}S\,$. Applying the DFT limit to a generalised diffeomorphism (with parameter $\,\Lambda^{\a M}$) acting on the scalar fields of the theory, one finds that $\,\Lambda^{-M}\,$ only\footnote{Importantly, no other fields entering the Lagrangian are affected by $\Lambda^{-M}$ transformations.} affects the gauge transformation of $\,\chi_0\,$
\begin{equation}
\delta_{\Lambda^-}\chi_0 = \partial_{M} \Lambda^{-M} \ ,
\end{equation}
and that $\,\chi_0\,$ transforms as a scalar with respect to $\Lambda^{+M}$ transformations. The quantity $\,\partial_{M} \Lambda^{-M}\,$ is the parameter of an axionic shift symmetry (both $\,x^\mu\,$ and $\,y^{M}\,$ dependent) while $\,\mathcal{D}_{\mu} \chi_0\,$ only involves $\,A_{\mu}{}^{-M}\,$ in the gauge connection
\begin{equation}
\cD_\mu \chi_0 = \partial_\mu \chi_0 -\partial_{M}A_{\mu}{}^{-M}  \ .
\end{equation}
As a result we can then gauge-fix the $\,\Lambda^{-M}\,$ transformations by setting $\,\chi_0 = 0\,$. This is the standard procedure for Peccei--Quinn symmetries that allows to remove from the Lagrangian the generalised $\theta$-term: $\,\chi_0 \, \eta_{MN}\Tr\cF^{+M}\wedge\cF^{+N}\,$. We thus arrive at 
\begin{equation}
\label{Dchi_gauge-fix}
\cD_\mu \chi_0 = -\partial_{M}A_\mu{}^{-M} \ ,
\end{equation}
and, since $\,A_{\mu}{}^{-M}\,$ are non-dynamical in the SL(2)-DFT action \eqref{eq:action-QFT_2}, we can integrate them away. Substituting (\ref{Dchi_gauge-fix}) into the field equations of the magnetic vectors (\ref{eq:duality-two-form}) one finds
\begin{equation}
\partial_{M} \Big( \tfrac{1}{6} \, \varepsilon^{\mu\nu\rho\sigma} \, \cH_{\nu\rho\sigma}  -  e\, e^{2\phi}\,g^{\mu\nu} \partial_{N} A_\nu{}^{-N}{}  \Big) = 0 \ .
\end{equation}
These equations are solved by setting 
\begin{equation}
\label{solving_duality}
\partial_{M}A_\mu{}^{-M} = e^{-2\phi}  \, (*\cH)_\mu + c_\mu  
\hspace{5mm} \textrm{ with } \hspace{5mm}
\partial_{M} c_\mu = 0 \ ,
\end{equation}
where $\,(*\cH)^\mu=\, \frac{1}{6} \, e^{-1}\,\varepsilon^{\mu\nu\rho\sigma} \, \cH_{\nu\rho\sigma}\,$ is the Hodge dual of $\,\mathcal{H}_{\nu\rho\sigma}\,$ and is a proper four-dimensional vector.

The last step in the dualisation process is to substitute (\ref{solving_duality}) into the relevant terms in the Lagrangian. These are the kinetic term for $\,\chi_0\,$ and $\,\widetilde{\mathcal{L}}_{\textrm{top}}\,$. Importantly, it can be shown that the axion $\,\chi_{0}\,$ drops out of the potential (\ref{eq:potQFT}) when taking the DFT limit.  Moreover, by noticing that only the component $\,[A_\mu,A_\nu]_{\textrm{S}}^{-M}\,$ of the S-bracket depends (linearly) on $A_\mu{}^{-M}$ in the DFT limit, it is straightforward to observe that magnetic vectors appear at most linearly in every term of the topological term (\ref{Ltilde_top}). Notice also that only $\,B_{\mu\nu}{}_{--}\,$ appears, and that the definition of $\,\Delta B_{\mu\nu}{}_{--}\,$ does not contain $\,\delta A_\mu{}^{-M}\,$. This means that we can simply use the variation \eqref{eq:kin-top-var} to deduce a compact expression for $\,\widetilde{\cL}_{\rm top}\,$ in the DFT limit.  After some algebra one arrives at 
\begin{equation}
\label{relevant_L}
\begin{array}{lll}
\cL_{\text{kin-}\chi_0} &=& -\frac12\, e\,e^{2\phi}\,g^{\mu\nu} (\partial_{M}A_\mu{}^{-M})(\partial_{N}A_\nu{}^{-N}) \ , \\[2mm]
\widetilde{\cL}_{\text{top}}  &=& \frac{1}{6} \, \varepsilon^{\mu\nu\rho\sigma} \, (\partial_{M}A_\mu{}^{-M}) \, \cH_{\nu\rho\sigma} \ .
\end{array}
\end{equation}
Upon substitution of (\ref{solving_duality}) into (\ref{relevant_L}), the integration constant $\,c_\mu\,$ only appears in a term $\,\propto c_\mu c^\mu\,$ and is thus set to vanish by its own field equation. The remaining terms combine into the kinetic term for $\,B_{\mu\nu}\,$, namely
\begin{equation}
\cL_{\text{kin-}\chi_0} + \widetilde{\cL}_{\text{top}} = -  e\, e^{-2\phi} \frac1{12} \, \cH^{\mu\nu\rho} \, \cH_{\mu\nu\rho}  \ .
\end{equation}

Lastly, in order to recover the DFT action in \cite{Hohm:2013nja} which is presented in the string frame, we perform a change of variables of the form
\begin{equation}
\label{eq:changvar}
 \tilde{g}_{\mu\nu} = e^{\phi} \, g_{\mu\nu}
\hspace{5mm} ,  \hspace{5mm} 
e^{2 d} = e^{\phi}  \ ,
\end{equation}
which in turn induces $\,\tilde{e} = e^{2\phi} \, e \,$. The transformations of $\,\tilde{e}_{\mu}{}^{a}\,$ and $\,e^{-2d}\,$ under generalised diffeomorphisms with parameter $\,\Lambda^{P} \equiv \Lambda^{+P} \,$ can be derived from (\ref{gen_diff_e_M})  and \eqref{eq:trans-dil} after using (\ref{eq:changvar}). They read
\begin{equation}
\delta_{\Lambda} \tilde{e}_{\mu}{}^{a} = \Lambda^{P} \, \partial_{P}\tilde{e}_{\mu}{}^{a} 
\hspace{5mm} \textrm{ and } \hspace{5mm}
\delta_{\Lambda} e^{-2d} = \Lambda^{P} \,  \partial_{P}e^{-2d}  + e^{-2d} \, \partial_{P}\Lambda^{P} =  \partial_{P}(e^{-2d} \Lambda^{P}) \ ,
\end{equation}
so that, as wanted, $\,\tilde{e}_{\mu}{}^{a}\,$ and $\,e^{-2d}\,$ respectively transform as a scalar and a scalar density under the $\Lambda$ transformations of DFT \cite{Hohm:2013nja}. Note that the transformation of the $\mathrm{SO}(6,6)$ scalar matrix $M^{MN}$ can be straightforwardly deduced from \eqref{gen_diff_e_M} and also matches the DFT expression. The density term in the transformation of $e^{-2d}$ is associated with an $\mathbb{R}^{+}_\text{DFT}$ which appears explicitly in the right column of Table~\ref{Table:duality_groups}, and which is a linear combination\footnote{As mentioned before, the correct weights in the DFT limit of the various fields under $\,\mathbb{R}^{+}_{\text{DFT}}\,$ were already assigned through the choice of the coefficients for the density terms in \eqref{gen_diff_e_M}.} of the original $\,\mathbb{R}^+\,$ in SL(2)-DFT and the $\mathbb{R}^{+} \subset\SL(2)$. Furthermore, the rescaling of the external metric is responsible for a shift of the modified external Ricci scalar, as is usual when moving from the Einstein to the string frame in four dimensions
\begin{equation}
\hat{R}(e)=e^{\phi}\hat{R}(\tilde{e})+\frac32\,e^{\phi}\,\tilde{g}^{\mu\nu}\,\mathcal{D}_\mu\phi\,\mathcal{D}_{\nu}\phi\,+3\,e^{\phi}\,\tilde{g}^{\mu\nu}\,\hat{\mathcal{D}}_\mu\mathcal{D}_\nu\phi\ .
\end{equation}
Here $\,\hat{\mathcal{D}}_{\mu}\,$ is the spacetime derivative covariantised with respect to  both external and internal generalised diffeomorphisms (\textit{i.e.} it contains generalised Christoffel symbols). When substituted into the action, the last term is integrated by part. In the process, one directly drops a total $\,\mathcal{D}_{\mu}\,$ derivative. This is allowed since it acts on a scalar density of weight 1 under $\,\mathbb{R}^+_{\text{DFT}}\,$. Note also that the rescaling \eqref{eq:changvar} has no effect on the $\,\mathcal{F}_{\mu\nu}{}^{+M}\,$ term in the modified Ricci scalar. After taking the DFT limit, dualising the axion $\,\chi_{0}\,$ into a tensor field $\,B_{\mu \nu}\,$ and moving to the string frame, the action \eqref{eq:action-QFT_2} then reduces to 
\begin{equation}
\label{eq:actionDFTres}
\begin{array}{lll}
S_{\text{\tiny{DFT}}} &=& \int d^4 x\,d^{12} y \,\tilde{e}\,e^{-2d}\,\Big[\,\hat{R}(\tilde{e})+4\,\mathcal{D}^\mu d\,\mathcal{D}_{\mu}d + \tfrac{1}{8}  \, \mathcal{D}^\mu M^{MN}\,\mathcal{D}_{\mu} M_{MN} \\[2mm]
&-& \frac{1}{12} \,  H_{\mu\nu\rho} \, H^{\mu\nu\rho}  - \frac14 \, M_{MN}\,\mathcal{F}_{\mu\nu}{}^{M}\mathcal{F}^{\mu\nu\,N} - V_{\text{\tiny{DFT}}}(d,M_{MN},\tilde{g})\,\Big] \ ,
\end{array}
\end{equation}
where the field strengths of the electric vectors read\footnote{Note that the last term from (\ref{F-modified}), \textit{i.e.} $\,- \frac12 \, \eta^{MN} B_{\mu\nu\,- N}\,$, is absent as $\,B_{\mu\nu\, \a M}\,$ are covariantly constrained compensating fields solving (\ref{eq:cov_cons-QFT}) as the internal derivatives (\ref{DFT_limit-partial}). This sets $\,B_{\mu\nu\,- M}=0\,$.}
\begin{equation}
\label{F-modified-DFT}
\mathcal{F}_{\mu\nu}{}^{M} \equiv \mathcal{F}_{\mu\nu}{}^{+M} = F_{\mu\nu}{}^{+ M}
- \eta^{MN}\partial_{N}B_{\mu\nu}  \ ,
\end{equation}
and where external space-time indices are now raised and lowered with the rescaled metric $\,\tilde{g}_{\mu \nu}\,$. Finally, the part of the Lagrangian containing the potential takes the form
\begin{equation}
\label{eq:DFTpotres}
\begin{array}{llll}
\mathcal{L}_{\text{Pot}}&=&-\tilde{e}\,e^{-2d}\,V_{\text{\tiny{DFT}}} \\[2mm]
&=& \tilde{e}\,e^{-2d}\Big[ \,\frac{1}{8}  \,  M^{MN} \,  (\partial_{M} M^{KL})(\partial_{N} M_{KL})  -\frac{1}{2}  \,  M^{MN} \,  (\partial_{N} M^{KL})(\partial_{L} M_{MK}) \\[2mm]
&& \,\,\,\,\,\,\,\,\,\,\,\,\,\,\,\,\,\,  - \, 2 \,  (\partial_{M}d)(\partial_{N}M^{MN}) + 4 \,  M^{MN}  \,  (\partial_{M}d)(\partial_{N}d)  \\[2mm]
&& \,\,\,\,\,\,\,\,\,\,\,\,\,\,\,\,\,\, + \, \frac14\,M^{MN}\,\partial_M \tilde{g}_{\mu\nu}\,\partial_N \tilde{g}^{\mu\nu}+\frac14\,M^{MN}\,\tilde{g}^{-1}(\partial_M\tilde{g})\,\tilde{g}^{-1}(\partial_N\tilde{g}) \\[2mm]
&& \,\,\,\,\,\,\,\,\,\,\,\,\,\,\,\,\,\, - \,2\,M^{MN}\,(\partial_M d)\,\tilde{g}^{-1}(\partial_N\tilde{g})+\frac12\,(\partial_M M^{MN})\,\tilde{g}^{-1}(\partial_N\tilde{g})\,\Big] \ .
\end{array}
\end{equation}
As previously stated, the axion $\,\chi_0\,$ cancels out in the above expression. Dropping a total derivative\footnote{Note that the second line of \eqref{eq:DFTpotres} can be rewritten as follows:
\begin{align}
&\tilde{e}\,e^{-2 d}\Big[\,-(\partial_M\partial_N M^{MN})-4\,M^{MN}\,(\partial_M d)(\partial_N d)+4\,\partial_M(M^{MN}\,\partial_N d)\nonumber \\
&\;\;\;\;\;\;\;\;-\tilde{e}^{-1}(\partial_M\tilde{e})[\partial_N M^{MN}-4M^{MN}\,\partial_N d]\,\Big] +\partial_M\Big(\,\tilde{e}\,e^{-2d}[\partial_N M^{MN}-4M^{MN}\,\partial_N d]\Big) \nonumber \  . 
\end{align}} 
and using $\,\tilde{e}^{-1}(\partial_M \tilde{e})=\tfrac12\,\tilde{g}^{-1}(\partial_{M} \tilde{g})\,$, the potential \eqref{eq:DFTpotres} can be expressed as
\begin{align}
\label{eq:potfinal}
\mathcal{L}_{\text{Pot}}=\tilde{e}\,e^{-2d}\Big[ \,\mathcal{R}(d,M)+\frac14\,M^{MN}\,\partial_M \tilde{g}_{\mu\nu}\,\partial_N \tilde{g}^{\mu\nu}+\frac14\,M^{MN}\,\tilde{g}^{-1}(\partial_M\tilde{g})\,\tilde{g}^{-1}(\partial_N\tilde{g})\,\Big] \ ,
\end{align}
where $\,\mathcal{R}(d,M)\,$ is the Ricci scalar for the internal doubled-space \cite{Hohm:2010pp}
\begin{align}
\label{eq:Riccint}
\mathcal{R}(d,M)=\,&\frac{1}{8}  \,  M^{MN} \,  (\partial_{M} M^{KL})(\partial_{N} M_{KL})  -\frac{1}{2}  \,  M^{MN} \,  (\partial_{N} M^{KL})(\partial_{L} M_{MK})-\partial_M\partial_N M^{MN}\nonumber\\
&-4\,M^{MN}\,(\partial_M d)(\partial_N d)+4\,(\partial_M M^{MN})\,(\partial_N d)+4\,M^{MN}\,(\partial_M\partial_N d) \ .
\end{align}
The potential \eqref{eq:potfinal} corresponds to the one derived in \cite{Hohm:2013nja} up to the last term.

\section{Gauge vectors and non-Abelian deformations}
\label{sec:gauge_vectors}

In this section we generalise the previous construction of SL(2)-DFT in two steps:
\begin{itemize}

\item[a)] Firstly, the SL(2)-DFT is extended to include $\,2 \, \times \, n\,$ gauge vectors. This theory does not descend from a truncation of $\textrm{E}_{7(7)}$-EFT as the $\,\SL(2)\times\O(6,6+n)\,$ duality group cannot be embedded into the exceptional duality group of maximal supergravity.

\item[b)] Secondly, this $\,\SL(2)\times\O(6,6+n)\,$ extended field theory is shown to admit deformations of its generalised Lie derivative. Such deformations are in parallel with the embedding tensor deformations of $\,\mathcal{N}=4\,$ gauged supergravity in four dimensions.

\end{itemize}

\subsection{$\SL(2)\times\O(6,6+n)$ extended field theory}

We discuss the SL(2) generalisation of the Abelian DFT constructed in \cite{Hohm:2011ex}. The latter is an ordinary DFT coupled to $\,\nv=n\,$ Abelian gauge vectors that features an enhanced $\,\O(6,6+n)\,$ duality group. In addition to the internal coordinates dual to momentum and winding as well as to the $\,n\,$ gauge vectors, the SL(2)-DFT considered here contains a second copy of such coordinates and therefore the full set of coordinates fills the $\,(\textbf{2},\textbf{12 + $n$})\,$ representation of the duality group $\,\SL(2)\times\O(6,6+n)\,$. 

The $\,\SL(2)\times\O(6,6+n)\,$ extended field theory with Abelian gauge vectors is still formally described by the action (\ref{eq:action-QFT}). The theory has generalised internal coordinates 
\begin{equation}
\label{extended_coords}
y^{\a M} = \left( \, y^{\a m} \,,\, y^{\a}{}_{m} \,,\, y^{\a A}   \right) \ ,
\end{equation}
where $\,\left( \, y^{\a m} \,,\, y^{\a}{}_{m}\right)\,$ with $\,m=1,...,6\,$ correspond to O(6,6) coordinates and $\,y^{\a A}\,$ with $\,A=1,...,n\,$ runs over the additional gauge vectors $\,A_{\mu}{}^{\a A}\,$. As in the previous sections, $\,\alpha=+,-\,$ denotes the SL(2) fundamental index. The structure tensor of the $\,\SL(2)\times\O(6,6+n)\,$ theory is still given by the expression in (\ref{Y_bos-QFT}), but this time $\,\eta_{MN}\,$ denotes the $\,{\textrm{O}(6,6+n)}$-invariant metric. When the O(6,6) block is expressed in light-cone coordinates, it takes the form
\begin{equation}
\label{metric_nv}
\eta_{MN} = \left(
\begin{array}{cc|c}
0 & \mathbb{I}_{6} & 0\\
\mathbb{I}_{6} & 0 & 0\\
\hline
0 & 0 & \delta_{AB}
\end{array}
\right) \ .
\end{equation}

It is important to mention that, despite the presence of the additional set of $\,2 \times n \,$ gauge vectors, the analysis of the solutions of the section constraints (\ref{sec_cons-QFT_2}) does not change. Any dependence of the fields and parameters on the extra $\,2\times n$ coordinates that must be introduced to fill the $\,(\mathbf 2,\,\textbf{12 + $n$})\,$ irrep of $\,\SL(2) \times \O(6,6+n)\,$ is forbidden by the section constraints, analogously to the Heterotic DFT case \cite{Hohm:2011ex}. This is a consequence of the $\,\delta_{AB}\,$ block in the metric (\ref{metric_nv}). The two solutions of the section constraints described before now correspond to chiral half-maximal supergravity in six dimensions coupled to $\,\nt=5+n\,$ tensor multiplets and half-maximal $(4+d)$-dimensional supergravity coupled to $\,\nv=6-d+n\,$ vector multiplets ($d \le 6$).

\subsection{Non-Abelian deformations of $\textrm{SL}(2)$-DFT}
\label{sec:X-deformation}

We now discuss the $\SL(2)$ generalisation of the non-Abelian DFT constructed in \cite{Hohm:2011ex}. The latter is an ordinary DFT coupled to $\,\nv=n\,$ non-Abelian gauge vectors that formally preserves $\,\textrm{O}(6, 6 + n)$, where $\,n\,$ is the dimension of the gauge group. To this end, we will introduce consistent deformations of the generalised diffeomorphisms in SL(2)-DFT. Importantly, and unlike in \cite{Hohm:2011ex}, we will study deformations of the full $\,\SL(2)\times\O(6,6+n)\,$ generalised diffeomorphisms, and not only of the vector sector. As a result, we will find non-Abelian structures both in the gauge and gravity sectors, the latter being associated with turning on background fluxes for the dilaton and the $B$-field in the Type~I/Heterotic theory.

\subsubsection{Deformed generalised Lie derivative}

Following the original construction in \Eseven-EFT \cite{Ciceri:2016dmd}, we first introduce a deformed generalised Lie derivative $\,\widetilde{\mathbb{L}}_{\Lambda}\,$. It acts on a vector $\,U^{\a M}\,$ of weight $\,\lambda_{U}\,$ as
\begin{equation}
\label{eq:Liederdef}
\widetilde{\mathbb{L}}_{\Lambda} U^{\a M} = \mathbb{L}_{\Lambda}U^{\a M} -  X_{\b N \g P}{}^{\a M} \Lambda^{\b N} \, U^{\g P} \ ,
\end{equation}
where $\,\mathbb{L}_{\Lambda}\,$ is the undeformed generalised Lie derivative defined in (\ref{L-QFT}), and where the deformation $\,X_{\a M \b N}{}^{\g P}\,$ is $\,\SL(2)\times \O(6,6+n)$-algebra valued such that $\,X_{\a M \b N}{}^{\g P}=\Theta_{\a M}{}^{\b' N' \g' P'}\, [t_{\b' N' \g' P'}]_{\b N}{}^{\g P}\,$. As in \cite{Ciceri:2016dmd}, the $X$ deformation is subject to a set of quadratic constraints necessary for the closure of the generalised diffeomorphisms algebra and of the Jacobi identity. In addition, the deformation is subject to linear (or representation) constraints which are required for the consistency of the deformed tensor hierarchy. These linear constraints allow the following decomposition of the $X$ deformation in terms of the constant irreducible representations $\,f_{\alpha MNP}=f_{\alpha [MNP]}\,$ and $\,\xi_{\alpha M}\,$ of the duality group 
\begin{equation}
\label{X_N4}
X_{\a M \b N}{}^{ \g P}  =  - \, \delta_{\b}^{\g} \, f_{\a MN}{}^{P} + \frac{1}{2} \left( \delta^{P}_{M} \, \delta_{\b}^{\g} \, \xi_{\a N} - \delta^{P}_{N} \, \delta_{\a}^{\g} \, \xi_{\b M} - \delta_{\b}^{\g} \, \xi_{\a}^{P} \eta_{MN} + \eps_{\a \b} \, \delta^{P}_{N} \, \xi_{\d M} \, \eps^{\d \g}  \right)     \ ,
\end{equation}
or, equivalently,
\begin{equation}
\label{Theta-tensor}
\Theta_{\a M}{}^{\b N \g P} = \frac{1}{2} \, \eps^{\b \g} \, \Big( f_{\a M}{}^{NP} +  \delta_{M}^{[N} \, \xi_{\a}{}^{P]} \Big) \, + \,  \frac{1}{12+n} \,  \delta_{\a}^{(\b} \, \xi^{\g)}{}_{M} \, \eta^{NP} \ .
\end{equation}
To make the forthcoming formulae lighter, it will prove convenient to introduce hatted index-pairs $\,\hat{M}=\alpha M\,$, $\,\hat{N}=\beta N\,$, etc. These can be understood as $\,\textrm{Sp}(24+2 n)\,$ fundamental indices which are raised and lowered with the symplectic invariant matrix $\,\Omega_{\hat{M}\hat{N}}=\eps_{\alpha \beta} \, \eta_{MN}\,$.  In terms of these indices, the representation constraints read\footnote{It is worth noticing that $\,X_{\hat{M}\hat{N}}{}^{\hat{M}} = (4+\frac{n}{2})\, \xi_{\hat{N}}\,$.}
\begin{equation}
\label{linear_symp}
X_{\hat{M}[\hat{N} \hat{P}]}= 0
\hspace{8mm} \textrm{ and } \hspace{8mm}
X_{(\hat{M}\hat{N}\hat{P})} = 0 \ .
\end{equation}
As explained in \cite{Schon:2006kz}, it is the second equation in (\ref{linear_symp}) that allows to write the gauge group generators as in (\ref{X_N4}), and leads to a consistent tensor hierarchy in $\,\cN=4\,$ gauged supergravity.

Closure of the deformed generalised diffeomorphisms algebra requires
\begin{equation}
\label{closure_def_eq}
\big[ \widetilde{\mathbb{L}}_{\Lambda} , \widetilde{\mathbb{L}}_{\Sigma} \big] W^{\hat{M}} = \widetilde{\mathbb{L}}_{[\Lambda,\Sigma]_{\textrm{X}}} W^{\hat{M}} \ ,
\end{equation}
where the X-bracket $\,[\cdot,\cdot]_{\textrm{X}}\,$ is defined as
\begin{equation}
\label{eq:SL2_Xbracket}
\big[\Lambda,\Sigma\big]_{\textrm{X}}^{\hat{M}} \equiv \frac12 \, (\widetilde{\mathbb{L}}_\Lambda \Sigma^{\hat{M}} - \widetilde{\mathbb{L}}_{\Sigma} \Lambda^{\hat{M}}) = \big[\Lambda,\Sigma\big]_{\textrm{X}}^{\hat{M}}   - X_{[\hat{N} \hat{P}]}{}^{\hat{M}}\, \Lambda^{\hat{N}} \, \Sigma^{\hat{P}} \ .
\end{equation}
The general analysis of the closure relation (\ref{closure_def_eq}) performed in \cite{Ciceri:2016dmd} shows that 
\begin{equation}
\label{closure_L}
\begin{array}{lll}
[ \widetilde{\mathbb{L}}_{\Lambda} , \widetilde{\mathbb{L}}_{\Sigma} ] W^{\hat{M}}  -  \widetilde{\mathbb{L}}_{[\Lambda,\Sigma]_{\textrm{X}}}W^{\hat{M}} & = & 
A^{\hat{M}}_{\hat{N}\hat{P}\hat{S}} \,\,  \Lambda^{\hat{N}} \Sigma^{\hat{P}} W^{\hat{S}} 
+ X_{[\hat{N}\hat{P}]}{}^{\hat{Q}} \,\, \Lambda^{\hat{N}} \Sigma^{\hat{P}} \partial_{\hat{Q}}W^{\hat{M}}  \\[2mm]
& + & B^{\hat{M}\hat{Q}}_{\hat{N}\hat{R}\hat{S}} \,\,  (\Lambda^{\hat{N}} \partial_{\hat{Q}}\Sigma^{\hat{R}} W^{\hat{S}} - \partial_{\hat{Q}}\Lambda^{\hat{R}} \Sigma^{\hat{N}} W^{\hat{S}}) \ ,
\end{array}
\end{equation}
where the section constraint $\,Y^{\hat{M}\hat{N}}{}_{\hat{P}\hat{Q}} \,\, \partial_{\hat{M}} \otimes \partial_{\hat{N}}=0\,$ has already been imposed, and where the tensors $\,A\,$ and $\,B\,$ take the form
\begin{equation}
\label{eq:AB_tens}
\begin{array}{lll}
A^{\hat{M}}_{\hat{N}\hat{P}\hat{S}} & = &  2 X_{[\hat{N}|\hat{Q}}{}^{\hat{M}} X_{\hat{P}]\hat{S}}{}^{\hat{Q}} - X_{\hat{Q}\hat{S}}{}^{\hat{M}} X_{[\hat{N}\hat{P}]}{}^{\hat{Q}} \ , \\[2mm]
B^{\hat{M}\hat{Q}}_{\hat{N}\hat{R}\hat{S}} & = &  X_{(\hat{N}\hat{R})}{}^{\hat{M}} \delta_{\hat{S}}^{\hat{Q}} - X_{\hat{N}\hat{S}}{}^{\hat{Q}} \delta_{\hat{R}}^{\hat{M}} \\[2mm]
&+& Y^{\hat{M}\hat{Q}}{}_{\hat{R}\hat{P}} X_{\hat{N}\hat{S}}{}^{\hat{P}} - Y^{\hat{P}\hat{Q}}{}_{\hat{R}\hat{S}} X_{\hat{N}\hat{P}}{}^{\hat{M}} + Y^{\hat{M}\hat{Q}}{}_{\hat{P}\hat{S}} X_{[\hat{N}\hat{R}]}{}^{\hat{P}} - \frac{1}{2} Y^{\hat{P}\hat{Q}}{}_{\hat{R}\hat{N}} X_{\hat{P}\hat{S}}{}^{\hat{M}} \ .
\end{array}
\end{equation}
The closure relation in (\ref{closure_L}) then requires
\begin{equation}
\label{ABX_A_conds}
A^{\hat{M}}_{\hat{N}\hat{P}\hat{S}}=0 
\hspace{6mm} , \hspace{6mm}
X_{[\hat{N}\hat{P}]}{}^{\hat{Q}} \, \partial_{\hat{Q}} = 0 
\hspace{6mm} \text{ and } \hspace{6mm}
B_{\hat{N}\hat{R}\hat{S}}^{\hat{M}\hat{Q}} \, \partial_{\hat{Q}} = 0 \ .
\end{equation}
The set of conditions (\ref{ABX_A_conds}) is not yet final. As for \Eseven-XFT \cite{Ciceri:2016dmd}, the deformed X-bracket in (\ref{eq:SL2_Xbracket}) does not define a Lie algebra since the Jacobi identity is not satisfied. Instead, finds
\begin{equation}
\label{Jacobiator-X-SL2}
\big[[\Lambda,\Sigma]_{\textrm{X}},\Gamma\big]_{\textrm{X}}+\text{cycl.} \,\, = \frac13 \, \big\{[\Lambda,\Sigma]_{\textrm{X}},\Gamma\big\}_{\textrm{X}} +\text{cycl.}  \ ,
\end{equation}
where the modified version of the symmetric bracket in (\ref{eq:sym_bra-QFT}) reads
\begin{equation}
\label{eq:sym_bra-XQFT}
\{\Lambda ,\Sigma \}^{\hat{M}}_{\textrm{X}}  \equiv  \frac12 \, \big(\widetilde{\mathbb{L}}_{\Lambda} \Sigma^{\hat{M}}+\widetilde{\mathbb{L}}_{\Sigma} \Lambda^{\hat{M}}) = \{\Lambda ,\Sigma \}^{\hat{M}}_{\textrm{S}} - X_{(\hat{N}\hat{P})}{}^{\hat{M}} \, \Lambda^{\hat{N}} \, \Sigma^{\hat{P}} \ .
\end{equation}
Consistency then requires that $\,\{\Lambda ,\Sigma \}^{\hat{M}}_{\textrm{X}}\,$ corresponds to a trivial gauge parameter such that $\,\widetilde{\mathbb{L}}_{\{\Lambda ,\Sigma \}_{\textrm{X}}}\,$ vanishes identically. Using again of the general results in \cite{Ciceri:2016dmd}, one has
\begin{equation}
\label{eq:Ltilde_trivial}
\begin{array}{lll}
\widetilde{\mathbb{L}}_{\{\Lambda ,\Sigma \}_{\textrm{X}}} U^{\hat{M}} & = &   C^{\hat{M}\hat{R}}_{\hat{S}\hat{P}\hat{Q}} \,\,  ( \Lambda^{\hat{Q}} \partial_{\hat{R}}\Sigma^{\hat{P}} U^{\hat{S}} + \partial_{\hat{R}} \Lambda^{\hat{P}} \Sigma^{\hat{Q}} U^{\hat{S}})  
 -  X_{(\hat{P}\hat{Q})}{}^{\hat{R}} \,\, \Lambda^{\hat{P}} \Sigma^{\hat{Q}} \, \partial_{\hat{R}}U^{\hat{M}} \\[2mm]
& + &  X_{(\hat{P}\hat{Q})}{}^{\hat{R}} \, X_{\hat{R}\hat{S}}{}^{\hat{M}}\, \Lambda^{\hat{P}} \Sigma^{\hat{Q}} U^{\hat{S}} \ ,
\end{array}
\end{equation}
where the tensor $C$ reads
\begin{equation}
\label{C_tensor-SL2}
C^{\hat{M}\hat{R}}_{\hat{S}\hat{P}\hat{Q}} = X_{(\hat{P}\hat{Q})}{}^{\hat{M}} \delta_{\hat{S}}^{\hat{R}}  - Y^{\hat{M}\hat{R}}{}_{\hat{T}\hat{S}} \, X_{(\hat{P}\hat{Q})}{}^{\hat{T}} - \frac{1}{2} \,  Y^{\hat{T}\hat{R}}{}_{\hat{P}\hat{Q}} \, X_{\hat{T}\hat{S}}{}^{\hat{M}} \ .
\end{equation}
This time the closure conditions (\ref{ABX_A_conds}) have been used. Therefore, the triviality of the modified symmetric bracket translates into the set of conditions
\begin{equation}
\label{Trivial_symmetric-SL2}
X_{(\hat{P}\hat{Q})}{}^{\hat{R}} \, X_{\hat{R}\hat{S}}{}^{\hat{M}} = 0 
\hspace{5mm} , \hspace{5mm}
X_{(\hat{P}\hat{Q})}{}^{\hat{R}} \,\, \partial_{\hat{R}} = 0 
\hspace{5mm} \text{ and } \hspace{5mm}
C^{\hat{M}\hat{R}}_{\hat{S}\hat{P}\hat{Q}} \,\,  \partial_{\hat{R}} = 0\ .
\end{equation}
Combining the various constraints necessary for the consistency of the gauge algebra, we obtain the following minimal set:
\begin{equation}
\label{consistency_constraints}
\begin{array}{rllcc}
Y^{\hat{M}\hat{N}}{}_{\hat{P}\hat{Q}} \,\, \partial_{\hat{M}} \otimes \partial_{\hat{N}} &=& 0 & \quad\text{( section constraint )} 
\\[2mm]
X_{\hat{M}\hat{N}}{}^{\hat{P}} \, \partial_{\hat{P}} &=& 0 & \quad\text{( \textrm{X-constraint} )}  \\[2mm]
\big( X_{(\hat{P}\hat{Q})}{}^{\hat{M}} \delta_{\hat{S}}^{\hat{R}}  - Y^{\hat{M}\hat{R}}{}_{\hat{T}\hat{S}} \, X_{(\hat{P}\hat{Q})}{}^{\hat{T}} - \frac{1}{2} \,  Y^{\hat{T}\hat{R}}{}_{\hat{P}\hat{Q}} \, X_{\hat{T}\hat{S}}{}^{\hat{M}} \big) \, \partial_{\hat{R}} &=& 0 & \quad\text{( \textrm{C-constraint} )} \\[2mm]
X_{\hat{M}\hat{P}}{}^{\hat{R}} X_{\hat{N}\hat{R}}{}^{\hat{Q}} - X_{\hat{N}\hat{P}}{}^{\hat{R}} X_{\hat{M}\hat{R}}{}^{\hat{Q}} + X_{\hat{M}\hat{N}}{}^{\hat{R}} X_{\hat{R}\hat{P}}{}^{\hat{Q}} &=& 0 & \quad\text{( quadratic constraint )} 
\end{array}
\end{equation}
Note that the B-constraint is absent as it can be shown to follow from the X-constraint. It is also important to notice at this point that contrarily to the $\mathrm{E}_{7(7)}$-EFT case, the C-constraint is no longer (at least fully) implied by the X-constraint.

We close this section by giving the expression of the various constraints in terms of the irreducible components $\,f_{\a MNP}\,$ and $\,\xi_{\a M}\,$ presented in (\ref{X_N4}). The section constraint reduces to the relations presented in (\ref{sec_cons-QFT_2}) while after some algebra, the X-constraint can be written as
\begin{equation}
\label{X-constraint_irreps}
\begin{array}{rll}
\xi_{\a}{}^{M} \, \partial_{\beta M} & = & 0 \ , \\[2mm]
\xi^{\a}{}_{(M} \, \partial_{|\a| N)} - \frac{1}{12+n} \, \eta_{MN}  \, \xi^{\a}{}^{P} \, \partial_{\a P}& = & 0 \ , \\[2mm]
f_{\a  MN}{}^{P} \, \partial_{\b  P} +  \xi_{\b [M} \, \partial_{|\a| N]} & = & 0  \ .
\end{array}
\end{equation}
The C-constraint imposes further restrictions. Assuming that the section constraints in (\ref{sec_cons-QFT_2}) and the X-constraint (\ref{X-constraint_irreps}) hold, then the C-constraint is satisfied provided that $\,\delta_{\a}^{\g} \, \eps^{\lambda \d} \, \eta_{N [ M} \,  C^{ \a N  \, \beta R}_{|\g | S \, | \lambda | P \, | \d | Q ]} \, \partial_{\beta  R}=0\,$. This gives the following extra restriction
\begin{equation}
\label{C-constraint_irreps}
\begin{array}{rll}
\eps^{\a \b} \, f_{\alpha}{}_{[MSP} \,\, \partial_{|\b| Q]} & = & 0 \ .
\end{array}
\end{equation}
As in gauged supergravity, the quadratic constraint in (\ref{consistency_constraints}) is the requirement that the gauge group generators $\,X_{\hat{M}}=(X_{\hat{M}})_{\hat{N}}{}^{\hat{P}}=X_{\hat{M}\hat{N}}{}^{\hat{P}}\,$ form a closed set and have commutation relations 
\begin{equation}
\label{X-algebra}
\big[ X_{\hat{M}}  ,  X_{\hat{N}} \big] = -X_{\hat{M}\hat{N}}{}^{\hat{P}} \, X_{\hat{P}}  \ .
\end{equation}
It decomposes as follows
\begin{equation}
\label{QC_N=4} 
\begin{array}{rll}
\xi_{\a M}\,\xi_{\b}{}^{M} & = &0 \ , \\[2mm]
\xi_{(\a}^{\phantom{a}P}\,f_{\b)PMN} & = & 0 \ , \\[2mm]
3\,f_{\a R[MN}\,f_{\b PQ]}^{\phantom{abcde}R}\,+\,2\,\xi_{(\a [M}\,f_{\b)NPQ]} & = & 0 \ , \\[2mm]
\eps^{\a \b}\left(\xi_{\a}^{\phantom{a}P}\,f_{\b PMN}\,+\,\xi_{\a M}\,\xi_{\b N}\right) & = & 0 \ , \\[2mm]
\eps^{\a \b}\left(f_{\a MNR}\,f_{\b PQ}^{\phantom{abcde}R}\,-\,\xi_{\a}^{\phantom{a}R}\,f_{\b R[M[P}\,\eta_{Q]N]}\,-\,\xi_{\a [M}\,f_{\b N]PQ}\,+\,\xi_{\a [P}\,f_{\b Q]MN}\right) & = & 0 \ .
\end{array}
\end{equation}
We will come back to the set of consistency constraints in (\ref{consistency_constraints}) when classifying the deformations compatible with the Type I/Heterotic solution of the section constraint.

\subsubsection{Structure of SL(2)-XFT}

Deformations of the generalised Lie derivative based on an embedding tensor like object $\,X\,$ were introduced in the context of \Eseven-EFT in \cite{Ciceri:2016dmd}. A set of modifications occurs at the level of the tensor hierarchy and the action induced by the $X$ deformation (\ref{X_N4}), although the field content of the theory remains unchanged. We will refer to the deformed theory as \mbox{SL(2)-XFT}. When taking the fields and parameters to be independent of the internal space coordinates $\,y^{\a M}\,$ the SL(2)-XFT reduces to $\,\cN=4\,$ gauged supergravity in four dimensions \cite{Schon:2006kz} and the $X$ deformation is identified with the embedding tensor. On the contrary, when the $X$ deformation is set to zero, the undeformed SL(2)-DFT is recovered.

The implementation of the $\,X\,$ deformation in the case of SL(2)-DFT is in direct analogy with the construction of the \Eseven-XFT in \cite{Ciceri:2016dmd}. For this reason we will only give a minimal presentation of the relevant structures in the presence of an $\,X\,$ deformation. Importantly, when restricted to $\,n=0$, the results presented here can be obtained from the \mbox{$\mathbb{Z}_2$-truncation} of the tensor hierarchy and action of the $\textrm{E}_{7(7)}$-XFT. The generalisation to arbitrary $\,n\,$ is then immediate and can be argued on the basis of covariance of the theory with respect to the $\,\textrm{SL}(2) \times \textrm{O}(6,6 \,+\,n)\,$ duality group.  The various modifications of the SL(2)-DFT tensor hierarchy presented in section~\ref{sec:Yang--Mills} are induced by the fact that the $(\textbf{2},\textbf{12 + $n$})$ vectors $\,A_{\mu}{}^{\alpha M}\,$ transform under modified generalised diffeomorphisms as
\begin{equation}
\label{eq:Avardef}
\delta_\Lambda A_{\mu}{}^{\alpha M} = \mathcal{D}_\mu\Lambda^{\alpha M}\equiv\,\big(\partial_\mu-\mathbb{\widetilde{L}}_{A_\mu}\big)\Lambda^{\alpha M} \ ,
\end{equation}
where $\,\mathcal{D}_\mu\,$ is now further covariantised with respect to the gauge symmetries generated by the $\,X\,$ deformation. As in gauged supergravity, the associated field strengths $\,\cF_{\m\n}{}^{\a M}\,$ are no longer covariant with respect to such gauge transformations, and must be modified with St\"uckelberg-like couplings to tensor fields of the form $\, \Omega^{\a M \b N} \, \Theta_{\b N}{}^{\g P \d Q} B_{\m \n \,\, \g P \d Q}\,$ where $\,B_{\m \n \,\, \g P \d Q} = \eps_{\g \d} \, B_{\m \n \,\, P Q} + \eta_{PQ} \, B_{\m \n \,\, \g\d}\,$.  After using (\ref{Theta-tensor}), one finds
\begin{equation}
\label{eq:FSdef}
\mathcal{F}_{\mu\nu}{}^{\alpha M} = F_{\mu\nu}{}^{\alpha M} + \widehat{\partial}^{\a M} [B_{\mu\nu}] + \eps^{\a\b} \, (f_{\b}{}^{MNP} + \eta^{MN} \, \xi_{\b}{}^{P}) \, B_{\m\n \, NP} \, + \, \eps^{\a\b} \, \xi^{\g M} \, B_{\m\n \, \b \g} \ ,
\end{equation}
which accounts for both the tensor hierarchy of SL(2)-DFT and the one of $\,\cN=4\,$ gauged supergravity. The modification of the vector and tensor gauge transformations (\ref{gauge_transfs}) induced by the $\,X\,$ deformation (more conveniently $\,\Theta\,$ in order to avoid traces over $\Gamma$-matrices) can be derived following the same steps as in \cite{Ciceri:2016dmd}. We will not present here the modified version of the tensor hierarchy, but it can be verified that
\begin{equation}
\label{eq:gaugetransdef}
\delta_\Lambda \,\mathcal{F}_{\mu\nu}{}^{\alpha M}=\widetilde{\mathbb{L}}_\Lambda \, \mathcal{F}_{\mu\nu}{}^{\alpha M}
\hspace{10mm} \textrm{ and } \hspace{10mm}
\delta_\Xi\,\mathcal{F}_{\mu\nu}{}^{\alpha M} = 0 \ .
\end{equation}

As for the SL(2)-DFT, the dynamics of SL(2)-XFT can be encoded into a gauge invariant pseudo-action supplemented by a set of twisted self-duality equations. 
The pseudo-action takes the same form as the SL(2)-DFT expressions \eqref{eq:tsd-equ} and \eqref{eq:action-QFT}, but with covariant derivatives and field strengths being now further covariantised with respect to the $\,X\,$ deformation as in \eqref{eq:Avardef} and \eqref{eq:FSdef}. From the gauge transformations \eqref{eq:gaugetransdef} of the field strengths, it should be clear that all the terms remain separately invariant under vector and tensor gauge transformations with the exception of the potential which requires a closer look. Once again in analogy to~\cite{Ciceri:2016dmd}, the potential in SL(2)-XFT can be expressed as the sum of three contributions
\begin{equation}
\label{eq:potX}
V_{\text{SL(2)-XFT}} (M,g,X ) = V_{\text{SL(2)-DFT}} (M,g) \, + \, V_{\text{cross}} (M,X) \, + \, V_{\text{SUGRA}} (M,X)   \ ,
\end{equation}
where the first term denotes the SL(2)-DFT potential presented in \eqref{eq:potQFT} while the second and third terms depend linearly and quadratically on the $\,X\,$ deformation, respectively. When expressed in terms of the irreducible pieces $\,f_{\a MNP}\,$ and $\,\xi_{\a M}\,$ these are given by
\begin{equation}
\label{eq:Vcross-XQFT}
\begin{array}{lll}
V_{\text{cross}} &=& -\frac12\,M^{\alpha\beta} M^{MN}M^{KL}\,f_{\alpha MK}{}^P \,\partial_{\beta N}M_{PL} \\[2mm]
& & -\frac12\,M^{\alpha\beta} M^{MN}M^{KL}\eta_{M[K}\,\xi_{\alpha S]}\,\eta^{PS}\,\partial_{\beta N} M_{PL} \\[2mm]
& & -M^{\alpha\beta} M^{MN}M^{\gamma\delta}\,\xi_{\gamma M}\,\partial_{\beta N}M_{\alpha\delta} \ ,
\end{array}
\end{equation}
and
\begin{equation}
\label{eq:Vq-XQFT}
\begin{array}{lll}
V_{\text{SUGRA}} & =& \frac{1}{12}\,f_{\alpha MNP}\,f_{\beta QRS}\,M^{\alpha\beta}M^{MQ}M^{NR}M^{PS}+\frac14\,f_{\alpha NM}{}^P\,f_{\beta PQ}{}^{N}\,M^{\alpha\beta}M^{MQ} \\[2mm]
& &-\frac{16}{9}\, f_{\alpha MNP}\,f_{\beta QRS}\,\varepsilon^{\alpha\beta}\, M^{MNPQRS} +\frac16\,f_{\alpha MNP}\,f_{\beta}{}^{MNP}\,M^{\alpha\beta} \\[2mm]
& &+12\,\xi_{\alpha M}\,\xi_{\beta N}\,M^{\alpha \beta}M^{MN} \ .
\end{array}
\end{equation}
As previously stated, when all the fields are independent of the internal coordinates, the first and second terms in \eqref{eq:potX} vanish while \eqref{eq:Vq-XQFT} reduces to the scalar potential of $\,\mathcal{N}=4\,$ gauged supergravity \cite{Schon:2006kz}. The potential in SL(2)-XFT can formally be derived by requiring invariance under generalised diffeomorphisms. The first term in \eqref{eq:potX} serves as the basis of the construction since one should reproduce the SL(2)-DFT potential by turning off the $\,X\,$ deformation. Due to the presence of the deformation in the generalised Lie derivative \eqref{eq:Liederdef}, the variation of this first term under generalised diffeomorphisms does not vanish as in SL(2)-DFT and gives terms which are linear in the deformation. In order to cancel these, one must add the second term in \eqref{eq:potX} which however also generates new terms that are quadratic in the deformation. These eventually cancel against the last term in \eqref{eq:potX}. This scheme ends here as the last term does not contain partial derivatives along the internal space and therefore does not generate variations of higher-order in the deformation. For this computation, it is crucial to recall that the $\,X\,$ deformation does not transform covariantly but as
\begin{equation}
\label{delta_X}
0=\delta_{\Lambda} X_{\hat{M}\hat{N}}{}^{\hat{P}} \neq \widetilde{\mathbb{L}}_{\Lambda} X_{\hat{M}\hat{N}}{}^{\hat{P}} =
 2 \, \partial_{[\hat{M}} \, \Lambda^{\hat{R}} \, X_{|\hat{R}|\hat{N}]}{}^{\hat{P}}  + Y^{\hat{P}\hat{Q}}{}_{\hat{R}\hat{N}} \, \partial_{\hat{Q}}\Lambda^{\hat{S}} \, X_{\hat{S}\hat{M}}{}^{\hat{R}} \ ,
\end{equation}
under deformed generalised diffeomorphisms \cite{Ciceri:2016dmd}.

A last remark can be made when $\,n=0\,$. In this case most of the $X$-dependent terms in the potential \eqref{eq:potX} can be systematically obtained by considering the \mbox{$\mathbb{Z}_2$-truncation} of the $\mathrm{E}_{7(7)}$-XFT potential in \cite{Ciceri:2016dmd}. Here one must however proceed with care as the truncated X- and quadratic constraints of $\mathrm{E}_{7(7)}$-XFT might be stronger than the constraints of SL(2)-XFT \eqref{consistency_constraints}, and therefore could implicitly prohibit the presence of certain terms originally present in \eqref{eq:potX}. In fact, it is already known from the supergravity analysis of \cite{Dibitetto:2011eu} that, after the $\mathbb{Z}_2$-truncation, the quadratic constraints of \Eseven-XFT correspond to the set in \eqref{QC_N=4} supplemented with two additional quadratic constraints (see eq.(\ref{QC_N=8_extra}) below). It can also be shown (see appendix \ref{sec:constraint_appendix}) that the truncated \mbox{X-constraint} of \Eseven-XFT is in one-to-one correspondance with the X- and \mbox{C-constraints} of SL(2)-XFT. For these reasons, the $\mathbb{Z}_2$-truncation of the potential in \Eseven-XFT must yield the full expression of the cross-term \eqref{eq:Vcross-XQFT} but only part of \eqref{eq:Vq-XQFT}. Indeed, due to the two extra quadratic constraints, the first term of the second line is restricted to its anti-self-dual part while the second term in the same line is absent.

\subsubsection{Deformations of the Type I/Heterotic theory}

Let us solve the section constraint in (\ref{consistency_constraints}) by allowing the fields and parameters of the theory to depend only on the Type~I/Heterotic $\, y^{m} \equiv y^{+m}\,$ internal coordinates in (\ref{extended_coords}), namely
\begin{equation}
\partial_{m} \equiv \partial_{+m} \neq 0  
\hspace{8mm} \textrm{ and } \hspace{8mm}
\partial_{+}{}^{m} = \partial_{-m}= \partial_{-}{}^{m} =\partial_{\alpha A} = 0 \ .
\end{equation}
An analysis of the X-constraint in (\ref{X-constraint_irreps}) reveals that the only deformations that are allowed are of the form
\begin{equation}
\xi_{+m} \,\,\, , \,\,\,  \xi_{+A}
\hspace{7mm} \textrm{ and } \hspace{7mm}
f_{\alpha mnp} \,\,\, , \,\,\,
f_{\alpha mn}{}^{C} \,\,\, , \,\,\,
f_{\alpha m}{}^{BC} \,\,\, , \,\,\,
f_{\alpha}{}^{ABC} \ .
\end{equation}
However the C-constraint in (\ref{C-constraint_irreps}) imposes $\,f_{- MNP}=0\,$, thus leaving a final set of deformations
\begin{equation}
\label{deformation_indep}
\xi_{+m} \,\,\, , \,\,\,  \xi_{+A}
\hspace{7mm} \textrm{ and } \hspace{7mm}
f_{+ mnp} \,\,\, , \,\,\,
f_{+ mn}{}^{C} \,\,\, , \,\,\,
f_{+ m}{}^{BC} \,\,\, , \,\,\,
f_{+}{}^{ABC} \ .
\end{equation}

The above parameters have an interpretation in the context of the Type~I/Heterotic theory. First, it is worth noticing that $\,\xi_{+A}\,$ is set to zero by the first quadratic constraint in (\ref{QC_N=4}). Then the remaining parameters in (\ref{deformation_indep}) have the following interpretation
\begin{equation}
\label{deformation_physics}
\begin{array}{lll}
\xi_{+m}  & : & \quad \text{dilaton flux}   \ , \\[2mm]
f_{+mnp} & : &  \quad  H_{mnp} \,\, \text{ flux } \,\, (\textrm{for $C_2$ in Type~I or $B_{2}$ in Heterotic})    \ , \\[2mm]
f_{+mn}{}^{C} & : &  \quad  F_{mn}{}^{C} \,\, \text{ gauge flux }   \ , \\[2mm]
f_{+m}{}^{BC}  & : & \quad   \text{\O(n) Scherk--Schwarz flux (compact)}   \ , \\[2mm]
f_{+}{}^{ABC}  & : &  \quad   \text{Yang--Mills gauge group in 10D}   \ .
\end{array}
\end{equation}
Amongst the above deformations only the Yang--Mills structure constants $\,f_{+}{}^{ABC}\,$ cannot be generated by field redefinitions in the undeformed SL(2)-DFT theory. The reason being that they correspond to a non-Abelian deformation already in ten dimensions. In contrast, the $\,F_{m}{}^{BC} \equiv f_{+m}{}^{BC}\,$ deformations can be obtained by an \O(n)-valued Scherk--Schwarz-like redefinition of (the internal components of) the ten-dimensional gauge vectors
\begin{equation}
A_{m}{}^{A}(x,y) \to A_{m}{}^{B}(x,y) \,  E_{B}{}^{A}(y)
\hspace{5mm} \textrm{ with } \hspace{5mm}
E_{B}{}^{A}(y)  \in  \O(n) \ .
\end{equation}
It is worth mentioning that the quadratic constraints (\ref{QC_N=4}) still impose further restrictions on the deformations (\ref{deformation_physics}). For example, in the absence of any other deformations, the $\,F_{mn}{}^{C}\equiv f_{+mn}{}^{C}\,$ are required to be invariant under the ten-dimensional gauge group specified by $\,f^{ABC}\equiv f_{+}{}^{ABC}\,$. In other words, only Abelian field strengths $\,F_{mn}{}^{C}\,$ can induce a deformation by themselves. This restriction is modified in the presence of other deformations. Also, when taking the DFT limit and restricting the deformation only to the Yang--Mills piece $\,f_{+}{}^{ABC}\neq\,0\,$, the potential \eqref{eq:potX} reduces to the potential in \cite{Hohm:2011ex} for the DFT formulation of Heterotic strings coupled to $\,\nv=n\,$ non-Abelian vector fields.

Except for the Yang--Mills structure constants $\,f^{ABC}\,$, all the $f$-type deformations in (\ref{deformation_physics}) can be generated as a Scherk--Schwarz-like redefinition of the vector fields $\,A_\mu{}^{\alpha M} (x,y) \to A_\mu{}^{\alpha N} (x,y) \,\,  E(y)_N{}^M \,$ with
\begin{equation}
\label{E-matrix}
E(y) = \exp 
\left(
\begin{array}{cc|c}
	0 & b_{mn}  &  a_{m}{}^{B} \\
	0 & 0 &  0 \\
	\hline
	0 & -a_{An} &   k_{A}{}^{B}
\end{array} \right)
\in \SOp(6,6+n) \ .
\end{equation}
The $\,E(y)\,$ matrix (\ref{E-matrix}) is the most general one satisfying the $E$-constraint of \cite{Ciceri:2016dmd}, namely $\,E_M{}^N \partial_{\alpha N}= \delta_{M}^{N} \, \partial_{\alpha N}\,$, after choosing the Type~I/Heterotic solution of the section constraints. The associated torsion yields the $f$-type deformations above. Schematically,
\begin{equation}
H_{(3)} \sim db_{(2)} + CS(a_{(1)}) 
\hspace{5mm} \textrm{ , } \hspace{5mm}
F_{(2)}{}^A \sim da_{(1)}{}^B  \, (e^{k}){}_{B}{}^{A} 
\hspace{6mm} \textrm{ and } \hspace{6mm}
F_{(1)}{}_{A}{}^B \sim d k_{A}{}^{B} \ ,
\end{equation}
where $\,CS(a_{(1)})\,$ is the non-Abelian Chern-Simons term entering the $\,H_{(3)}\,$ field strength in $\,\cN=1\,$ ten-dimensional supergravity. Notice that, while $\,b_{(2)}\,$ and $\,a_{(1)}{}^A\,$ can be regarded as background values for scalar fields in the theory, the algebra-valued $\,k_{A}{}^{B} \in \mathfrak{so}(n)\,$ cannot and simply induces an \SO(n) redefinition of the gauge vectors.

\section{Scherk--Schwarz reductions and de Roo--Wagemans angles}
\label{sec:Scherk-Schwarz}

Thus far, one of the most successful applications of extended field theories has been the derivation of consistent reduction ans\"atze of 11D/10D supergravities on non-trivial internal spaces by performing generalised Scherk--Schwarz (SS) reductions. While most of the results are in the context of exceptional field theories \cite{Lee:2014mla,Hohm:2014qga,Baguet:2015sma,Cassani:2016ncu,Malek:2016bpu}, there are also interesting constructions in DFT \cite{Baguet:2015iou}. However, generalised SS reductions of DFT \cite{Geissbuhler:2011mx,Grana:2012rr} only produce electric gaugings of $\,\cN=4\,$ supergravity: non-trivial de Roo--Wagemans angles \cite{deRoo:1985jh} cannot be generated due to the absence of the $\,\textrm{SL}(2)\,$ factor in the duality group. The resulting scalar potential cannot accommodate de Sitter (dS) or anti-de Sitter (AdS) vacua but only Minkowski or domain-wall solutions. In other words, full moduli stabilisation including the SL(2) dilaton $\,S\,$ in (\ref{eq:action-QFT_2}) is \textit{not} possible in ordinary DFT.

The SL(2)-DFT constructed here includes the relevant SL(2) factor in the duality group and potentially allows for generalised SS reductions producing $\,\cN=4\,$ gaugings at non-trivial SL(2) de Roo--Wagemans angles. However such gaugings at SL(2) angles turn to require a non-trivial dependence of the fields on both $\,y^{+}\,$ and $\,y^{-}\,$ types of coordinates simultaneously, thus violating the section constraints (\ref{sec_cons-QFT_2}). This issue is the SL(2) analog of the violation of the $\,\textrm{O}(d,d+n)\,$ section constraint in DFT. Although the construction of SL(2)-DFT strongly relies on imposing these constraints, we will still proceed and look at the classes of $\,\cN=4\,$ gaugings with $\,n=0\,$ that are accessible as generalised SS reductions when they are relaxed. Similarly to what has been done in DFT \cite{Geissbuhler:2013uka}, developing a flux formulation of SL(2)-DFT would help to understand this and other related issues. This goes beyond the scope of the paper and will be investigated somewhere else.

\subsection{Generalised frames and torsion}

Our starting point is a generalised frame matrix $\,(E^{-1})_{\underline{\alpha M}}{}^{\beta N}(y) \in \bbR^+ \times  \textrm{SL(2)}\times \textrm{O}(6,6)\,$ for the extended internal space\footnote{The frame $\,E_{\alpha M}{}^{\underline{\beta N}}\,$ could still be $\,(x,y)\,$ dependent if we regarded it as the generalised frame in a frame formulation of SL(2)-DFT. We are not considering this possibility here.} taking the general form
\begin{equation}
\label{twist_matrix}
(E^{-1})_{\underline{\alpha M}}{}^{\beta N}  =  e^{-\lambda} \, (e^{-1})_{\underline{\alpha}}{}^{\beta} \, (U^{-1})_{\underline M}{}^N \ ,
\end{equation}
where $\,e^{\lambda(y)}\in \bbR^+\,$, $\,e_{\alpha}{}^{\underline\alpha}(y) \in \SL(2,\bbR)\,$ and $\,U_M{}^{\underline M}(y) \in \SO(6,6)\,$. From now on we will denote $\,(E^{-1})_{\underline{\alpha M}}{}^{\beta N}  \equiv E_{\underline{\alpha M}}{}^{\beta N}\,$, and similarly for $\,(U^{-1})_{\underline M}{}^N\,$ and $\,(e^{-1})_{\underline{\alpha}}{}^{\beta}\,$, whenever we write indices explicitly. In a Scherk--Schwarz like reduction of SL(2)-DFT, the frame (\ref{twist_matrix}) is used to factorise the internal space $\,y^{\a M}\,$ dependence of the fields. Consequently, quantities with underlined indices correspond to four-dimensional ($x^{\mu}$ dependent) ones.

Applying a generalised diffeomorphism (\ref{L-QFT_0}) on a vector field $\,E_{\underline{\b N}}\,$ with parameter $\,E_{\underline{\a M}}\,$, where $\,\underline{\a M}\,$ and $\,\underline{\b N}\,$ must be understood as labels, one encounters
\begin{equation}
\mathbb{L}_{E_{\underline{\a M}}} E_{\underline{\b N}} = - X_{\underline{\a M \b N}}{}^{\underline{\g P}} \,\, E_{\underline{\g P}} \ .
\end{equation}
Following the procedure in exceptional generalised geometry \cite{Aldazabal:2013mya,Cederwall:2013naa}, the torsion $\,X_{\underline{\a M \b N}}{}^{\underline{\g P}}\,$ can be written as
\begin{equation}
\label{X-torsion}
	X_{\underline{\alpha M}\,\underline{\beta N}}{}^{\underline{\gamma P}} \,=\, 
	 W_{\underline{\alpha M}\,\underline{\beta N}}{}^{\underline{\gamma P}} 
	\,-\, W_{\underline{\beta N}\,\underline{\alpha M}}{}^{\underline{\gamma P}} 
	\,+\, Y^{\underline{\gamma P}\,\underline{\delta Q}}{}_{\underline{\lambda R}\,\underline{\beta N}} 
	 W_{\underline{\delta Q}\,\underline{\alpha M}}{}^{\underline{\lambda R}} \ ,
\end{equation}
in terms of the Weitzenb\"ock connection
\begin{equation}
W_{\underline{\alpha M}\,\underline{\beta N}}{}^{\underline{\gamma P}} = 
e^{-\lambda} e_{\underline \alpha}{}^{\alpha} U_{\underline M}{}^M 
\left[ \delta_{\underline \beta}{}^{\underline \gamma} \, (U^{-1}\partial_{\alpha M} U)_{\underline N}{}^{\underline P} \,+\,   \delta_{\underline N}{}^{\underline P} \, (e^{-1}\partial_{\alpha M} e)_{\underline \beta}{}^{\underline \gamma}  
 	 +  \delta_{\underline \beta}{}^{\underline \gamma} \,  \delta_{\underline N}{}^{\underline P} \,  \partial_{\alpha M} \lambda \right] \ .
\end{equation}
The torsion (\ref{X-torsion}) can be decomposed into the same irreducible pieces as the embedding tensor of a (trombone) gauging of $\,\cN=4\,$ supergravity, namely
\begin{equation}
\label{irrep_torsion}
X_{\underline{\alpha M}\,\underline{\beta N} \,\underline{\gamma P}} =
	-\eps_{\underline{\beta\gamma}} f_{\underline{\alpha\,MNP}}
	-\eps_{\underline{\beta\gamma}} \eta_{\underline{M}[\underline{N}} (\xi_{|\underline{\alpha}|\underline{P}]} { +2 \, \vartheta_{|\underline{\alpha}|\underline{P}]} })
	-\eps_{\underline{\alpha}(\underline{\beta}} \xi_{\underline{\gamma})\underline{M}} \eta_{\underline{NP}} 
 - { \vartheta_{\underline{\alpha M}}} \eps_{\underline{\beta\gamma}} \eta_{\underline{NP}}  \ .
\end{equation}
In (\ref{irrep_torsion}) we have included the trombone gauging parameter \footnote{Notice that the trace $\,X_{\underline{\alpha M} \underline{\beta N}}{}^{\underline{\alpha M}} \neq0\,$ even when the trombone component vanishes. This differs from the maximally supersymmetric case.} $\,\vartheta_{\underline{\alpha M}}\,$ which is not present in the embedding tensor deformation \eqref{X_N4} of the $\,\cN=4\,$ supergravity action \cite{Schon:2006kz}. The first two terms in \eqref{irrep_torsion} gauge a subalgebra of $\,\SO(6,6)\,$, whereas the last two terms gauge respectively a subalgebra of \SL(2) and the trombone $\bbR^+$.
The expressions for the irreducible components in the torsion are given by:\footnote{One could in principle redefine $\xi_{\underline{\alpha M}}$ by terms proportional to $\vartheta_{\underline{\alpha M}}$ (and/or vice-versa) and appropriately modify the last three terms in \eqref{irrep_torsion}.
Our definitions are unambiguous in that we identify $\,\xi_{\underline{\alpha M}}\,$  with the source of \SL(2) gauging and $\,\vartheta_{\underline{\alpha M}}\,$ with the trombone one.} 
\begin{equation}
\label{gauging_N=4_trombone}
\begin{array}{rll}
f_{\underline{\alpha MNP}} &=& -3 \, e^{-\lambda} \, e_{\underline \alpha}{}^{\alpha} \, \eta_{\underline{Q}[\underline{M}} \,   U_{\underline N}{}^N  \, U_{\underline P]}{}^P \,\,  \partial_{\alpha N} U_{P}{}^{\underline{Q}}   \ , \\[2mm]
\xi_{\underline{\alpha M}} &=& e^{-\lambda} U_{\underline M}{}^M \partial_{\alpha M} e_{\underline \alpha}{}^{\alpha} - e^{-\lambda} e_{\underline \alpha}{}^{\alpha} \partial_{\alpha M} U_{\underline M}{}^M  
  + e^{-\lambda} e_{\underline \alpha}{}^{\alpha} U_{\underline M}{}^M \partial_{\alpha M} \lambda \ , \\[2mm]
\vartheta_{\underline{\alpha M}} &=& \tfrac12 e^{-\lambda} \partial_{\alpha M} ( U_{\underline M}{}^M e_{\underline \alpha}{}^{\alpha} ) -\tfrac32 e^{-\lambda} e_{\underline \alpha}{}^{\alpha} U_{\underline M}{}^M  \partial_{\alpha M} \lambda   \ .
\end{array}
\end{equation}
A generalised Scherk--Schwarz reduction requires these three objects to be constant.
Requiring no trombone gaugings, \textit{i.e.} $\,\vartheta_{\underline{\alpha M}}=0\,$, corresponds to a generalised unimodularity condition for the SS ansatz, which guarantees  consistency of the reduction not only at the level of the EOM's but also at the level of the actions (at least as long as the internal space is compact).

\subsubsection*{DFT limit and electric gaugings}
 
In order to make contact with some of the results found in the DFT literature we must impose the DFT limit (\ref{DFT_limit-partial}) so that $\,\partial_{-P} E_{\underline{\alpha M}}{}^{\beta N} =0\,$. As a consequence, only $\,e_{\underline\alpha}{}^+\,$ appears in the torsion pieces (\ref{gauging_N=4_trombone}). We will also assume the unimodularity condition $\,\vartheta_{\underline{\alpha M}}=0\,$. The requirement of constant $\,\xi_{\underline{\alpha M}}\,$ and $\,f_{\underline{\alpha MNP}}\,$ then implies $\,e_{\underline{\,+\,}}{}^+ \propto e_{\underline{\,-\,}}{}^+\,$ with a coordinate-independent proportionality constant. Applying then a constant \SL(2,\bbR) transformation in order to set $\,e_{\underline{\,-\,}}{}^+ = 0\,$\footnote{This is a duality transformation in the truncated four-dimensional theory, \textit{i.e.} the dualisation acts on the `flat' index $\underline\alpha$ and does not affect the internal derivatives.}, one sees that all four-dimensional $\,\cN=4\,$ gauged supergravities that can be obtained from (locally) geometric generalised Scherk--Schwarz reductions of ten-dimensional $\,\cN=1\,$ supergravity, or even from locally non-geometric reductions of DFT, only give rise to electric gaugings. Namely, gaugings that satisfy $\,f_{\underline{-MNP}}=\xi_{\underline{-M}}=0\,$, possibly up to a duality redefinition.

Following the above reasoning we now recover the explicit expressions for the torsion in ref~\cite{Geissbuhler:2011mx}. We will assume dependence on $\,y^{+M}\,$ coordinates only and restrict the \SL(2,\bbR) twist matrix as
\begin{equation}
\label{e_for_Geissbuhler}
e_{\underline \alpha}{}^\alpha =\begin{pmatrix}
	e^{\lambda_{2}} & e^{\lambda_{2}} f(y^{+M})\\ 0 & e^{-{\lambda_{2}}}
\end{pmatrix} \ .
\end{equation}
There is no loss of generality in such a restriction as long as we impose unimodularity, which we will at due time. The function $\,f(y^{+M})\,$ is arbitrary and drops out entirely from the torsion. Then, all $\,\underline\alpha = -\,$ components of the torsion irrep's vanish and the other ones reduce to
\begin{equation}
\label{Torsion_Geissbuhler}
\begin{array}{rlll}
f_{\underline{+ MNP}} &=& -3 \, e^{(\lambda_{2}-\lambda)}\,  \eta_{\underline{Q}[\underline{M}} \,   U_{\underline N}{}^N  \, U_{\underline P]}{}^P \,  \partial_{+N} U_{P}{}^{\underline{Q}} & ,\\[2mm]
\xi_{\underline{+ M}} &=&  e^{(\lambda_{2}-\lambda)} \, \left[  U_{\underline M}{}^M \partial_{+M} (\lambda + \lambda_{2}) - \,\partial_{+M} U_{\underline M}{}^M  \right] & ,\\[2mm]
-2 \, \vartheta_{\underline{+ M}} &=& e^{(\lambda_{2}-\lambda)} \, \left[  U_{\underline M}{}^M \partial_{+M} (3 \lambda -\lambda_{2}) - \,\partial_{+M} U_{\underline M}{}^M  \right] & .
\end{array}
\end{equation}
Performing a bit of algebra we notice that once we set to zero the trombone component, $\,\vartheta_{\underline{+ M}}=0\,$, there are some equivalent ways to write $\,\xi_{\underline{+M}}\,$:
\begin{equation}
\label{Relations_Geissbuhler_2}
\xi_{\underline{+ M}} =  2 \,    U_{\underline M}{}^M \partial_{+M} (e^{(\lambda_{2}-\lambda)}) 
= e^{(\lambda_{2}-\lambda)} \, \left[  - 2 \,  \partial_{+M}  U_{\underline M}{}^M + 4 \, U_{\underline M}{}^M  \,  \partial_{+ M} \lambda \right]  \ .
\end{equation}
These two relations were identified in \cite{Geissbuhler:2011mx} as necessary conditions for the Scherk--Schwarz reduction of DFT to produce an $\cN=4$ gauged supergravity. The first one is needed for the (external) three-form field strength $\,H_{\mu\nu\rho}\,$ obtained upon reduction to match the gauged supergravity form \cite{Schon:2006kz}. This is
\begin{equation}
\begin{array}{llll}
H_{\mu \nu \rho} &=& 3 \, \partial_{[\mu} B_{\nu \rho]} - 3 {A_{[\mu}}^{\underline{M}} \, B_{\nu \rho]} \,\,\, 2 \, U_{\underline M}{}^M \, \partial_{+M}(e^{(\lambda_{2}-\lambda)}) + ...  \\[2mm]
&\overset{!}{=}& 3 \, \partial_{[\mu} B_{\nu \rho]} - 3 {A_{[\mu}}^{\underline{M}} \, B_{\nu \rho]} \,\,\, \xi_{\underline{+M}} + ...  \ .
\end{array}
\end{equation}
The second one is needed to recover the scalar potential of the $\,\cN=4\,$ gauged supergravity. Finally, the identification between the twist parameters here (left) and in ref.~\cite{Geissbuhler:2011mx} (right) reads: $\,U_{\underline{M}}{}^M=E_{\underline{M}}{}^M\,$, $\,{\lambda=d}\,$ and $\,(\lambda_{2}-\lambda)=\frac{\gamma}{2}\,$.

\subsection{$\textrm{SO}(3)^{(4-p)} \times \textrm{U}(1)^{3p}\,$ gaugings at SL(2) angles}

In this section we present twist matrices (\ref{twist_matrix}) whose associated torsion reproduces the embedding tensor of families of $\,\textrm{SO}(3)^{(4-p)} \times \textrm{U}(1)^{3p}\,$ gaugings of $\,\cN=4\,$ supergravity with $\,p=0,...,4\,$.\footnote{No fundamental matter is charged under the \U(1) factors.} These include the most general family of $\,\textrm{SO}(4) \times \textrm{SO}(4)\,$ gaugings ($p=0$) studied in \cite{Inverso:2015viq}. To this end, we will construct generalised frames with $\,\lambda=0\,$ and $\,{e}_{\underline\alpha}{}^{\alpha}=\delta_{\underline\alpha}{}^{\alpha}\,$, namely
\begin{equation}
\label{twist_matrix_SO4xSO4}
(E^{-1})_{\underline{\alpha M}}{}^{\beta N}  =   \delta_{\underline{\alpha}}{}^{\beta} \, (U^{-1})_{\underline M}{}^N \ ,
\end{equation}
where $\,U  \in \textrm{SO}(6,6)\,$ depends on both $\,y^{+M}\,$ and $\,y^{-M}\,$ coordinates, thus violating the section constraints (\ref{sec_cons-QFT_2}). 
The form of the frame in (\ref{twist_matrix_SO4xSO4}) implies that the unimodularity condition $\,\vartheta_{\underline{\alpha M}}=0\,$ translates into $\,\partial_{\a M}U_{\underline{M}}{}^{M}=0\,$ and automatically implies $\,\xi_{\underline{\alpha M}}=0\,$. When using light-cone coordinates, the $\,U\,$ twist matrix in \eqref{twist_matrix_SO4xSO4} can be \mbox{parameterised as}
\begin{equation}
\begin{array}{llll}
U_{M}{}^{\underline N} (y^{\a M}) &=& 
\left(  
\begin{array}{cc}
 \mathbb{I}_{6} & 0_{6} \\
 \beta & \mathbb{I}_{6}
 \end{array}
 \right)
 \left(  
\begin{array}{cc}
 \mathbb{I}_{6} & b \\
 0_{6} & \mathbb{I}_{6}
 \end{array}
 \right)
 \left(  
\begin{array}{cc}
 u & 0_{6} \\
 0_{6} & u^{-t}
 \end{array}
 \right) & ,\\[7mm]
 &=&
  \left(  
\begin{array}{cc}
 u_{m}{}^{\underline n} & b_{m p} \, (u^{-t})^{p}{}_{\underline n} \\
 \beta^{mp} \, u_{p}{}^{\underline n} & (u^{-t})^{m}{}_{\underline n} + \beta^{mp} \, b_{pq} \, (u^{-t})^{q}{}_{\underline n}
 \end{array}
 \right) & ,
 \end{array}
\end{equation}
with $\,y^{\alpha M}=(y^{\alpha m},y^{\alpha \bar{m}})\,$ and $\,m=1,...,6\,$. For the sake of simplicity, we will consider sub-classes of twist matrices of the form
\begin{equation}
\label{U_SO33xSO33}
U \in \textrm{SO}(3,3)_{(1)} \times \textrm{SO}(3,3)_{(2)} \subset \textrm{SO}(6,6) \ .
\end{equation}
This translates into a further splitting of coordinates of the form $\,y^{\alpha m}=(y^{\alpha a},y^{\alpha i})\,$, $\,{y^{\alpha \bar{m}}=(y^{\alpha \bar{a}},y^{\alpha \bar{i}})}\,$ with $\,a=1,2,3\,$, $\,i=4,5,6\,$,  and a block-diagonal structure of the twist parameters 
\begin{equation}
\label{twisting-param}
\begin{array}{llll}
\hspace{-1mm}
\beta^{mn}=
\left(  
\begin{array}{cc}
 (\beta_{(1)})^{ab} & 0_{3} \\
 0_{3} & (\beta_{(2)})^{ij}
 \end{array}
 \right)
\hspace{0mm} , \hspace{0mm}
b_{mn}=
\left(  
\begin{array}{cc}
 (b_{(1)})_{ab} & 0_{3} \\
 0_{3} & (b_{(2)})_{ij}
 \end{array}
 \right)
\hspace{0mm} , \hspace{0mm}
u_{m}{}^{\underline n}=
\left(  
\begin{array}{cc}
 (u_{(1)})_{a}{}^{\underline b} & 0_{3} \\
 0_{3} & (u_{(2)})_{i}{}^{\underline j}
 \end{array}
 \right) ,
\end{array}
\end{equation}
where the $\,_{(1),(2)}\,$ labels refer to the $\,\textrm{SO}(3,3)_{(1),(2)}\,$ factors, respectively. We refer the reader to \cite{Dibitetto:2012rk} for an account on $\,\textrm{SO}(3,3)\,$ twist matrices in the context of generalised SS reductions of DFT to 7D half-maximal gauged supergravity.

The general families of $\,\textrm{SO}(3)^{(4-p)} \times \textrm{U}(1)^{3p}\,$ gaugings of $\,\cN=4\,$ supergravity are obtained from twisting parameters (\ref{twisting-param}) of the form
\begin{equation}
\label{twist_matrices_pieces_1}
\begin{array}{lll}
u_{(1),(2)} &=&
\left(
\begin{array}{ccc}
 1 & 0 & 0 \\[2mm]
 0 & \frac{1}{2} \, (\cos Y_{(1),(2)} + \cos \widetilde{Y}_{(1),(2)}) &  - \frac{1}{2} \, (\sin Y_{(1),(2)}+\sin \widetilde{Y}_{(1),(2)}) \\[2mm]
0  & \frac{1}{2} \, (\sin Y_{(1),(2)} + \sin \widetilde{Y}_{(1),(2)})  & \frac{1}{2} \, (\cos Y_{(1),(2)} + \cos \widetilde{Y}_{(1),(2)}) 
\end{array}
\right) \ , 
\end{array}
\end{equation}
\begin{equation}
\label{twist_matrices_pieces_2}
\begin{array}{lll}
b_{(1),(2)} &=&
\left(
\begin{array}{ccc}
0 & 0 & 0 \\
0 & 0 & \frac{1}{2} \, \sin(Y_{(1),(2)} - \widetilde{Y}_{(1),(2)} ) \\
0 & - \frac{1}{2} \, \sin(Y_{(1),(2)} - \widetilde{Y}_{(1),(2)} )   & 0  \\
\end{array}
\right) \ , 
\end{array}
\end{equation}
and
\begin{equation}
\label{twist_matrices_pieces_3}
\begin{array}{lll}
\beta_{(1),(2)} & = &
\left(
\begin{array}{ccc}
0 & 0 & 0 \\
0 & 0 &  \, \tan \left( \frac{1}{2} \, (Y_{(1),(2)} - \widetilde{Y}_{(1),(2)} ) \right) \\
0 & -  \tan \left( \frac{1}{2} \, (Y_{(1),(2)} - \widetilde{Y}_{(1),(2)} ) \right)  & 0  \\
\end{array}
\right) \ ,
\end{array}
\end{equation}
which depend on four linear combinations of coordinates given by
\begin{equation}
\label{Y_coord_SO(4)xSO(4)}
\begin{array}{cclc}
Y_{(1)}&=&  (\omega^{+}_{1} - h^{+}_{1}) \,  (y^{+1}-y^{+\bar{1}}) + (\omega^{-}_{1}- h^{-}_{1}) \, (y^{-1}-y^{-\bar{1}}) & , \\[2mm]
\widetilde{Y}_{(1)}&=& (\omega^{+}_{1}+h^{+}_{1}) \,  (y^{+1}+y^{+\bar{1}})  + (\omega^{-}_{1}+h^{-}_{1}) \,  (y^{-1}+y^{-\bar{1}})  & , \\[4mm]
Y_{(2)}&=& (\omega^{+}_{2}-h^{+}_{2}) \,  ( y^{+4}-y^{+\bar{4}} ) + (\omega^{-}_{2}-h^{-}_{2}) \, (y^{-4}-y^{-\bar{4}}) & ,\\[2mm]
\widetilde{Y}_{(2)}&=&  (\omega^{+}_{2}+h^{+}_{2}) \,   ( y^{+4}+y^{+\bar{4}} )  + (\omega^{-}_{2}+h^{-}_{2}) \, (y^{-4}+y^{-\bar{4}}) & .
\end{array}
\end{equation}
These gaugings are specified by eight arbitrary parameters that activate sixteen components inside the $\,f_{\underline{\alpha MNP}}\,$ piece of the torsion:
\begin{equation}
\label{chain_fluxes_1}
\begin{array}{llllllll}
f_{\underline{+ abc}} = h^{+}_{1} 
& \,\, , \,\, &
f_{\underline{+ ab\bar{c}}} = \omega^{+}_{1}
& \,\, , \,\, &
f_{\underline{+ \bar{a}\bar{b} c}} =  h^{+}_{1}
& \,\, , \,\, &
f_{\underline{ + \bar{a}\bar{b}\bar{c} }} =  \omega^{+}_{1}  & ,   \\[2mm]
f_{\underline{+ ijk}}= h^{+}_{2}  
& \,\, , \,\, &
f_{\underline{+ ij\bar{k}}} =  \omega^{+}_{2}  
& \,\, , \,\, &
f_{\underline{+ \bar{i}\bar{j} k}} = h^{+}_{2} 
& \,\, , \,\, &
f_{\underline{+ \bar{i}\bar{j}\bar{k}}}=  \omega^{+}_{2} &  ,   \\[4mm]
f_{\underline{- abc}} =  h^{-}_{1}
& \,\, , \,\, &
f_{\underline{- ab\bar{c}}} = \omega^{-}_{1} 
& \,\, , \,\, &
f_{\underline{- \bar{a}\bar{b} c}} = h^{-}_{1}
& \,\, , \,\, &
f_{\underline{- \bar{a}\bar{b}\bar{c}}} = \omega^{-}_{1}  & , \\[2mm]
f_{\underline{- ijk}}=  h^{-}_{2} 
& \,\, , \,\, &
f_{\underline{- ij\bar{k}}} =  \omega^{-}_{2}
& \,\, , \,\, &
f_{\underline{- \bar{i}\bar{j} k}} =  h^{-}_{2}
& \,\, , \,\, &
f_{\underline{- \bar{i}\bar{j}\bar{k}}}=  \omega^{-}_{2} & .
\end{array}
\end{equation}
The eight arbitrary parameters can be mapped to four gauge couplings and four SL(2) orientations, one pair for each $\,\textrm{SO}(3)\,$ or $\,\textrm{U}(1)^{3}\,$ factor of the gauge group.
The twist matrix $\,U\,$ constructed from (\ref{twist_matrices_pieces_1})-(\ref{twist_matrices_pieces_3}) satisfies $\,\partial_{\a M}U_{\underline{M}}{}^{M}=0\,$, which in turn implies $\,\xi_{\underline{\a M}}=\vartheta_{\underline{\a M}}=0\,$.

Let us take a closer look at the (purely $\,f\,$) four-dimensional gauge algebra determined by the commutation relations $\, [ \, X_{\underline{\alpha M}} \, , \, X_{\underline{\beta N}} ] =  f_{\underline{\alpha MN}}{}^{\underline{P}} \, X_{\underline{\beta P}}\,$. 
Moving temporarily to conventions where $\,\eta_{\underline{MN}}=\textrm{diag}(-\mathbb{I}_{6},\mathbb{I}_{6})\,$,
an analysis of the components of the embedding tensor $\,\Theta_{\underline{\alpha M}}{}^{\underline{\beta N \gamma P}}=\frac{1}{2} \, \eps^{\underline{\beta \gamma}}\, f_{\underline{\alpha M}}{}^{\underline{NP}} \,\,$ shows that these families of $\,\cN=4\,$ gaugings involve $\,\textrm{SO}(6,6)\,$ \mbox{generators} $\,t_{\underline{MN}}\,$ and vector fields $\,A_{\mu}{}^{\underline{\alpha M}}\,$ of the form
\begin{equation}
\label{SO(4)xSO(4)_parameters_1}
\textrm{SO}(3,3)_{(1)} \left\lbrace
\hspace{2mm}
\begin{array}{llllll}
t_{\underline{ab}} & \hspace{-1mm} : \hspace{1mm} &    (\omega^{+}_{1}-h^{+}_{1}) \,  \eps_{\underline{abc}} \, A_{\mu}{}^{\underline{+c}} & \hspace{-2mm} + & (\omega^{-}_{1}-h^{-}_{1}) \,  \eps_{\underline{abc}} \, A_{\mu}{}^{\underline{-c}} & , \\[2mm]
t_{\underline{(6+a)(6+b)}} & \hspace{-1mm} : \hspace{1mm} &    (\omega^{+}_{1}+ h^{+}_{1}) \,  \eps_{\underline{abc}} \, A_{\mu}{}^{\underline{+(6+c)}} & \hspace{-2mm} + & (\omega^{-}_{1}+h^{-}_{1}) \,  \eps_{\underline{abc}} \, A_{\mu}{}^{\underline{-(6+c)}} & ,
\end{array} \right.
\end{equation}
\begin{equation}
\label{SO(4)xSO(4)_parameters_2}
\textrm{SO}(3,3)_{(2)} \left\lbrace
\hspace{2mm}
\begin{array}{llllll}
t_{\underline{ij}} & \hspace{-1mm} : \hspace{1mm} &    (\omega^{+}_{2}-h^{+}_{2}) \,  \eps_{\underline{ijk}} \, A_{\mu}{}^{\underline{+k}} &\hspace{-2mm}+ & (\omega^{-}_{2}-h^{-}_{2}) \,  \eps_{\underline{ijk}} \, A_{\mu}{}^{\underline{-k}} & , \\[2mm]
t_{\underline{(6+i)(6+j)}} & \hspace{-1mm} : \hspace{1mm} &    (\omega^{+}_{2}+ h^{+}_{2}) \,  \eps_{\underline{ijk}} \, A_{\mu}{}^{\underline{+(6+k)}} &\hspace{-2mm} + & (\omega^{-}_{2}+h^{-}_{2}) \,  \eps_{\underline{ijk}} \, A_{\mu}{}^{\underline{-(6+k)}} & .
\end{array} \right.  
\end{equation}
Each of the sets of generators $\,t_{\underline{ab}}\,$, $\,t_{\underline{(6+a)(6+b)}}\,$, $\,t_{\underline{ij}}\,$ and $\,t_{\underline{(6+i)(6+j)}}\,$ corresponds to an SO(3) factor inside $\,\textrm{SO}(3,3) \times \textrm{SO}(3,3)\,$. By taking identifications amongst the parameters in (\ref{SO(4)xSO(4)_parameters_1}) and (\ref{SO(4)xSO(4)_parameters_2}), it is possible to decouple some of these SO(3)'s to obtain $\,{\textrm{SO}(3)^{(4-p)} \times \textrm{U}(1)^{3p}}\,$ gaugings with $\,p=0,...,4\,$. For any value of $\,p\,$, the $\,\cN=4\,$ quadratic constraints in (\ref{QC_N=4}) are satisfied. The gauging parameters in (\ref{chain_fluxes_1}) correspond then to a consistent superposition of $\,f_{\underline{\,+\,}}\,$ and $\,f_{\underline{-}}\,$ configurations, each of which contains two copies of a three-dimensional chain $\,H \rightarrow \omega \rightarrow Q \rightarrow R\,$ of non-geometric T-dual fluxes \cite{Shelton:2005cf}
\begin{equation}
\label{chain_fluxes_2}
\hspace{-1.5mm}
\begin{array}{llllllll}
f_{\underline{+ abc}} = {H^{\textrm{(+)}}}_{\underline{abc}}
& \,\, , \,\, &
f_{\underline{+ ab\bar{c}}} = {\omega^{\textrm{(+)}}}_{\underline{ab}}{}^{\underline{c}}
& \,\, , \,\, &
f_{\underline{+ \bar{a}\bar{b} c}} = {Q^{\textrm{(+)} \underline{ab}}}{}_{\underline{c}}
& \,\, , \,\, &
f_{\underline{ + \bar{a}\bar{b}\bar{c} }} =  R^{\textrm{(+)} \underline{abc}} & \hspace{-2mm} , \\[2mm]
f_{\underline{+ ijk}}= {H^{\textrm{(+)}}}_{\underline{ijk}}
& \,\, , \,\, &
f_{\underline{+ ij\bar{k}}} = {\omega^{\textrm{(+)}}}_{\underline{ij}}{}^{\underline{k}}
& \,\, , \,\, &
f_{\underline{+ \bar{i}\bar{j} k}} = {Q^{\textrm{(+)} \underline{ij}}}{}_{\underline{k}}
& \,\, , \,\, &
f_{\underline{+ \bar{i}\bar{j}\bar{k}}}= R^{\textrm{(+)} \underline{ijk}} & \hspace{-2mm} ,   \\[4mm]
f_{\underline{- abc}} = {H^{\textrm{(-)}}}_{\underline{abc}}
& \,\, , \,\, &
f_{\underline{- ab\bar{c}}} = {\omega^{\textrm{(-)}}}_{\underline{ab}}{}^{\underline{c}} 
& \,\, , \,\, &
f_{\underline{- \bar{a}\bar{b} c}} =  {Q^{\textrm{(-)} \underline{ab}}}{}_{\underline{c}}
& \,\, , \,\, &
f_{\underline{- \bar{a}\bar{b}\bar{c}}} = R^{\textrm{(-)} \underline{abc}} & \hspace{-2mm} , \\[2mm]
f_{\underline{- ijk}}= {H^{\textrm{(-)}}}_{\underline{ijk}}
& \,\, , \,\, &
f_{\underline{- ij\bar{k}}} =  {\omega^{\textrm{(-)}}}_{\underline{ij}}{}^{\underline{k}}
& \,\, , \,\, &
f_{\underline{- \bar{i}\bar{j} k}} = {Q^{\textrm{(-)} \underline{ij}}}{}_{\underline{k}} 
& \,\, , \,\, &
f_{\underline{- \bar{i}\bar{j}\bar{k}}}= R^{\textrm{(-)} \underline{ijk}} & \hspace{-2mm} .
\end{array}
\end{equation}
Hence, a higher-dimensional interpretation in terms of Type I/Heterotic T-folds \mbox{\cite{Hull:2004in,Hull:2006va}} could generically be available when $\,f_{\underline{-}}=0\,$ (or $\,f_{\underline{+}}=0\,$). 


\subsubsection*{Section constraint violating terms and non-geometry}

Section constraint violating terms have been an indicator of non-geometry in the DFT literature \cite{Dibitetto:2012rk}. More concretely, when working with a frame formulation of DFT \cite{Geissbuhler:2011mx} (see also \cite{Hohm:2011ex,Grana:2012rr}), a section constraint violating term of the form
\begin{equation}
\label{ff_(3,1)-term-DFT}
\frac{1}{6} \, e^{\phi(x)} \, f_{\underline{MNP}} \, f^{\underline{MNP}}  \ ,
\end{equation}
was introduced in order to reproduce the scalar potential of $\,\cN=4\,$ (electrically) gauged supergravity upon generalised Scherk--Schwarz reductions. The term (\ref{ff_(3,1)-term-DFT}) is just the $\,{(\a,\b)=(+,+)}\,$ component of the SL(2)-covariant expression
\begin{equation}
\label{ff_(3,1)-term}
\frac{1}{6} \, M^{\underline{\a \b}}(x) \, f_{\underline{\a MNP}} \, f_{\underline{\b}}{}^{\underline{MNP}}  \ .
\end{equation}
The contraction $\,f_{\underline{\a MNP}} \, f_{\underline{\b}}{}^{\underline{MNP}}\,$ was identified with one of the two additional quadratic constraints that must be imposed on an $\,\cN=4\,$ gauging with $\,n=0\,$ for it to be liftable to an $\,\cN=8\,$ one. More concretely, these two additional constraints read \cite{Dibitetto:2011eu}
\begin{equation}
\label{QC_N=8_extra} 
\left. f_{\underline{\a} [\underline{MNP}} \, f^{\underline{\alpha}}{}_{\underline{QRS}]} \,\, \right|_{\textrm{SD}} = 0 
\hspace{10mm} \textrm{ and } \hspace{10mm}
f_{\underline{\a MNP}} \, {f_{\underline{\b}}}^{\underline{MNP}} = 0  \ ,
\end{equation}
where SD stands for the self-dual part of the $\,\textrm{SO}(6,6)\,$ six-form. Note that if $\,f=f_{\underline{+}}\,$ then the contractions in (\ref{QC_N=8_extra}) reproduce the unique section constraint violating term \eqref{ff_(3,1)-term-DFT} of DFT. It is also worth emphasising that the second constraint in \eqref{QC_N=8_extra} can be formally extended to arbitrary $\,n\,$ whereas the first one is defined only if $\,n=0\,$. It is only in this case that the field content of the Type I/Heterotic theory can be mapped to the universal sector of the Type II theories.  We can adopt the same criterion as in DFT and use the violation of the constraints in (\ref{QC_N=8_extra}) as an indicator of non-geometry. Note that the reverse is not true: satisfying (\ref{QC_N=8_extra}) does \textit{not} guarantee the existence of a higher-dimensional description of the corresponding gauging, as we will see in a moment. 

It will prove convenient to introduce two-dimensional flux vectors $\,\vec{h}^{\pm}\equiv (h^{\pm}_{1},h^{\pm}_{2})\,$ and $\,{\vec{\omega}^{\pm}\equiv (\omega^{\pm}_{1},\omega^{\pm}_{2})}\,$. In terms of these, the explicit computation of the additional quadratic constraints (\ref{QC_N=8_extra}) in the case of the $\,{\textrm{SO}(3)^{(4-p)} \times \textrm{U}(1)^{3p}}\,$ gaugings gives
\begin{equation}
\label{QC_extra_SO(4)xSO(4)_1}
\vec{h}^{+} \times \,  \vec{h}^{-} = 0
\hspace{3mm} \textrm{ , } \hspace{3mm}
\vec{\omega}^{+} \times  \,  \vec{\omega}^{-} = 0 
\hspace{5mm} \textrm{and} \hspace{5mm}
\vec{h}^{+} \cdot \, \vec{\omega}^{-} =  \vec{h}^{-} \cdot \, \vec{\omega}^{+} \ ,
\end{equation}
coming from the first (SD) condition, as well as 
\begin{equation}
\label{QC_extra_SO(4)xSO(4)_2}
\vec{h}^{+} \cdot \,  \vec{\omega}^{+} = 0
\hspace{3mm} \textrm{ , } \hspace{3mm}
\vec{h}^{-} \cdot \,  \vec{\omega}^{-} = 0
\hspace{5mm} \textrm{and} \hspace{5mm}
\vec{h}^{+} \cdot \,  \vec{\omega}^{-} = - \vec{h}^{-} \cdot \,  \vec{\omega}^{+}  \ ,
\end{equation}
coming from the second condition.  In the Type I/Heterotic solution of the section constraints, these additional constraints are not automatically satisfied due to the presence of (T-dual) non-geometric $\,Q\,$ and $\,R\,$ fluxes. Importantly, moduli stabilisation is not yet possible in this setup due to the absence of relative SL(2) orientations between the gauge factors.

\subsubsection*{$\textrm{SO}(4) \times \textrm{SO}(4)$ gaugings and $S^3\times S^3$ reduction ans\"atze}

As an example, let us look at the family of $\,\textrm{SO}(4) \times \textrm{SO}(4)\,$ gaugings $(p=0)$ which depends on the eight parameters in (\ref{chain_fluxes_1}). The counting of parameters agrees with the $\,\cN=4\,$ results of \cite{Inverso:2015viq}.\footnote{The dictionary to the parameterisation used in \cite{Inverso:2015viq} reads:
\begin{equation*}
\begin{array}{llll}
h_1^+ + \omega_1^+\equiv \tfrac{1}{\sqrt2} h_1\cos\beta_1& \hspace{5mm} , \hspace{5mm} &
h_1^- + \omega_1^-\equiv \tfrac{1}{\sqrt2} h_1\sin\beta_1 & , \\[3pt]
h_1^+ - \omega_1^+\equiv \tfrac{1}{\sqrt2} g_0\cos\alpha_0 & \hspace{5mm} , \hspace{5mm} &
h_1^- - \omega_1^-\equiv -\tfrac{1}{\sqrt2} g_0\sin\alpha_0 & , \\[3pt]
h_2^+ + \omega_2^+\equiv \tfrac{1}{\sqrt2} h_2\cos\beta_2 & \hspace{5mm} , \hspace{5mm} &
h_2^- + \omega_2^-\equiv \tfrac{1}{\sqrt2} h_2\sin\beta_2 & , \\[3pt]
h_2^+ - \omega_2^+\equiv \tfrac{1}{\sqrt2} g \cos\alpha & \hspace{5mm} , \hspace{5mm} &
h_2^- - \omega_2^-\equiv -\tfrac{1}{\sqrt2} g \sin\alpha & .
\end{array}
\end{equation*}}

A first interesting subclass of $\,\textrm{SO}(4) \times \textrm{SO}(4)\,$ gaugings is given by the choice of parameters
\begin{equation}
\label{family_2}
\vec{h}^{-} = \vec{\omega}^{-} = 0 \ .
\end{equation}
In this case the gaugings are purely electric  and can be interpreted as Type~I/Heterotic backgrounds with (T-dual) non-geometric fluxes. Of course, an analogous family with only magnetic fluxes exists. The set of additional quadratic constraints in (\ref{QC_extra_SO(4)xSO(4)_1}) and (\ref{QC_extra_SO(4)xSO(4)_2}) gives just one relation
\begin{equation}
\label{relation_electric}
\vec{h}^{+} \cdot \,  \vec{\omega}^{+} = 0 \ ,
\end{equation}
coming from the latter. According to the criterion for non-geometry stated before, a higher-dimensional interpretation of these electric $\,\textrm{SO}(4) \times \textrm{SO}(4)\,$ gaugings is only possible when (\ref{relation_electric}) holds. By further setting  $\,\vec{\omega}^{+} = 0\,$, the relation (\ref{relation_electric}) is satisfied and the frame (\ref{twist_matrix}) depends on the coordinates
\begin{equation}
\label{Y_coord_SO(4)xSO(4)_henning}
\begin{array}{llll}
Y_{(1)} =   - h^{+}_{1} \,  (y^{+1}-y^{+\bar{1}}) & \hspace{3mm} , \hspace{3mm} &
\widetilde{Y}_{(1)}=  h^{+}_{1} \,  (y^{+1}+y^{+\bar{1}})   & , \\[2mm]
Y_{(2)} = -h^{+}_{2} \,  ( y^{+4}-y^{+\bar{4}} ) & \hspace{3mm} , \hspace{3mm} &
\widetilde{Y}_{(2)} = h^{+}_{2} \,   ( y^{+4}+y^{+\bar{4}} ) & .
\end{array}
\end{equation}
One then has a realisation of the gaugings in terms of Type~I/Heterotic fluxes (\ref{chain_fluxes_2}) of the form
\begin{equation}
\label{S3xS3_solution}
H^{(+)}{}_{\underline{abc}} = Q^{(+)}{}^{\underline{ab}}{}_{\underline{c}} = h^{+}_{1}
\hspace{8mm} \textrm{ and } \hspace{8mm}
H^{(+)}{}_{\underline{ijk}} = Q^{(+)}{}^{\underline{ij}}{}_{\underline{k}} = h^{+}_{2}  \ .
\end{equation}
While the SS ansatz we provide is still a non-geometric toroidal reduction\footnote{Notice in particular that the internal space metric resulting from our ansatz is always flat.}, this case has recently been uplifted to $\,\cN=1\,$ ten-dimensional supergravity on $\,{\textrm{S}^3 \times \textrm{S}^3}\,$ in \cite{Baguet:2015iou}, giving one more example of a globally geometric compactification beyond the toroidal setup that induces non-geometric $Q$-fluxes. In addition, there is a T-dual solution of (\ref{relation_electric}) with $\,\vec{h}^{+}=0\,$ which is described in terms of fluxes $\,\omega^{(+)}{}_{\underline{ab}}{}^{\underline{c}} = R^{(+)}{}^{\underline{abc}} = \omega^{+}_{1}\,$ and $\,{\omega^{(+)}{}_{\underline{ij}}{}^{\underline{k}} = R^{(+)}{}^{\underline{ijk}} = \omega^{+}_{2}}\,$. The most general solution of (\ref{relation_electric}) contains three arbitrary parameters (two moduli and one overall phase) and involves all types of T-dual fluxes.
It is also straightforward to check that two copies of the section constraint violating $S^3$ generalised frames discussed in \cite{Lee:2015xga} can be combined into an \SO(6,6) non-geometric frame reproducing the full set of electrically gauged $\SO(4) \times \SO(4)$ gaugings.
All the twist matrices based on $S^3$ mentioned here however require a non-trivial $\lambda$ function and $e_{\underline\alpha}{}^\alpha$ matrix, as a consequence of the non-trivial warping of the resulting backgrounds. 
This makes it difficult, if not impossible, to introduce further modifications of these ans\"atze that can induce magnetic couplings and moduli stabilisation in the resulting gauging.

A second interesting subclass of $\,\textrm{SO}(4) \times \textrm{SO}(4)\,$ gaugings is given by the choice of parameters
\begin{equation}
\label{family_1}
\begin{array}{llllllll}
h^{+}_{1} = \dfrac{1+ \sin 2 \varpi}{2 \, \sqrt{2}}
& \,\, , \,\, &
\omega^{+}_{1} = -\dfrac{1-\sin 2 \varpi}{2 \, \sqrt{2}} 
& \,\, , \,\, &
h^{+}_{2} =   \dfrac{\cos 2 \varpi}{2 \, \sqrt{2}}
& \,\, , \,\, &
\omega^{+}_{2} =  \dfrac{\cos 2 \varpi}{2 \, \sqrt{2}}   & ,\\[4mm]
h^{-}_{2} = \dfrac{1-\sin 2 \varpi}{2 \, \sqrt{2}}
& \,\, , \,\, &
\omega^{-}_{2} = -\dfrac{1+ \sin 2 \varpi}{2 \, \sqrt{2}}
& \,\, , \,\, &
h^{-}_{1} = \dfrac{\cos 2 \varpi}{2 \, \sqrt{2}}  
& \,\, , \,\, &
\omega^{-}_{1} =  \dfrac{\cos 2 \varpi}{2 \, \sqrt{2}}  & .
\end{array}
\end{equation}
This one-parameter\footnote{We are denoting the parameter $\varpi$ instead of $\omega$ \cite{Dall'Agata:2012bb} in order to avoid confusion with the metric flux.} family of $\,\textrm{SO}(4) \times \textrm{SO}(4)\,$ gaugings of $\,\cN=4\,$ supergravity corresponds to the $\mathbb{Z}_{2}$-truncation of the one-parameter family of $\,\textrm{SO}(8)\,$ gaugings of $\,\cN=8\,$ supergravity presented in \cite{Dall'Agata:2012bb}. As such, they satisfy the additional quadratic constraints in (\ref{QC_extra_SO(4)xSO(4)_1}) and (\ref{QC_extra_SO(4)xSO(4)_2}) for any value of $\,\varpi\,$. The existence of an $\,\cN=1\,$ ten-dimensional origin of these $\,\cN=4\,$ gaugings has been less explored. The case $\,\varpi=0\,$ of course corresponds to a truncation to half-maximal supergravity of eleven-dimensional supergravity on $S^7$ \cite{deWit:1982ig}.
This is not the ansatz we provide here, which is instead toroidal with a coordinate dependence of the form
\begin{equation}
\label{Y_coord_SO(4)xSO(4)_w=0}
\begin{array}{llll}
Y_{(1)}=  -\frac{1}{\sqrt{2}} \,  (y^{+1}-y^{+\bar{1}}) & \hspace{3mm} , \hspace{3mm} & 
\widetilde{Y}_{(1)}= \frac{1}{\sqrt{2}} \,  (y^{-1}+y^{-\bar{1}})  & , \\[2mm]
Y_{(2)}=-\frac{1}{\sqrt{2}} \,  (y^{-4}-y^{-\bar{4}}) & \hspace{3mm} , \hspace{3mm} &
\widetilde{Y}_{(2)}= \frac{1}{\sqrt{2}} \,   ( y^{+4}+y^{+\bar{4}} )   & .
\end{array}
\end{equation}
This $\,\cN=4\,$ gauging allows for full moduli stabilisation \cite{Dibitetto:2012ia}, which prevents it from having a Type I/Heterotic description. 
It would be interesting to investigate the relation between this gauging and the one resulting from a Type IIB orientifold reduction on $\,S^{3} \times S^{3}\,$ with O3-planes, although this setup cannot be directly accommodated within SL(2)-DFT.
The case $\,\varpi \neq 0\,$ seems even more challenging as there are no-go results against a higher-dimensional lift of the $\textrm{SO}(8)\,$ gaugings to Type~II or eleven-dimensional supergravity \cite{deWit:2013ija,Lee:2015xga}.

The two subclasses of $\,\textrm{SO}(4) \times \textrm{SO}(4)\,$ gaugings we have just discussed satisfy the set of additional quadratic constraints in  (\ref{QC_extra_SO(4)xSO(4)_1}) and (\ref{QC_extra_SO(4)xSO(4)_2}). This implies that they can also be obtained from generalised Scherk--Schwarz reductions of \Eseven-EFT. 
On the contrary, genuinely $\,\cN=4\,$ gaugings not satisfying (\ref{QC_extra_SO(4)xSO(4)_1}) and (\ref{QC_extra_SO(4)xSO(4)_2}) cannot be obtained in this way but, due to the larger number of gauging parameters they contain, they represent a more promising arena for phenomenological applications like the study of moduli stabilisation in non-geometric flux backgrounds.

\subsubsection*{Remarks on moduli stabilisation} 

Let us briefly come back to the issue of moduli stabilisation in generalised Scherk--Schwarz reductions of SL(2)-DFT. We have already argued that moduli stabilisation requires non-trivial de Roo--Wagemans angles, and these a violation of the section constraints (\ref{sec_cons-QFT_2}) as the frame (\ref{twist_matrix}) must simultaneously depend on both $\,y^{+}\,$ and $\,y^{-}\,$ coordinates. The violation of the section constraints clashes with the consistency of the SL(2)-DFT, which requires them to hold at several stages in its construction. Building upon previous results in the literature obtained in a frame formulation of DFT \cite{Hohm:2011ex,Geissbuhler:2011mx,Grana:2012rr,Geissbuhler:2013uka,Cho:2015lha} and EFT \cite{Musaev:2013rq,Aldazabal:2013mya,Aldazabal:2013via}, relaxing the section constraints would require the introduction of additional section constraint violating terms in the action in order to restore invariance under gauge transformations. Such terms would encode the presence of sources in the background \cite{Dibitetto:2011eu,Aldazabal:2011yz,Geissbuhler:2013uka}. Adopting a Type~I/Heterotic description, these would include NS-branes (see \cite{Andriot:2014uda} and references therein) as well as their SL(2) duals. Only when adding sources, the full scalar potential of $\,\cN=4\,$ gauged supergravity could arise upon a generalised Scherk--Schwarz reduction of SL(2)-DFT. Their contributions to the potential, which are \textit{a priori} related to contractions like (\ref{QC_N=8_extra}) (if $n=0$), play a central role in the moduli dynamics as they induce specific moduli couplings that are crucial to achieve stabilisation \cite{Hertzberg:2007wc}.

We will postpone to future work the construction of the frame formulation of SL(2)-DFT and the addition of section constraint violating terms to the action. 
Nevertheless, motivated by its phenomenological relevance, let us assume for the time being that such a formulation becomes available. 
Then, starting from it and performing a generalised Scherk--Schwarz reduction based on the twist matrix $\,U\,$ in (\ref{twist_matrices_pieces_1})-(\ref{twist_matrices_pieces_3}), one would obtain an actual $\,\cN=4\,$ scalar potential. For the sake of simplicity, we will focus on the $\textrm{SO}(3)$-invariant subsector of the $\,\cN=4\,$ theory which preserves $\,\cN=1\,$ supersymmetry \cite{Dibitetto:2011gm}. This $\,\cN=1\,$ subsector contains three chiral fields $S$ (axion-dilaton), $T$ (overall K\"ahler modulus) and $U$ (complex structure modulus) parameterising three copies of the scalar manifold $\,\mathcal{M}_{\textrm{scalar}}=\textrm{SL}(2)/\textrm{SO}(2)\,$, and it is usually referred to as STU-model in the literature. The scalar potential can then be obtained from a K\"ahler potential $\,K\,$ and a flux-induced superpotential $\,W\,$ of the form
\begin{equation}
\begin{array}{lll}
K & = & -\log[-i (S-\bar{S})]  - 3\,\log[-i (T-\bar{T})] - 3\,\log[-i (U-\bar{U})]  \\[3mm]
W &=& \,\,\,  ( h^{+}_{2}-h^{+}_{1} \, U^3 ) + 3 \, T (\omega^{+}_{1} \, U^2 + \omega^{+}_{2} \,  U ) + 3 \,  T^2  ( h^{+}_{2}  \, U^2 - h^{+}_{1} \, U )  + T^3 ( \omega^{+}_{1} + \omega^{+}_{2} \,  U^3 ) 
 \\[2mm]
&-& 
\hspace{-3mm} S  \Big[ \, (h^{-}_{2} - h^{-}_{1} \,  U^3 ) + 3 \, T (\omega^{-}_{1} \,  U^2+\omega^{-}_{2} \,  U )   + 3 \, T^2  (h^{-}_{2} \,  U^2 - h^{-}_{1} \,  U )   + T^3  (\omega^{-}_{1} + \omega^{-}_{2} \, U^3)  \,  \Big]
\end{array}
\end{equation}
by using standard $\,\cN=1\,$ formulae. Similar STU-models have been investigated in the context of Type~I/Heterotic flux compactifications. Following the notation of \cite{Aldazabal:2006up}, the superpotential takes the form of an integral over the internal space 
\begin{equation}
W=\hspace{-1mm}\int_{\textrm{M}_{6}}  \hspace{-2mm} \big[  ( H^{(+)} -S H^{(-)})  + (\omega^{(+)}- S \omega^{(-)}) J_{c} + (Q^{(+)} - S Q^{(-)} ) J_{c}^{(2)} + (R^{(+)} - S R^{(-)}) J_{c}^{(3)}  \big] \wedge \Omega \ ,
\end{equation}
where $\,J_{c}\,$ is the complexified K\"ahler form and $\,\Omega\,$ is the holomorphic three-form of $\,\textrm{M}_{6}\,$. Only the terms induced by fluxes $\,H^{(+)}\,$ and $\,\omega^{(+)}\,$ can be understood from higher dimensions as gauge and metric fluxes \cite{Gurrieri:2004dt,Derendinger:2004jn,deCarlos:2005kh}. Importantly, note the presence in $\,W\,$ of terms linear in $\,S\,$ which are induced by non-geometric Type~I/Heterotic fluxes of $\,f_{\underline{-}}\,$ type. These are needed to stabilise the axion-dilaton modulus. Various AdS, dS and Minkowski vacua have been found in this type of STU-models \cite{Dibitetto:2011gm}.



\section*{Acknowledgments}

We would like to thank Bernard de Wit, Henning Samtleben and Mario Trigiante for conversations. GD, AG and GI thank the Galileo Galilei Institute for Theoretical Physics for the hospitality and the INFN for partial support during the final stages of this work. The work of FC is supported by the ERC Advanced Grant no. 246974, ``Supersymmetry: a window to non-perturbative physics''. GD is supported by the Swedish Research Council (VR). JJF-M is supported by Seneca Foundation and JSPS grant FY2016. AG is partially supported by a Marina Solvay fellowship and by F.R.S.-FNRS through the conventions PDRT.1025.14 and IISN-4.4503.15. GI is partially supported by FCT/Portugal through a CAMGSD post-doc fellowship. AG and GI are also partially supported by the ERC Advanced Grant no. 246974, ``Supersymmetry: a window to non-perturbative physics''.

\newpage

\appendix

\section{$\mathbb{Z}_{2}$-truncation: from EFT to SL(2)-DFT ${(n=0)}$}
\label{app:Z2-truncation}

In this appendix we collect the details of the group theoretical $\mathbb{Z}_{2}$-truncation of $\textrm{E}_{7(7)}$-EFT to an $\textrm{SL}(2) \times \textrm{O}(6,6)$ extended field theory, \textit{i.e.} an SL(2)-DFT with $n=0$.

\subsection{Notation and conventions}
\label{sec:conventions}

We adopt the NorthWest-SouthEast (NW-SE) conventions of ref.~\cite{Hohm:2013uia} to rise and lower $\textrm{E}_{7(7)}$ fundamental indices $\,\cM,\cN=1,...,56\,$ with the $\textrm{Sp}(56)$-invariant  skew-symmetric $\,\Omega_{\cM \cN}\,$ matrix, \textit{e.g.}~$\,U_{\cM}=U^{\cN}\,\Omega_{\cN \cM}\,$. In order to $\mathbb{Z}_{2}$-truncate the $\textrm{E}_{7(7)}$-EFT we will make use of the decomposition of different representations of $\,\textrm{E}_{7(7)}\,$ under its $\,\textrm{SL}(2) \times \textrm{SO}(6,6)\,$ maximal subgroup. Of special interest are the following branching rules
\beqa
\label{irrep_56_app}
\textbf{56} & \longrightarrow  & (\textbf{2},\textbf{12})\,+\,(\textbf{1},\textbf{32}) \ , \\[2mm]
\label{irrep_133_app}
\textbf{133} & \longrightarrow  & (\textbf{1},\textbf{66})\,+\,(\textbf{3},\textbf{1})\,+\,(\textbf{2},\textbf{32'}) \ , \\[2mm]
\label{irrep_912_app}
\textbf{912} & \longrightarrow  & (\textbf{2},\textbf{12})\,+\,(\textbf{2},\textbf{220})\,+\,(\textbf{1},\textbf{352'}) \,+\,(\textbf{3},\textbf{32}) \ , 
\eeqa
where $\textbf{32}$ and $\textbf{32'}$ respectively denote left- and right-handed Majorana-Weyl (M-W) spinorial representations of $\textrm{SO}(6,6)$ and similarly for the other spinorial irrep's\footnote{See the appendix in \cite{Dibitetto:2011eu} for conventions about M-W spinorial irrep's of $\,\textrm{SO}(6,6)\,$.}. The decomposition of the $\textbf{56}$ in (\ref{irrep_56_app}) amounts to the index splitting $\,\cM=(\alpha,M) \,\oplus\, \hat{\m}\,$, where $\,\alpha=\pm\,$ is an electric-magnetic $\,\textrm{SL}(2)\,$ index, $\,M=1,\dots,12\,$ refers to an $\,\textrm{SO}(6,6)\,$ vector index and $\,\hat{\m}=1,\dots,32\,$ denotes a M-W left-handed spinorial index. Analogously, an index $\,\dm=1,\dots,32\,$ will denote a M-W right-handed spinor. To carry out the truncation one has to apply a discrete $\,\mathbb{Z}_{2}$-projection\footnote{In a string theory realisation of maximal supergravity, this $\,\mathbb{Z}_{2}$-projection corresponds to orientifolding the theory (see section~\ref{sec:orientifold}).} 
\beq
\begin{array}{cccl}
\mathbb{Z}_{2} \,\,: \hspace{5mm}  &      \textrm{E}_{7(7)}    & \longrightarrow &  \textrm{SL}(2) \times \textrm{SO}(6,6) 
\end{array}
\eeq
under which different $\,\textrm{SL}(2) \times \textrm{SO}(6,6)\,$ indices acquire a parity. In particular, the bosonic indices $\,\alpha\,$ and $\,M\,$ are even whereas the spinorial indices $\,\hat{\m}\,$ and $\,\dm\,$ become odd. The $\mathbb{Z}_{2}$-truncation keeps only states which are parity even. As a result, the skew-symmetric $\Omega_{\cM \cN}$ matrix becomes block-diagonal with bosonic and spinorial blocks
\beq
\label{Omega_app}
\Omega_{\cM \cN} =
\left( 
\begin{array}{c|c}
\Omega_{\a M \b N} & 0 \\[1mm] 
\hline
\\[-4mm]
0 & \Omega_{\hat{\m} \hat{\n}} 
\end{array}
\right)
= 
\left( 
\begin{array}{c|c}
\eps_{\a \b} \, \eta_{MN} & 0 \\[1mm] 
\hline
\\[-4mm]
0 & \mathcal{C}_{\hat{\m} \hat{\n}} 
\end{array}
\right) \ .
\eeq
It is worth observing that the bosonic part involves the Levi-Civita tensor $\,\varepsilon_{\alpha \beta}\,$ (with $\,\varepsilon_{+ -}=1$) associated to the $\textrm{SL}(2)$ factor as well as the $\,\textrm{SO}(6,6)$-invariant metric $\,\eta_{M N}$, whereas the spinorial part contains the $\,\textrm{SO}(6,6)$-invariant charge conjugation matrix $\,\mathcal{C}_{\hat{\m} \hat{\n}}$.

We denote $\,K^{IJ}\,$, with $I,J=1,...,133$ being adjoint $\,\textrm{E}_{7(7)}\,$ indices, the inverse of the $\textrm{E}_{7(7)}$ Killing-Cartan metric
\beq
\label{KC_metric_app}
K_{IJ}=\textrm{Tr}(t_{I} \, t_{J})=[t_{I}]_{\cM \cN} \, [t_{J}]_{\cP \cQ} \,\,  \Omega^{\cP \cN} \, \Omega^{\cM \cQ} \ ,
\eeq
which, in turn, also depends on the $\,[t_{I}]_{\cM \cN}\,$ symmetric generators of $\,\textrm{E}_{7(7)}\,$ in the fundamental representation. By virtue of the decomposition (\ref{irrep_133_app}), the general form of the $\textrm{SL}(2) \times \textrm{SO}(6,6)$ generators in the $(\textbf{2},\textbf{12})$ and $(\textbf{1},\textbf{32})$ representations are given by
\beq
\label{A2D6_gen_app}
\begin{array}{cclc}
\left[t_{\a M \b N}\right]_{\g P \d Q} & = & \varepsilon_{\a \b} \, \varepsilon_{\g \d} \, \left[t_{MN}\right]_{PQ} + \eta_{MN}\, \, \eta_{PQ} \, \left[t_{\a \b}\right]_{\g \d} & , \\[3mm]
\left[t_{\a M \b N}\right]_{\hat{\m} \hat{\n}} & = &\dfrac{1}{4} \, \varepsilon_{\a \b} \, \left[\g_{MN}\right]_{\hat{\m} \hat{\n}} & .
\end{array}
\eeq
Using the above expressions for the generators, the $\textrm{E}_{7(7)}$ Killing-Cartan metric (\ref{KC_metric_app}) induces an $\,\textrm{SL}(2) \times \textrm{SO}(6,6)\,$ metric\footnote{The $\,\textrm{SL}(2) \times \textrm{SO}(6,6)\,$ metric computed from the generators $\,\left[t_{\a M \b N}\right]_{\g P \d Q}\,$ in (\ref{A2D6_gen_app}) reads
\begin{equation}
K^{\textrm{SL}(2)\times \textrm{SO}(6,6)}_{\a M \b N , \g P \d Q}  =  2  \, \eps_{\a \b} \, \eps_{\g \d} \,  K_{MN , PQ} + 12 \, \eta_{MN} \, \eta_{PQ} \,  K_{\a \b, \g \d} \ ,
\end{equation}
and differs from the expression in (\ref{Killin-Cartan-A2D6_app}) because of the contribution of the spinor representation to the SO(6,6) trace.
} and its inverse\footnote{We have taken $\,[t_{\a \b}]^{\g \d} = \delta_{\a}^{(\g}\delta_{\b}^{\d)}\,$, $\,[t^{\a \b}]_{\g \d} = -\delta^{\a}_{(\g}\delta^{\b}_{\d)}\,$, $\,[t_{\a \b}]_{\g \d} = -K_{\a \b , \g \d}\,$ and $\,[t^{\a \b}]^{\g \d} = K^{\a \b , \g \d}\,$, as well as $\,[t_{M N}]^{P Q} = \delta_{MN}^{PQ}\,$, $\,[t^{M N}]_{P Q} = -\delta^{MN}_{PQ}\,$,  $\,[t_{M N}]_{P Q} = - K_{MN,PQ}\,$ and $\,[t^{M N}]^{P Q} = K^{MN,PQ}\,$. This is consistent with the definitions
\beq
\begin{array}{ccc}
K_{\a \b , \g \d} \equiv  \eps_{\a (\g} \, \eps_{\d) \b} \hspace{10mm} & \textrm{ and }&  \hspace{10mm} K_{M N , P Q} \equiv - \eta_{M [P} \,\eta_{Q] N} \ , \\[3mm]
K^{\a \b , \g \d} \equiv  \eps^{\a (\g} \, \eps^{\d) \b} \hspace{10mm} & \textrm{ and }&  \hspace{10mm} K^{M N , P Q} \equiv - \eta^{M [P} \,\eta^{Q] N} \ ,
\end{array}
\eeq
of the $\,\textrm{SL}(2)\,$ and $\,\textrm{SO}(6,6)\,$ metrics and their inverses. In particular, note that $\,K^{++,--}=[t^{++}]^{--}=-1\,$.}
of the form
\beq
\label{Killin-Cartan-A2D6_app}
\begin{array}{cclc}
K_{\a M \b N , \g P \d Q} & = & \dfrac{12}{2}  \, \eps_{\a \b} \, \eps_{\g \d} \,  K_{MN , PQ} + 12 \, \eta_{MN} \, \eta_{PQ} \,  K_{\a \b, \g \d}  &  , \\[2mm]
K^{\a M \b N , \g P \d Q} & = & \dfrac{1}{2 \times 12}  \, \eps^{\a \b} \, \eps^{\g \d} \,  K^{MN , PQ} + \dfrac{1}{(12)^3} \, \eta^{MN} \, \eta^{PQ} \,  K^{\a \b, \g \d}  &  .
\end{array}
\eeq
The latter can be used to obtain the following expression for the $\,\textrm{SL}(2) \times \textrm{SO}(6,6)\,$ generators 
\beq
\label{A2D6-gen-up_app}
\begin{array}{cclc}
[t^{\a M \b N}]^{\g P \d Q} & = & \dfrac{1}{12} \, \varepsilon^{\a \b} \, \varepsilon^{\g \d} \, [t^{MN}]^{PQ} + \dfrac{1}{(12)^2} \, \eta^{MN}\, \, \eta^{PQ} \, [t^{\a \b}]^{\g \d} \ ,
\end{array}
\eeq
that appears at several places in the main text. When considering the extension to $\,\SO(6,6+n)\,$ in section~\ref{sec:gauge_vectors}, the expressions in (\ref{Killin-Cartan-A2D6_app}) and (\ref{A2D6-gen-up_app}) are still valid after replacing the factors of $\,12\,$ by $\,12+n\,$.

\subsection{Structure tensor, generalised Lie derivative and section constraints}
\label{sec:Y&L}

Our starting point is the structure $Y$-tensor of the $\textrm{E}_{7(7)}$-EFT \cite{Hohm:2013uia} which has the form
\be
\label{Y-EFT_app}
Y^{\cM\cN}{}_{\cP\cQ} = -12 \,K^{IJ}\, [t_{I}]^{\cM\cN} \,  [t_{J}]_{\cP\cQ} - \dfrac{1}{2} \, \Omega^{\cM\cN}  \Omega_{\cP\cQ} \ ,
\ee
and specifies a generalised Lie derivative with a gauge parameter $\,\Lambda^{\cM}\,$ of the form
\be
\label{L-EFT_app}
\begin{array}{ccl}
\mathbb{L}_{\Lambda} U^{\cM} &=& \Lambda^{\cN} \partial_{\cN} U^{\cM} - U^{\cN} \partial_{\cN} \Lambda^{\cM} + Y^{\cM\cN}{}_{\cP\cQ} \, \partial_{\cN} \Lambda^{\cP} \, U^{\cQ} + (\lambda_{U}-\omega) \partial_{\cN} \Lambda^{\cN} U^{\cM}  \ .
\end{array}
\ee

Using the definitions \eqref{A2D6_gen_app} and \eqref{Killin-Cartan-A2D6_app} in the previous section, as well as the index decomposition \eqref{irrep_56_app}, an explicit computation of the $\mathbb{Z}_{2}$-even components of the structure tensor (\ref{Y-EFT_app}) yields
\begin{equation}
\label{Y-QFT_app}
\begin{array}{llll}
Y^{\a M \b N}{}_{\g P \d Q} &=&  \delta_{\d}^{\a} \, \delta_{\g}^{\b}  \, \eta^{MN} \, \eta_{PQ} + 2 \,  \varepsilon^{\a \b} \, \varepsilon_{\g \d} \, \delta^{MN}_{PQ}  & , \\[2mm]
Y^{\a M \b N}{}_{\hat{\rho} \hat{\sigma}} &=& - \frac{1}{2} \, \varepsilon^{\a\b}  \left( \,\, \eta^{MN}\,  \mathcal{C}_{\hat{\rho} \hat{\sigma}} + [\gamma^{MN}]_{\hat{\rho} \hat{\sigma}} \,\, \right) & , \\[2mm]
Y^{\hat{\mu} \hat{\nu}}{}_{\g P \d Q} &=& - \frac{1}{2} \, \varepsilon_{\g\d}  \left( \,\, \eta_{PQ}\,  \mathcal{C}^{\hat{\mu} \hat{\nu}} - [\gamma_{PQ}]^{\hat{\mu} \hat{\nu}} \,\, \right) & , \\[2mm]
Y^{\hat{\mu} \hat{\nu}}{}_{\hat{\rho} \hat{\sigma}} &=& - \frac{1}{8} \,  [\gamma_{MN}]^{\hat{\mu} \hat{\nu}} \,  [\gamma^{MN}]_{\hat{\rho} \hat{\sigma}}  - \frac{1}{2} \,  \mathcal{C}^{\hat{\mu} \hat{\nu}} \,  \mathcal{C}_{\hat{\rho} \hat{\sigma}}  & .
\end{array}
\end{equation}
Of particular importance is the component
\begin{equation}
\label{Y_bos-QFT_app}
\begin{array}{llll}
Y^{\a M \b N}{}_{\g P \d Q} &=&  \delta_{\d}^{\a} \, \delta_{\g}^{\b}  \, \eta^{MN} \, \eta_{PQ} + 2 \,  \varepsilon^{\a \b} \, \varepsilon_{\g \d} \, \delta^{MN}_{PQ}  & ,
\end{array}
\end{equation}
which plays the role of structure tensor in SL(2)-DFT when only a dependence on the $\,y^{\a M}\,$ coordinates is allowed. In this case the generalised Lie derivative with parameter $\,\Lambda^{\a M}\,$ can be obtained from (\ref{L-EFT_app}) using (\ref{Y_bos-QFT_app}), and takes the form
\be
\label{L-QFT_app}
\begin{array}{ccl}
\mathbb{L}_{\Lambda} U^{\a M} &=& \Lambda^{\b N} \partial_{\b N} U^{\a M} - U^{\b N} \partial_{\b N} \Lambda^{\a M} + \eta^{MN} \, \eta_{PQ} \, \partial_{\b N} \Lambda^{\b P} \, U^{\a Q} \\[2mm] 
&+& 2 \, \varepsilon^{\a \b} \, \varepsilon_{\g \d} \, \partial_{\b N} \Lambda^{\g[M} \, U^{|\d |N]} + (\lambda_{U}-\omega) \partial_{\b N} \Lambda^{\b N} U^{\a M}  \ .
\end{array}
\ee

The section constraints in SL(2)-DFT can be obtained in a similar fashion by decomposing the one of $\textrm{E}_{7(7)}$-EFT. Starting from \cite{Hohm:2013uia}
\begin{equation}
\label{sec_cons_EFT_app}
Y^{\cM\cN}{}_{\cP\cQ} \, \partial_{\cM} \otimes \partial_{\cN}=0 \ ,
\end{equation}
and allowing only a dependence on the $\,y^{\a M}\,$ coordinates, one finds
\be
\label{sec_cons-QFT_app}
\begin{array}{rlrl}
\Omega^{\cM\cN} \partial_{\cM} \otimes \partial_{\cN} = 0 & \longrightarrow &\varepsilon^{\alpha\beta} \, \eta^{MN} \, \partial_{\a M} \otimes  \partial_{\b N} = 0 & , \\[2mm]
[t_{I}]^{\cM\cN} \partial_{\cM} \otimes \partial_{\cN} = 0 & \longrightarrow & \varepsilon^{\a \b}  \, \partial_{\a [M |} \otimes  \partial_{\b | N]} = 0 & , \\[2mm]
 &  & \eta^{MN} \, \partial_{(\a | M |} \otimes  \partial_{\b)  N} = 0 & ,
\end{array}
\ee
corresponding to $(\textbf{1},\textbf{1})$, $(\textbf{1},\textbf{66})$ and $(\textbf{3},\textbf{1})$ irrep's of $\,\textrm{SL}(2) \times \textrm{SO}(6,6)\,$, respectively. They can be more concisely expressed as
\begin{equation}
\label{sec_cons-QFT_2_app}
\eta^{MN} \, \partial_{\a M} \otimes  \partial_{\b  N} = 0
\hspace{10mm} \textrm{and} \hspace{10mm}
\varepsilon^{\a \b}  \, \partial_{\a [M |} \otimes  \partial_{\b | N]} = 0 \ .
\end{equation}
In addition to (\ref{sec_cons_EFT_app}), the remaining constraints needed for the closure of the generalised Lie derivative in the $\textrm{E}_{7(7)}$-EFT (see ref.~\cite{Berman:2012vc} for a general study of closure constraints)
\begin{equation}
\label{sec_cond_additional_EFT_app}
\hspace{-1.1mm}
\begin{array}{rlll}
\big( Y^{\cal{M}(\cal{P}}{}_{\cal{TQ}} \, Y^{\cal{T}|\cal{N})}{}_{\cal{RS}} - Y^{\cal{M}(\cal{P}}{}_{\cal{RS}} \, \delta^{\cal{N})}_{\cal{Q}} \big)  (\partial_{\cal{P}} \partial_{\cal{N}}) =  0  \\[2mm]
\big(Y^{\cal{MN}}{}_{\cal{TQ}} \, Y^{\cal{TP}}{}_{[\cal{SR}]} + 2  Y^{\cal{MN}}{}_{[\cal{R}|\cal{T}|} \, Y^{\cal{TP}}{}_{\cal{S}]\cal{Q}} - Y^{\cal{MN}}{}_{[\cal{RS}]} \, \delta^{\cal{P}}_{\cal{Q}} - 2  Y^{\cal{MN}}{}_{[\cal{S}|\cal{Q}|} \, \delta^{\cal{P}}_{\cal{R}]} \big)  \, \partial_{(\cal{N}} \otimes \partial_{\cal{P})} =  0  \\[2mm]
\big(Y^{\cal{MN}}{}_{\cal{TQ}} \, Y^{\cal{TP}}{}_{(\cal{SR})} + 2  Y^{\cal{MN}}{}_{(\cal{R}|\cal{T}|} \, Y^{\cal{TP}}{}_{\cal{S})\cal{Q}} - Y^{\cal{MN}}{}_{(\cal{RS})} \, \delta^{\cal{P}}_{\cal{Q}} - 2  Y^{\cal{MN}}{}_{(\cal{S}|\cal{Q}|} \, \delta^{\cal{P}}_{\cal{R})} \big)   \, \partial_{[\cal{N}} \otimes \partial_{\cal{P}]} =  0 
\end{array}
\end{equation}
are also satisfied when $\,\cM=\alpha M\,$, $\,\cN=\beta N\,$, etc., provided \eqref{sec_cons-QFT_2_app} holds. This can be seen as a crosscheck of the SL(2)-DFT structure tensor (\ref{Y_bos-QFT_app}) obtained upon truncation.

\subsection{Truncating the \Eseven-EFT action}
\label{sec:A2D6_action}

We will continue our program and obtain the bosonic action of SL(2)-DFT by $\mathbb{Z}_{2}$-truncating the one of $\textrm{E}_{7(7)}$-EFT. Following ref.~\cite{Hohm:2013uia}, the starting bosonic action reads
\begin{equation}
\label{eq:action}
\begin{array}{lll}
S_{\text{\tiny{$\textrm{E}_{7(7)}$-EFT}}} &=&  \displaystyle\int d^4 x\,d^{56} y\,e\,\big[\,\hat{R}\,+\,\tfrac{1}{48} \, g^{\mu\nu}\,\mathcal{D}_\mu\mathcal{M}^{\cM\cN}\,\mathcal{D}_{\nu}\mathcal{M}_{\cM\cN} -\tfrac18 \, \mathcal{M}_{\cM\cN}\,\mathcal{F}^{\mu\nu \cM}\mathcal{F}_{\mu\nu}{}^\cN  \\[2mm]
& & \quad \quad \quad \quad \quad  \quad  + \, e^{-1}\,\mathcal{L}_{\text{top}}-V_{\text{\tiny{$\textrm{E}_{7(7)}$-EFT}}}(\mathcal{M},g)\,\big] \, .
\end{array}
\end{equation}
We will proceed with the truncation of each piece in the above action separately in order to obtain the SL(2)-DFT action
\begin{equation}
\label{eq:action-QFT_app}
\begin{array}{lll}
S_{\text{\tiny{SL(2)-DFT}}} &=&  \displaystyle\int d^4 x\,d^{24} y \,e\,\big[\,\hat{R}\,+\,\tfrac{1}{4} \, g^{\mu\nu}\,\mathcal{D}_\mu M^{\a\b}\,\mathcal{D}_{\nu} M_{\a\b}  + \tfrac{1}{8} \, g^{\mu\nu} \, \mathcal{D}_\mu M^{MN}\,\mathcal{D}_{\nu} M_{MN} \\[2mm]
& &  -\tfrac18 \, M_{\a \b} \, M_{MN} \,\mathcal{F}^{\mu\nu \,  \a M}\mathcal{F}_{\mu\nu}{}^{\b N}  + \, e^{-1}\,\mathcal{L}_{\text{top}}-V_{\text{\tiny{SL(2)-DFT}}}(M,g)\,\big] \, .
\end{array}
\end{equation}

\subsubsection*{Einstein, kinetic and topological terms}

\noindent $\bullet$ The $\mathbb{Z}_{2}$-truncation of the Einstein term reads
\begin{equation}
\begin{array}{lcl}
\hat{R}_{\mu\nu}{}^{ab}=\,R_{\mu\nu}{}^{ab}[\omega]+\mathcal{F}_{\mu\nu}{}^\cM \,e^{a\rho}\,\partial_\cM e_\rho{}^b & \rightarrow & 
\hat{R}_{\mu\nu}{}^{ab}=\,R_{\mu\nu}{}^{ab}[\omega]+\mathcal{F}_{\mu\nu}{}^{\a M} \,e^{a\rho}\,\partial_{\a M} e_\rho{}^b \ .
\end{array}
\end{equation}

\noindent $\bullet$ The $\mathbb{Z}_{2}$-truncation of the kinetic terms of the scalars proceeds as for the supergravity case studied in ref.~\cite{Dibitetto:2011eu}. Its action on the scalar coset of \Eseven-EFT reads
\beq
\begin{array}{cccl}
\mathbb{Z}_{2} \,\,: \hspace{5mm}  &      \dfrac{\textrm{E}_{7(7)}}{\textrm{SU}(8)}    & \longrightarrow &  \dfrac{\textrm{SL}(2)}{\textrm{SO}(2)} \times \dfrac{\textrm{SO}(6,6)}{\textrm{SO}(6) \times \textrm{SO}(6)} \ ,
\end{array}
\eeq
and reduces the number of scalar fields in the truncated theory from $\,70\,$ to $\,2+36\,$. The parameterisation of the $\,\textrm{E}_{7(7)}/\textrm{SU}(8)\,$ coset is given by a symmetric $\,\mathcal{M}_{\cM \cN}\,$ matrix which, after the truncation, becomes block-diagonal,
\beq
\label{E7_scalars}
\mathcal{M}_{\cM \cN} = 
\left( 
\begin{array}{c|c}
\cM_{\a M \b N} & 0 \\[1mm] 
\hline
\\[-4mm]
0 & \cM_{\hat{\m} \hat{\n}} 
\end{array}
\right)
=
\left( 
\begin{array}{c|c}
M_{\a \b} \, M_{MN} & 0 \\[1mm] 
\hline
\\[-4mm]
0 & \dfrac{1}{6!} \, M_{MNPQRS} \, \left[ \g^{MNPQRS} \right]_{\hat{\m} \hat{\n}} 
\end{array}
\right) \ ,
\eeq
with a bosonic $\,\cM_{\a M \b N}\,$ and a spinorial $\,\cM_{\hat{\m} \hat{\n}}\,$ block. The former contains the $\,\textrm{SL}(2)\,$ and the $\,\textrm{SO}(6,6)\,$ scalars $\,M_{\a \b}\,$ and $\,M_{MN}\,$ of the SL(2)-DFT whereas the latter now involves a contraction with the $\,[\gamma^{MNPQRS}]_{\hat{\m} \hat{\n}}\,$ anti-self-dual (ASD) matrix. This time it is contracted with the $\,\textrm{SO}(6,6)\,$ six-form
\beq 
\label{M6}
M_{MNPQRS} \equiv \varepsilon_{mnpqrs}\mathcal{V}_{M}^{\phantom{M}m}\mathcal{V}_{N}^{\phantom{M}n}\mathcal{V}_{P}^{\phantom{M}p}\mathcal{V}_{Q}^{\phantom{M}q}\mathcal{V}_{R}^{\phantom{M}r}\mathcal{V}_{S}^{\phantom{M}s} \ ,
\eeq
where $\,\mathcal{V}\,$ denotes an $\,\textrm{SO}(6,6)/\textrm{SO}(6) \times \textrm{SO}(6)\,$ Zw\"{o}lfbein such that $\,M=\mathcal{V}\,\mathcal{V}^{T}\,$ and the index $\,m\,$ only runs over the six time-like directions \cite{Schon:2006kz}. 

\noindent The truncation of the kinetic term for the scalars proceeds as follows
\begin{equation}
\begin{array}{lcl}
\frac{1}{48} \, g^{\mu\nu}\,\mathcal{D}_\mu\mathcal{M}^{\cM\cN}\,\mathcal{D}_{\nu}\mathcal{M}_{\cM\cN} & \rightarrow & \frac{1}{48} \, g^{\mu\nu}\, \left( \mathcal{D}_\mu\mathcal{M}^{\a M \b N}\,\mathcal{D}_{\nu}\mathcal{M}_{\a M \b N} + \mathcal{D}_\mu\mathcal{M}^{\hat{\rho} \hat{\sigma}}\,\mathcal{D}_{\nu}\mathcal{M}_{\hat{\rho} \hat{\sigma}} \right) \\[2mm]
&=& \frac{1}{4} \, g^{\mu\nu}\,\mathcal{D}_\mu M^{\a\b}\,\mathcal{D}_{\nu} M_{\a\b}  + \frac{1}{8} \, g^{\mu\nu} \, \mathcal{D}_\mu M^{MN}\,\mathcal{D}_{\nu} M_{MN} \ .
\end{array}
\end{equation}
As noticed in \cite{Dibitetto:2011eu}, the spinorial contribution to the trace is crucial in order to recover the right normalisation of the kinetic term of the $\,\SO(6,6)\,$ scalars. 
\\[-2mm]

\noindent $\bullet$ The $\mathbb{Z}_{2}$-truncation of the kinetic terms of the vectors reads
\begin{equation}
\begin{array}{lcl}
-\frac18 \, \mathcal{M}_{\cM\cN}\,\mathcal{F}^{\mu\nu \cM}\mathcal{F}_{\mu\nu}{}^\cN & \rightarrow & 
-\frac18 \, \mathcal{M}_{\a M \b N}\,\mathcal{F}^{\mu\nu \,  \a M}\mathcal{F}_{\mu\nu}{}^{\b N}
 \\[2mm]
&=& -\frac18 \, M_{\a \b} \, M_{MN} \,\mathcal{F}^{\mu\nu \,  \a M}\mathcal{F}_{\mu\nu}{}^{\b N} \ ,
\end{array}
\end{equation}
where the field strengths $\,\mathcal{F}_{\mu\nu}{}^{\a M}\,$ are obtained upon truncation of the $\textrm{E}_{7(7)}$-EFT ones \cite{Hohm:2013uia} and read
\begin{equation}
\label{Fmunu-QFT_app}
\mathcal{F}_{\mu\nu}{}^{\a M} = F_{\mu\nu}{}^{\a M} \,+\, 2 \, \varepsilon^{\a \b}\, \eta^{M P} \, \eta^{NQ} \, \partial_{\b N} B_{\mu \nu \, PQ} \,+\,  \eta^{MN} \, \varepsilon^{\a \g}\, \varepsilon^{\b \d} \partial_{\b N} B_{\mu \nu \, \g\d} \, - \,  \tfrac{1}{2} \,  \varepsilon^{\a \b}\eta^{MN} B_{\mu\nu \, \b N} \ .
\end{equation}
The above field strengths contain tensor fields  in the $(\textbf{1},\textbf{66})\oplus (\textbf{3},\textbf{1})$ and $(\textbf{2},\textbf{12})$ given by
\begin{equation}
\label{B_adjoint_app}
B_{\mu\nu \, \a M \b N} = \varepsilon_{\a\b} \, B_{\mu \nu \, MN} + \eta_{MN} \,  B_{\mu \nu \, \alpha \beta}
\hspace{10mm} \textrm{and} \hspace{10mm}
B_{\mu\nu \, \a M} \ ,
\end{equation}
which satisfy $\,B_{\mu \nu \, MN}=-B_{\mu \nu \, NM}\,$ and $\,B_{\mu \nu \, \alpha \beta}=B_{\mu \nu \, \beta \alpha}\,$. The tensor fields enter the field strengths (\ref{Fmunu-QFT_app}) in the form of trivial parameters of the SL(2)-DFT (see section~\ref{sec:Yang--Mills}).
\\[-2mm]

\noindent $\bullet$ The $\mathbb{Z}_{2}$-truncation of the topological term reads
\begin{equation}
\begin{array}{lcl}
\varepsilon^{\mu\nu\rho\sigma\tau}\mathcal{F}_{\mu\nu}{}^\cM\,\mathcal{D}_{\rho}\mathcal{F}_{\sigma\tau\,\cM} & \rightarrow & 
\varepsilon^{\mu\nu\rho\sigma\tau} \, \varepsilon_{\b\a} \, \eta_{MN}  \, \mathcal{F}_{\mu\nu}{}^{\a M}\,\mathcal{D}_{\rho}\mathcal{F}_{\sigma\tau}{}^{\b N}  \ .
\end{array}
\end{equation}

\subsubsection*{Scalar potential}

The potential in SL(2)-DFT can be also obtained by $\mathbb{Z}_{2}$-truncating the potential in $\textrm{E}_{7(7)}$-EFT \cite{Hohm:2013uia}
\begin{equation}
\label{eq:potEFT_app}
\begin{array}{lll}
V_{\text{\tiny{$\textrm{E}_{7(7)}$-EFT}}}(\cM,g) &=&\,-\frac{1}{48}\,\mathcal{M}^{\cM\cN}\,\partial_{\cM}\mathcal{M}^{\cK\cL}\,\partial_{\cN}\mathcal{M}_{\cK\cL}+\frac12\,\mathcal{M}^{\cM\cN}\,\partial_{\cM}\mathcal{M}^{\cK\cL}\,\partial_{\cL}\mathcal{M}_{\cN\cK} \\[2mm]
&&\,-\frac12 \,g^{-1}\partial_\cM g\,\partial_\cN\mathcal{M}^{\cM\cN}-\frac14\,\mathcal{M}^{\cM\cN}\,g^{-1}\partial_\cM g\,g^{-1}\partial_{\cN} g\\[2mm]
&& \, -\frac14\,\mathcal{M}^{\cM\cN}\,\partial_\cM g^{\mu\nu}\,\partial_\cN g_{\mu\nu} \ .
\end{array}
\end{equation}
We will look at each term in the above potential separately. The first term yields
\begin{equation}
\begin{array}{lll}
-\frac{1}{48}\,\mathcal{M}^{\cM\cN}\,\partial_{\cM}\mathcal{M}^{\cK\cL}\,\partial_{\cN}\mathcal{M}_{\cK\cL} & \rightarrow &  -M^{\a \b} M^{MN} \big[ \frac{1}{4} \,  (\partial_{\a M} M^{\g\d})(\partial_{\b N} M_{\g\d})  \\[2mm]
&&  \quad\quad\quad\quad\quad \,\,\,\,\, + \, \frac{1}{8} \,  (\partial_{\a M} M^{PQ})(\partial_{\b N} M_{PQ})   \big] \ ,
\end{array}
\end{equation}
where, as for the case of the scalar kinetic terms, the spinorial contribution to the trace is important in order to get the coefficient $\frac{1}{8}\,$. 
The second term yields
\begin{equation}
\begin{array}{lll}
\frac12\,\mathcal{M}^{\cM\cN}\,\partial_{\cM}\mathcal{M}^{\cK\cL}\,\partial_{\cL}\mathcal{M}_{\cN\cK} & \rightarrow &  
\tfrac{1}{2} \, \big[ M^{\a \b} M^{MN}  (\partial_{\a M} M^{\g\d})(\partial_{\d N} M_{\b\g})  \\[2mm]
&&\quad \, + \, M^{\a \b} M^{MN}  (\partial_{\a M} M^{PQ})(\partial_{\b Q} M_{NP})  \\[2mm]
&&\quad \, + \, M^{MN} M^{PQ}  (\partial_{\a M} M^{\a \d})(\partial_{\d Q} M_{NP})  \\[2mm]
&&\quad \, + \, M^{\a \b} M^{\g \d}  (\partial_{\a M} M^{MQ})(\partial_{\d Q} M_{\b \g}) \big]  \ .
\end{array}
\end{equation}
The third, fourth and fifth terms ($g_{\mu\nu}$-dependent) yield
\begin{equation}
\begin{array}{rll}
-\frac12 \,g^{-1}\partial_\cM g\,\partial_\cN\mathcal{M}^{\cM\cN}
& \rightarrow &  
-\frac12 \,g^{-1} \, (\partial_{\a M} g) \,\big[ (\partial_{\b N} M^{\a \b}) M^{MN} + (\partial_{\b N} M^{MN}) M^{\a \b}  \big]  , \\[2mm]
-\frac14\,\mathcal{M}^{\cM\cN}\,g^{-1}\partial_\cM g\,g^{-1}\partial_{\cN} g
& \rightarrow & 
-\frac14 \, M^{\a \b} \, M^{MN} \, g^{-1} (\partial_{\a M} g) \, g^{-1} (\partial_{\b N} g) \ , \\[2mm]
-\frac14\,\mathcal{M}^{\cM\cN}\,\partial_\cM g^{\mu\nu}\,\partial_\cN g_{\mu\nu} 
& \rightarrow &  
-\frac14 \, M^{\a \b} \, M^{MN} \, (\partial_{\a M} g^{\mu\nu}) \, (\partial_{\b N} g_{\mu\nu}) \ .
\end{array}
\end{equation}
Bringing all the terms together we get the expression of the SL(2)-DFT potential which takes the form
\begin{equation}
\label{eq:potQFT_app}
\begin{array}{lll}
V_{\text{\tiny{SL(2)-DFT}}}(M,g) &=& 
M^{\a \b} M^{MN} \big[ - \frac{1}{4} \,  (\partial_{\a M} M^{\g\d})(\partial_{\b N} M_{\g\d}) -  \frac{1}{8} \,  (\partial_{\a M} M^{PQ})(\partial_{\b N} M_{PQ})    \\[2mm]
&& \quad\quad\quad\quad\quad\,\, + \, \tfrac{1}{2} \,  (\partial_{\a M} M^{\g\d})(\partial_{\d N} M_{\b\g})   +  \tfrac{1}{2} (\partial_{\a M} M^{PQ})(\partial_{\b Q} M_{NP}) \big]  \\[2mm]
&+& \tfrac{1}{2}  \, M^{MN} M^{PQ}  (\partial_{\a M} M^{\a \d})(\partial_{\d Q} M_{NP})  +  \tfrac{1}{2} \, M^{\a \b} M^{\g \d}  (\partial_{\a M} M^{MQ})(\partial_{\d Q} M_{\b \g}) \\[2mm]
&-& \frac12 \,g^{-1} \, (\partial_{\a M} g) \,\big[ (\partial_{\b N} M^{\a \b}) M^{MN} + (\partial_{\b N} M^{MN}) M^{\a \b}  \big] \\[2mm]
&-& \frac14 \, M^{\a \b} \, M^{MN} \,  \big[ g^{-1} (\partial_{\a M} g) \, g^{-1} (\partial_{\b N} g) \, + \, (\partial_{\a M} g^{\mu\nu}) \, (\partial_{\b N} g_{\mu\nu}) \big] \ .
\end{array}
\end{equation}

\subsection{Deformations and constraints in SL(2)-DFT}
\label{sec:constraint_appendix}

The $X$ deformation was introduced in the context of \Eseven-EFT where $\,X_{\mathcal{MN}}{}^{\mathcal{P}} \in \textbf{912}\,$ was shown to be subject to so-called $X$ and $C$ constraints of the form \cite{Ciceri:2016dmd}
\begin{equation}
\label{X&C_const_app}
\begin{array}{rll}
X_{\cM\cN}{}^\cP \partial_\cP &=& 0 \ ,\\[2mm]
C^\cM_{\cS\cP\cQ}  \equiv 
X_{(\cP\cQ)}{}^\cM \partial_\cS - Y^{\cM\cR}{}_{\cT\cS} X_{(\cP\cQ)}{}^\cT \partial_\cR - \frac12 Y^{\cT\cR}{}_{\cP\cQ} X_{\cT\cS}{}^\cM \partial_\cR  &=& 0 \ .
\end{array}
\end{equation}
In \Eseven-EFT the C-constraint is redundant as it is implied by the X-constraint, \textit{i.e.} $\,X_{\textrm{EFT}} \Rightarrow C_{\textrm{EFT}}\,$. The same constraints formally appear also in SL(2)-DFT for $\,X_{\hat{M}\hat{N}}{}^{\hat{P}}\in (\textbf{2},\textbf{220}) + (\textbf{2},\textbf{12})\,$ just by replacing $\,\cM \rightarrow \hat{M}=\alpha M\,$, $\,\cN \rightarrow {\hat{N}=\beta N}\,$, etc. However, a detailed analysis of such constraints in this case reveals that the C-constraint is no longer implied by the \mbox{X-constraint}, \textit{i.e.} $\,X_{\textrm{SL(2)-DFT}} \nRightarrow C_{\textrm{SL(2)-DFT}}\,$. Here we will show that the two SL(2)-DFT conditions ($X$ and $C$) descend from the $X$-condition of \Eseven-EFT \emph{and viceversa}, 
\begin{equation}
\label{bi_implication}
\begin{array}{ccc}
X_{\textrm{EFT}}
& \Leftrightarrow &
X_{\textrm{SL(2)-DFT}} \hspace{3mm} \textrm{ and }\hspace{3mm} C_{\textrm{SL(2)-DFT}} \ ,
\end{array}
\end{equation}
when assuming that the section constraint of \Eseven-EFT holds with $\,\partial_{(\mathbf1,\mathbf{32})} = 0\,$, and that $\,X_{\mathcal{MN}}{}^{\mathcal{P}}\,$ only contains $\,f_{\alpha MNP}\,$ and $\,\xi_{\alpha M}\,$ irreducible pieces when decomposed under (\ref{irrep_912_app}) \cite{Dibitetto:2011eu}, namely, no trombone \cite{LeDiffon:2008sh,LeDiffon:2011wt} or spinorial deformations \cite{Dibitetto:2014sfa}.

The first direction of the double implication in (\ref{bi_implication}) is straightforward to prove. It was shown  in \cite{Ciceri:2016dmd} that $\,X_{\textrm{EFT}} \Rightarrow C_{\textrm{EFT}}\,$. Moreover, under the assumptions discussed above, $\,X_{\textrm{EFT}} \Rightarrow X_{\textrm{SL(2)-DFT}}\,$ and $\,C_{\textrm{EFT}} \Rightarrow C_{\textrm{SL(2)-DFT}}\,$  just by setting $\,\cM \to {\hat{M} = \alpha M}\,$, etc. Therefore, one has that
\begin{equation}
\label{implication==>}
\begin{array}{ccc}
X_{\textrm{EFT}}
& \Rightarrow &
X_{\textrm{SL(2)-DFT}} \hspace{3mm} \textrm{ and }\hspace{3mm} C_{\textrm{SL(2)-DFT}} \ .
\end{array}
\end{equation}
To prove the reverse implication in (\ref{bi_implication}) we just need to focus on the contribution
\begin{equation}
\label{special_contribution}
X_{\hat{\mu}\hat{\nu}}{}^{\gamma P} \,\, \partial_{\gamma P} = 0\,,
\end{equation}
to the X-constraint of \Eseven-EFT, and use the decomposition \cite{Dibitetto:2011eu}
\begin{formula}
X_{\hat{\m} \, \hat{\n}}{}^{\a M} =\ &  \dfrac{1}{8} \, \eps^{\alpha \gamma} \, f_{\gamma PQ}{}^{M} \, \left[ \g^{PQ} \right]_{\hat{\m} \hat{\n}} + \dfrac{1}{24} \, \eps^{\alpha \gamma} \, f_{\g PQ}{}^{N} \, \left[ {\g_{N}}^{MPQ} \right]_{\hat{\m} \hat{\n}} \\[3mm]
 &+  \dfrac{1}{8} \, \eps^{\alpha \gamma} \, \xi_{\g N} \, \left[ {\g}^{MN} \right]_{\hat{\m} \hat{\n}} - \dfrac{1}{8} \, \eps^{\alpha \gamma} \, \xi_{\g}^{M} \, \cC_{\hat{\m} \hat{\n}} \ .
\end{formula}
The $\gamma$'s and $\,\cC\,$ are orthogonal to each other, so we can decompose (\ref{special_contribution}) into three constraints
\begin{equation}
\begin{array}{rrrr}
\gamma_{MN}^{\hat{\mu}\hat{\nu}} :  \hspace{5mm}& 
( f^\alpha{}_{MN}{}^P -\xi^\alpha{}^{\vphantom\alpha}_{[M}\delta_{N]}^P ) \partial_{\alpha P} &=& 0 \ , \\[2mm]
\cC^{\hat{\mu}\hat{\nu}} : \hspace{5mm}  &
\xi^{\alpha M} \partial_{\alpha M} &=& 0 \ ,\\[2mm]
\gamma_{MNPQ}^{\hat{\mu}\hat{\nu}} : \hspace{5mm} &
f^{\vphantom\alpha}_{\alpha[MNP} \partial^\alpha_{Q]} &=& 0 \ .
\end{array}
\end{equation}
The first two constraints correspond to the X-constraint of SL(2)-DFT in (\ref{X-constraint_irreps}) upon appropriate contractions. The last one is precisely the projection of the C-constraint of SL(2)-DFT in (\ref{C-constraint_irreps}). Therefore, 
\begin{equation}
\label{implication<==}
\begin{array}{ccc}
X_{\textrm{SL(2)-DFT}} \hspace{3mm} \textrm{ and }\hspace{3mm} C_{\textrm{SL(2)-DFT}}
& \Rightarrow &
X_{\textrm{EFT}} \ ,
\end{array}
\end{equation}
under the assumptions discussed before.

%
%

\small

\bibliography{references}
\bibliographystyle{JHEP} 

\end{document}